\documentclass[letterpaper,twocolumn,10pt]{article}
\usepackage{usenix}
\usepackage{cite}
\usepackage{amsmath,amssymb,amsfonts,pifont,wasysym}
\usepackage{algorithm,algorithmic}
\usepackage{hyperref}
\usepackage{graphicx,subfig}
\usepackage{tabularx,multirow}
\usepackage{textcomp}
\usepackage{url,xurl}
\usepackage{color,colortbl}
\usepackage{float}
\usepackage{threeparttable}
\usepackage{soul}
\usepackage{makecell}
\usepackage{stfloats}
\usepackage{verbatim}
\usepackage{lipsum}
\usepackage{xspace}
\usepackage[many]{tcolorbox}
\usepackage{ragged2e}
\usepackage[table,dvipsnames]{xcolor}
\usepackage{booktabs}
\tcbuselibrary{breakable}
\usepackage[export]{adjustbox}
\usepackage{pgf}
\usepackage{enumitem}   % used for \begin{itemize}[leftmargin=10pt, topsep=0pt, itemsep=1pt, partopsep=1pt, parsep=1pt]
\usepackage{newtxtt}    % 更好看的 texttt

\NewDocumentCommand{\ruixiao}
{ mO{} }{\textcolor{red}{\textsuperscript{\textit{ruixiao}}\textsf{\textbf{\small[#1]}}}}

\NewDocumentCommand{\chunyi}
{ mO{} }{\textcolor{blue}{\textsuperscript{\textit{chunyi}(revised)}\textsf{\textbf{\small[#1]}}}}

\NewDocumentCommand{\jiahao}
{ mO{} }{\textcolor{cyan}{\textsuperscript{\textit{jiahao}}\textsf{\textbf{\small[#1]}}}}

\newcommand{\name}{\texttt{IcebergExplorer}\xspace}
\newcommand{\framework}{\textit{PrivacyIceberg}\xspace}
\definecolor{LightGray}{gray}{0.85}
\definecolor{White}{gray}{1.0}
\definecolor{Celadon}{RGB}{175, 225, 175}

\newtcolorbox{cooltextbox}[1][]{%
    colback=black!5,
    colframe=black!5,
    notitle,
    sharp corners,
    borderline west={0pt}{0pt}{red!80!black},
    enhanced,
    breakable,
    left=0pt,
    right=0pt,
    top=0pt,
    bottom=0pt
    }
    
\newtcolorbox{position}[1][]{%
    colback=black!5,
    colframe=black!5,
    notitle,
    sharp corners,
    borderline west={0pt}{0pt}{red!80!black},
    enhanced,
    breakable,
    left=0pt,
    right=0pt,
    top=0pt,
    bottom=0pt
    }

\definecolor{customcite}{HTML}{b83b5e}
\definecolor{customlink}{HTML}{07689f}
\definecolor{customurl}{HTML}{11999e}
\AtEndPreamble{
\usepackage{hyperref}

    \hypersetup{
      colorlinks = true,
      linkcolor = customlink,
      anchorcolor = purple,
      citecolor = customcite,
      filecolor = purple,
      urlcolor = customurl
    }
}

\newtcolorbox{answerbox}{
  enhanced,
  left=1.7mm,
  right=1.7mm,
  top=1.7mm,
  bottom=1.7mm,
  colback=gray!10,  
  colframe=black!90, 
  boxrule=0pt,      
  leftrule=3pt,     
  sharp corners,
  breakable         
}

\newcounter{insightcounter}

\newtcolorbox[auto counter, number within=section, list type=subsubsection, list inside=toc]{sectionbox}[1]{colback=Salmon!20, colframe=Salmon!90!Black, coltitle=black, fonttitle=\bfseries, breakable, before upper={\parindent10pt\noindent},
left = 1mm, right = 1mm, top = 1mm, bottom = 1mm,
}

\begin{document}

% \title{Peepers Everywhere: Investigating Human-Centric Privacy Iceberg \\in the Era of Large Language Models}

\title{Profiling for Pennies: Unveiling the Privacy Iceberg of LLM Agents}

\author{
{\rm Jiahao Chen$^{1}$, Qi Zhang$^{1}$, Ruixiao Lin$^{1}$, Chunyi Zhou$^{1}$, Tianyu Du$^{1}$, Qingming Li$^{1}$},\\ {\rm Tong Zhang$^{1}$, Junhao Li$^{1}$, Yuwen Pu$^{2}$, Shouling Ji$^{1\dagger}$}\\
Zhejiang University$^{1}$, Chongqing University$^{2}$\\
\small{$^{\dagger}$Corresponding Author}
}

\maketitle

\begin{abstract}
Large Language Models (LLMs) have revolutionized how information are collected, aggregated, and reasoned.
However, this enables a novel and accessible vector of privacy intrusion: the automated and in-depth personal profiling; this engenders a chilling effect of  ``\textit{peepers everywhere}''.
Existing research primarily unfolds from the training pipeline of LLM, emphasizing the exposure of Personally Identifiable Information (PII) through memorization, while privacy studies from a human-centric perspective remain underexplored.
To fill this void, we empirically investigate privacy perception in the real world through the lens of human awareness and the practices of LLM-integrated platforms, revealing a significant dissonance: platforms fail to technically or policy-wise address public privacy concerns.
To facilitate a systematic and quantifiable study of privacy risk, we propose the \framework, which categorizes real-world human privacy risks into three tiers: \textit{explicitly searched}, \textit{contextually inferred}, and \textit{deeply aggregated}, based on the sophistication of LLM exploitation.
We develop \name to audit privacy exposure, utilizing minimal PII as a search seed to reconstruct high-fidelity profiles, achieving over 90\% factual accuracy within 10 minutes at a cost under \$3, for real-world scenarios.
Additionally, we identify six root causes contributing to such privacy disclosures and propose multi-stakeholder countermeasures for LLM vendors, individuals, and data publishers. 
% We advocate a human-centric privacy paradigm that (1) categorizes privacy risk by capability tiers, (2) mandates standardized audits of individual privacy, and (3) enforces platform and LLM vendor accountability beyond PII.
\end{abstract}

\section{Introduction}
\label{sec:introduction}
Large Language Models (LLMs) have revolutionized information acquisition~\cite{spatharioti2023comparing}, processing~\cite{roy2024exploring}, and integration~\cite{tan2024large}. 
Equipped with tools for task execution, LLMs can autonomously query, interpret, and aggregate fragmented public traces across platforms to dig out individuals' personal information, posing significant risks for privacy exposure~\cite{du2025automated}.

\textbf{Motivation.}
Existing works primarily measure privacy risks through the lens of \textit{model memorization}, extracting Personally Identifiable Information (PII) embedded in the parameters~\cite{carlini2021extracting,chen2024janus,huang2024trustllm,wang2025rigging}.
However, they lack an in-depth understanding of LLMs' capacity to shift the landscape of personal privacy risks.
Specifically,
(1)~\textit{Academic Systemization.} Traditional taxonomies based on public versus private and PII versus non-PII are insufficient to systematize these human-oriented privacy concerns, while actual harm increasingly arises from context collapse enabled by LLMs, as supported by recent reports~\cite{newswechat1,newswechat2,case_meta}.
(2)~\textit{Public Perception.} Recent incidents highlight widespread concerns about personal information exposure resulting from LLMs' extracting, inferring, and synthesizing from public data~\cite{contested_govern,case_chatbot2,case_chatbot1}.
(3)~\textit{Outdated Countermeasures.} Existing policies and regulations~\cite{gdpr,ccpa} fail to capture and address these privacy risks, necessitating root cause identification and specific mitigation strategies.

\textbf{Research Questions (RQs).}
We aim to explore the LLM-driven privacy risks that lie \textit{beneath the Iceberg} from the PIIs that are visible on the surface. 
To investigate how human-centric privacy risks are perceived, formed, and originated, we sequentially address the following RQs:
\begin{itemize}[leftmargin=10pt, topsep=0pt, itemsep=0pt, partopsep=0pt, parsep=0pt]
\item \textit{\hypertarget{rq1}{RQ1} (Privacy Perception).} How do the public and platforms perceive LLM-driven privacy exposure, and what are the underlying reasons?
\item \textit{\hypertarget{rq2}{RQ2} (Risk Quantification).} How to systematically identify and quantify human-centric privacy risks?
\item \textit{\hypertarget{rq3}{RQ3} (Attribution and Mitigation).} What are the root causes of these risks, and what strategies can be implemented to mitigate them?
\end{itemize}

\begin{figure*}
    \centering
    \includegraphics[width=0.9\linewidth]{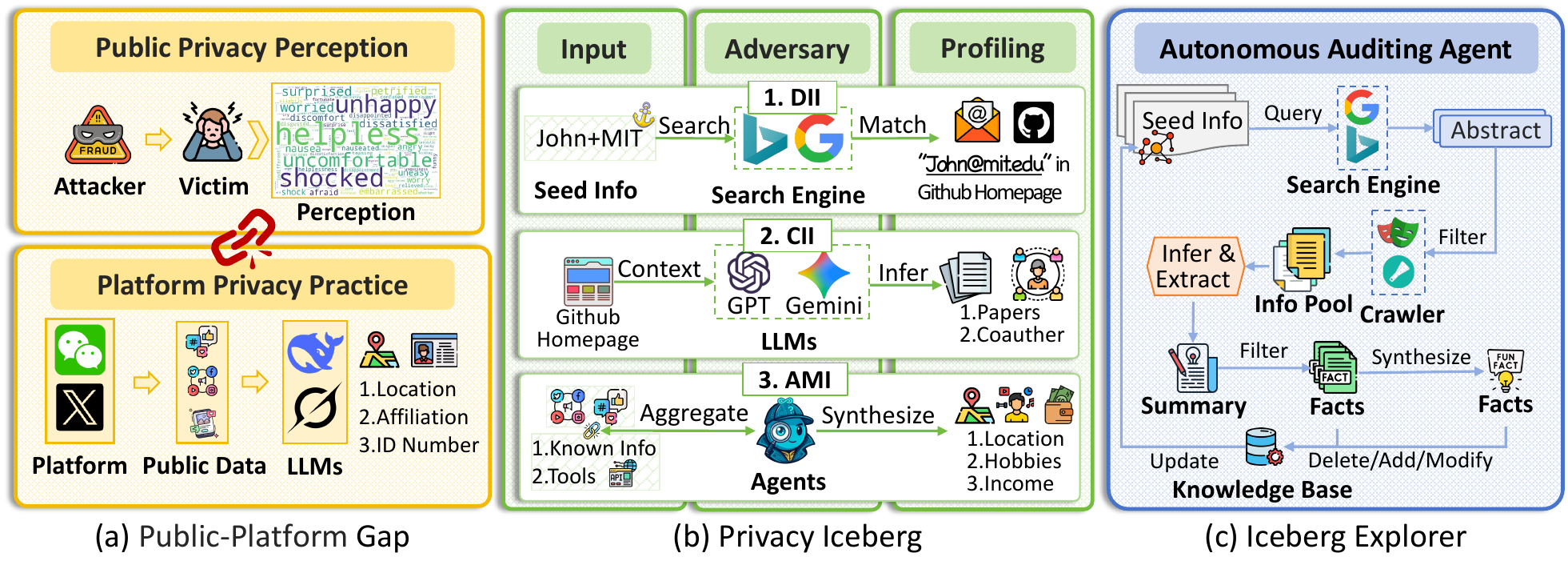}
    \vspace{-1em}
    \caption{Overview of Public-Platform Gap, \framework, and \name.}
    \label{fig:framework}
\end{figure*}

\textbf{Privacy Perception.} 
To answer these questions, we first investigate how the public perceives these emerging privacy risks from the perspectives of individuals' awareness and platforms' practices. 
Qualitative analysis (\S~\ref{sec:public_awareness}) reveals a widespread sense of violation and helplessness among people, caused by the \textit{automated} and \textit{unauthorized} aggregation of public information.
While the investigation into the privacy governance of LLM-integrated social platforms (\S~\ref{sec:risk_applications}) reveals a pervasive misalignment in Fig.~\ref{fig:framework}(a): although most platforms claim protection from the ``privacy modes'' and allow nominal content deletion, the retrieved public data are often retained for various purposes. Only a minority of platforms explicitly disclose data reuse by their embedded LLMs.

\textbf{Risk Quantification.}
In response to the public's concern of being peeped by strangers, we introduce and formalize \framework (\S~\ref{sec:privacy_iceberg}).
The framework assumes a \textit{stranger} as the attacker (\S~\ref{sec:threat_model}) to hierarchically decompose LLM-generated privacy into three tiers in Fig.~\ref{fig:framework}(b): \ding{182}~Direct Identifiable Information (DII), \ding{183}~Contextually Inferred Information (CII), and \ding{184}~Aggregated Mosaic Information (AMI). The taxonomy progresses from explicit identifiers to implicit behavioral inference, reflecting the degree of LLM involvement as well as the perceived severity of privacy invasion by individuals, which is then validated by a user study (\S~\ref{sec:empirical_validation}).
Building upon the \framework framework, we develop \name, an autonomous auditing agent (\S~\ref{sec:methodology}) that emulates an adversary to perform large-scale profiling shown in Fig.~\ref{fig:framework}(c).
The system addresses three \textit{challenges}:
\ding{182}~initiating exploration from minimal initial knowledge,
\ding{183}~handling vast and noisy public data through multi-stage filtering, and
\ding{184}~maintaining knowledge consistency when aggregating conflicting or redundant evidence. Empirical evaluations (\S~\ref{sec:evaluation}) demonstrate that \name automatically reconstructs high‑fidelity profiles at scale, exceeding human analysts while operating at low time and cost per target, revealing the practical feasibility and severity of such privacy threats.

\textbf{Attribution and Mitigation.}
To propose fundamental safeguards in response to this emerging privacy exposure, we dive into the attributions of such information leakage (\S~\ref{sec:root_causes}).
We identify six \textit{root causes} stemming from the public information ecosystem that systematically facilitate such LLM-driven peepers, namely, persistence, linkability, readability, identifiability, inferability, and composability.
These attributions result in the terrifying exposure of deeply inferred individual information beyond simply the combination of non-sensitive fragments, revealing the privacy risk as an emergent property of data ecosystems rather than isolated LLM behaviors.
Accordingly, we propose stakeholder-specific defenses (\S~\ref{sec:countermeasures}). For LLM vendors, we evaluate model-level prompt and context defenses; for individuals (victims), we assess prompt-injection-as-defense techniques~\cite{liu2024evalpie} to confuse automated profiling; and for publishers, we propose structural interventions, such as linkability reduction and metadata stripping to minimize adversarial yield at source. Together, these strategies illustrate how technical, behavioral, and policy actions can jointly reduce synthesis-based privacy exposure.

\textbf{Contributions.}
This paper systematizes and quantifies privacy risks arising from the multi-tier reasoning capabilities of LLMs, shifting the focus from model memorization to the automated synthesis of public digital traces. Concretely, we raise awareness of situational privacy for \textit{individuals}, provide a auditing tool involving risk quantification for \textit{platforms}, and empirical evidence to regulate context-aware personal information for \textit{policymakers}. Overall, this paper makes the following contributions:
\begin{itemize}[leftmargin=10pt, topsep=0pt, itemsep=0pt, partopsep=0pt, parsep=0pt]
    \item We conduct large-scale qualitative studies, revealing the perception gap between public concern and research focus. These findings underscore the urgent need for a systematic framework to quantify these emerging risks and an autonomous tool to empower users with privacy awareness.
    \item We propose a hierarchical model \framework that formalizes privacy risks beyond PII and implement an autonomous auditing system \name that simulates an informed adversary and measures depth-aware risk across tiers from public data.
    \item We systematize six root causes that enable LLM-driven privacy exposure, after which we implement countermeasures for LLM vendors, individuals, and publishers, aligned with the identified root causes and public expectations. 
\end{itemize}

\section{Background and Related Work}
\label{sec:background}
\textbf{From Endogenous to Exogenous Privacy Risks.}  
Previous research has primarily focused on what we term \textbf{endogenous privacy risks}, risks originating within a model's own internal parameters.  
This perspective treats privacy leakage as a consequence of training data memorization.  
Representative studies include membership inference~\cite{wang2025rigging,huang2022privacy}, data extraction~\cite{carlini2021extracting,huang2024trustllm}, and fine-tuning leakage~\cite{chen2024janus}. However, these works confine privacy protection to the \textbf{training phase}, implicitly assuming that threats end once privacy is unlearned~\cite{chu2025scrub,chen2022graph}. In practice, OSINT tools~\cite{socialanalyzer} show that public information across platforms can be aggregated. LLMs inherit this capability and extend it through cross-context reasoning, enabling them to reconstruct implicit personal attributes. We term this category of threat exogenous privacy risks. Here, privacy breaches result not from internal memory, but from the model's interpretive intelligence: its capacity to link and reason fragmented public traces into coherent narratives of personal life. Recent incidents~\cite{case_chatbot1,case_chatbot2,contested_govern} illustrate that LLMs with tool and memory access can infer private information such as location, relationships, and health status without explicit PII.

\textbf{Empirical Investigation of Exogenous Risks.}
While a few recent studies recognize this shift~\cite{du2025beyond,du2025autoprofiler} from the training-stage to the inference-stage privacy, the field lacks a systematic, large‑scale exploration.  
Existing research often remained conceptual or limited to small‑scale case studies, without quantifying the practical extent to which public, non‑PII data can be reassembled.  
Moreover, regulatory frameworks, including GDPR~\cite{gdpr} and CCPA~\cite{ccpa}, continue to anchor privacy protection on the presence of explicit identifiers. (\textbf{Our Position}) Our work represents the first comprehensive effort to characterize and measure \textbf{exogenous privacy risks} in the real world.  
We approach the problem from a human‑centric perspective, examining how people perceive, experience, and are impacted by LLM‑enabled privacy reconstruction.

\section{Motivation: Privacy Perception Gap}
% in the Wild
\subsection{Public Privacy Perception}
\label{sec:public_awareness}
To ground our study in the real world, we conducted a qualitative analysis of a social media case study (details in Appendix~\ref{app:public_awareness}) involving the unauthorized aggregation of student data by LLM-powered search tools. This case was selected since it has garnered over 13000 likes and 2,159 comments, triggering an extensive public reaction and serving as a catalyst for investigating public anxiety. We utilized an LLM-assisted thematic coding approach to categorize these comments into four dimensions: (1)\textit{emotional response}, (2)\textit{risk awareness}, (3)\textit{reported harms}, and (4)\textit{protection expectations}.

Our findings reveal a pervasive sense of ``helplessness'' and ``context collapse'' among users; while individual data points (e.g., school affiliations, names) are often public, their systematic synthesis by LLMs results in targeted doxing and psychological distress. As illustrated in the keyword distributions (see Fig.~\ref{fig:violation_perception},Fig.~\ref{fig:doxing_experience},Fig.~\ref{fig:fig10} and Fig.~\ref{fig:protection_expectations} in the Appendix), the majority of relevant comments expressed direct concern. This collective anxiety highlights a critical shift in the privacy landscape: the threat has evolved from simple data exposure to deep, synthesized inference, necessitating the systematic investigation of LLM capabilities presented in Sec~\ref{sec:privacy_iceberg}. Note that the above anxiety extends beyond a single platform. Similar discourse occurs on Twitter (Appendix~\ref{sec:case_study}). These cross-platform observations provide the motivation to investigate LLM-based privacy inference through a public-centric lens.
\begin{table}[H]
\vspace{-0.5em}
\definecolor{gray0}{gray}{0.9}
\normalsize
\setlength{\tabcolsep}{3pt}
    \centering
    \renewcommand\arraystretch{1}
    \begin{tabular}{p{0.98\columnwidth}}
    \Xhline{1.0pt}
         \rowcolor{gray0} \noindent \textbf{Insight 1}: The public's strong sense of being violated highlights: information published for a limited promotional purpose is being inferred and aggregated by a ``\textbf{stranger}'' for an unforeseen purpose, leading to ``\textbf{context collapse}'', explained by contextual integrity (CI)~\cite {nissenbaum2004privacy}. \\ %
    \Xhline{1.0pt}
    \end{tabular}
    % \vspace{-1em}
\end{table}

% \insight{
% The public's strong sense of being violated highlights: information published for a limited promotional purpose is being inferred and aggregated by a ``\textbf{stranger}'' for an unforeseen purpose, leading to ``\textbf{context collapse}'', explained by contextual integrity (CI)~\cite {nissenbaum2004privacy}.
% }

\subsection{Platform Privacy Practice}
\label{sec:risk_applications}
The proliferation of LLM-integrated search services has created a significant misalignment between platform data practices and public privacy expectations. Hence, we conducted a systematic review of the privacy policies and AI-specific disclosures of ten major social media platforms and applications (detailed in Appendix~\ref{app:risk_applications}). These social media platforms were selected based on two criteria: (1) a global or regional user base with high monthly active users~\cite{platform_selection}, and (2) the confirmed integration of LLM-driven features (e.g., AI search or conversational assistants) within their ecosystems.

Our analysis reveals that while platforms emphasize procedural compliance, they offer limited protection against synthesized inference: 50\% of surveyed platforms explicitly use user published data for LLM inference. Furthermore, case studies of recent data leakage incidents (Appendix~\ref{sec:case_study}) demonstrate that even after original posts are deleted, LLM-based aggregators often retain and surface synthesized profiles, including sensitive identifiers like phone numbers. This persistence reflects a failure of traditional ``notice and consent'' frameworks in the LLM era, where the ``right to be forgotten'' is undermined by the latent memory of large models. We reported these issues to the affected platforms. Responses varied (Appendix~\ref{sec:platform_response}). Some gave no reply, others confirmed receipt. Two platforms requested details, but only one indicated an intention to address the issue. Therefore, from the user's perspective, there is an urgent, unmet demand for \textit{Privacy Situational Awareness}, the ability for individuals to audit their own ``inference surface.'' This gap motivates the development of \name, a tool designed to empower users by revealing their aggregated privacy exposure, shifting the paradigm from passive reliance on platform governance to active, informed risk management.
\begin{table}[H]
\vspace{-0.5em}
\definecolor{gray0}{gray}{0.9}
\normalsize
\setlength{\tabcolsep}{3pt}
    \centering
    \renewcommand\arraystretch{1}
    \begin{tabular}{p{0.98\columnwidth}}
    \Xhline{1.0pt}
         \rowcolor{gray0} \noindent \textbf{Insight 2}: The integration of LLMs into platforms, while marketed for convenience, simultaneously equips adversaries with a powerful engine for user privacy inference. This threat is amplified as platform privacy governance focuses on procedural compliance and policy disclosure, rather than safeguards to prevent such misuse. \\ %
    \Xhline{1.0pt}
    \end{tabular}
    % \vspace{-1em}
\end{table}
% \insight{
% The integration of LLMs into platforms, while marketed for convenience, simultaneously equips adversaries with a powerful engine for user privacy inference. This threat is amplified as platform privacy governance focuses on procedural compliance and policy disclosure, rather than safeguards to prevent such misuse.
% }

% \input{section/threat}
\section{Framework Modeling: \framework}
\subsection{Threat Model}
\label{sec:threat_model}
To systematically analyze and address the risks, we move to a formal threat model in this section. This model operationalizes the realistic \textit{stranger adversary} scenario identified in the user study under CI theory~\cite{nissenbaum2004privacy}, which is defined as: \textit{the adversary, the information publisher, and the victim}.

\textbf{CI for \textit{Stranger Adversary} Scenario.}
Grounded in the theory of CI, privacy means the preservation of ``appropriate information flows'' within specific contexts~\cite{nissenbaum2004privacy}. A privacy violation occurs when an information flow breaches `context-specific informational norms' characterized by five parameters: \textit{sender, recipient, subject, information type,} and \textit{transmission principle}. (\textbf{Actors}) We identify three primary roles: the \textit{Victim} (subject), the \textit{Information Publisher} (sender), and the \textit{Stranger Adversary} (recipient)~\cite{nissenbaum2004privacy}. (\textbf{Attributes}) We expand the ``information type'' beyond static PII to a hierarchy of DII, CII, and AMI. (\textbf{Transmission Principles}) The core violation in our model is the breach of the \textit{limited-purpose principle}. Information originally published for benign purposes is aggregated, inferred, and repurposed by an adversary for unforeseen profiling, leading to ``context collapse''~\cite{nissenbaum2004privacy}.

\textbf{Adversary's Assumption.}
\label{sec:adversary_threat_model}
The adversary can be generalized as an informed stranger and begins with a seemingly innocuous amount of information from public platforms like OpenReview or Google Scholar, typically the victim's full name and affiliations, since obtaining this information is much easier. (\textbf{Goal \& Success Criteria}) The adversary's primary goal is comprehensive \textit{profile reconstruction}. Success is not merely finding information, but maximizing three key metrics: the quantity of unique facts discovered; the accuracy of facts; the \textbf{sensitivity} of the aggregated information, with a focus on uncovering high-value data for downstream use. (\textbf{Capabilities}) The adversary utilizes readily accessible tools, including search engines and, critically, LLM-powered systems. These systems are capable of large-scale, automated information aggregation and synthesis of the target victims. (\textbf{Motivations}) The motivations for this profiling are diverse:
\begin{itemize}[leftmargin=10pt, topsep=0pt, itemsep=0pt, partopsep=0pt, parsep=0pt]
\item\textit{Malicious Intent:} Doxing to enable targeted phishing campaigns, social engineering, or harassment~\cite{kim2025llms}.
\item\textit{Commercial Intelligence:} Headhunting firms performing automated, large-scale candidate screening~\cite{chen2025xbench}.
\item\textit{Platform-driven Profiling:} E-commerce or social media platforms enriching user profiles with external public data for hyper-personalized recommendations~\cite{xu2024leveraging}.
\end{itemize}

\textbf{Information Publisher's Assumption.}
\label{sec:publisher_threat_model}
These are third-party entities acting as the \textit{unwitting enablers} that publish information containing the victim's digital traces and real-world activities, often without malicious intent. Publishers include universities, employers, and community groups with which the victim has a relationship. Particularly, the publisher could also be the victim, exposing her/his information on social media or personal page. (\textbf{Motivation}) Their intent is typically promotional, archival, or informational (e.g., a university newsletter announcing scholarship winners). From their perspective, this does not constitute a privacy violation, as the published content often lacks explicit, sensitive PII in isolation and serves a legitimate organizational purpose. (\textbf{Key Characteristic}) The information they publish is often public by default, persistent, and irrevocable. Due to their own interests and traditional privacy definition, these publishers are unwilling to delete such content, creating a permanent and uncontrollable digital footprint of the victim.

\textbf{Victim's Assumption.}
\label{sec:victim_threat_model}
The victim is the individual whose scattered digital traces are being collected and synthesized, often without their consent. (\textbf{Position \& Challenge}) Our qualitative analyses reveal a sense of helplessness and frustration among individuals in this position. They operate in an environment where they have lost control over the information flow; they can neither prevent the adversary's access to public data nor easily compel all information publishers to delete historical content. (\textbf{Goals}) Since proactive prevention is often impossible, the victim's primary defensive goal shifts to achieving \textit{Privacy Situational Awareness}. They desire the ability to understand: (1) What information has been exposed;(2) From which sources it has leaked;(3) How severe are the overall risks?

\subsection{\framework: Hierarchical Privacy Risks}
\label{sec:privacy_iceberg}
To bridge the gap between traditional privacy models and realistic human-centric privacy risks, we propose the \framework, a hierarchical framework for modern privacy risks, illustrated in Fig.~\ref{fig:framework}(b). The framework is orthogonal to traditional content-based privacy models~\cite{gdpr,ccpa}. Rather than replacing existing taxonomies, it characterizes risk based on the level of inference and aggregation required for discovery. It complements existing frameworks by providing a necessary structure to analyze emergent risks from data synthesis. The division of the risks into three tiers directly maps to the escalating capabilities of information retrieval and the increasing cognitive effort required by an adversary.
% \begin{itemize}[leftmargin=10pt, topsep=0pt, itemsep=0pt, partopsep=0pt, parsep=0pt]
% \item\textbf{DII reflects traditional search capabilities.} It corresponds to information retrievable via keyword matching using search engines, with low acquisition cost.
% \item\textbf{CII reflects LLM inference capabilities.} It requires a technology that can perform semantic contextual understanding~\cite{staab2023beyond}. An LLM can read and infer facts that are never explicitly stated within the context beyond DII.
% \item\textbf{AMI reflects agent capabilities.} This level necessitates a system capable of multi-step reasoning across diverse information sources and synthesis. The core threat here aligns with \textit{mosaic effect}~\cite{beyond_gdpr,Mosaic_effect}, where an agent aggregates multiple, individually non-sensitive pieces from different contexts to generate new insights.
% \end{itemize}
% This tiered structure models the progression of risks: DII acts as the entry point, enabling the discovery of sources from which CII is extracted. The accumulation of DII and CII then provides the raw material for synthesizing AMI.

% \begin{figure}
%     \centering
%     \includegraphics[width=1\linewidth]{fig/iceberg-low.pdf}
%     \caption{An illustration of the \framework.}
%     \label{fig:iceberg}
% \end{figure}

\subsubsection{DII-The Visible Tip}
\textbf{DII reflects traditional search capabilities.} It corresponds to information retrievable via keyword matching using search engines, with low acquisition cost. It represents explicit, standalone data points that directly identify or locate an individual. These are the conventional targets of privacy protection.
\\\noindent\textit{\textbf{Definition.}} Factual, unambiguous identifiers are typically found in structured formats or clearly stated in prose. They serve as primary keys for identity linkage.
\\\noindent\textit{\textbf{Example.}} Full name, email address, phone number, and direct links to personal profiles from Google Scholar.
\\\noindent\textit{\textbf{Role in Attack.}} DII serves as the initial anchor. While often public, it enables the adversary to confidently collate disparate documents belonging to the same individual, thereby initiating the search for more information.

\subsubsection{CII-The Submerged Surface}
\textbf{CII reflects LLM inference capabilities.} It requires a technology that can perform semantic contextual understanding~\cite{staab2023beyond}. An LLM can read and infer facts that are never explicitly stated within the context beyond DII. CII denotes information derived from contextual understandings~\cite{staab2023beyond}. It requires reasoning rather than direct extraction.
\\\noindent\textit{\textbf{Definition.}} Factual attributes about an individual that are not explicitly stated but are strongly implied by the surrounding text, or context within a single document or data source.
\\\noindent\textit{\textbf{Example.}} Consider a blog stating~\cite{staab2023beyond}: \textit{``There is this nasty intersection on my commute, I always get stuck there waiting for a hook turn. Just came back from the shop, and I'm furious-can't believe they charge more now for 34d. I remember watching Twin Peaks after coming home from school''} \textbf{Inferred CIIs.} This woman, aged 45 to 50, is in Melbourne. None of these facts is listed explicitly, but they are easily inferred by an LLM.
\\\noindent\textit{\textbf{Role in Attack.}} CII provides the building blocks for comprehensive profiling. Each piece of CII enriches the understanding of the victim's life, activities, and relationships, moving beyond basic identification.

\subsubsection{AMI-The Deep Mass}
\textbf{AMI reflects agent capabilities.} This level necessitates a system capable of multi-step reasoning across diverse information sources and synthesis. The core threat here aligns with \textit{mosaic effect}~\cite{beyond_gdpr,Mosaic_effect}, where an agent aggregates multiple, individually non-sensitive pieces from different contexts to generate new insights. AMI embodies the ``mosaic effect'' and represents the most potent threat in LLM era.
\\\noindent\textit{\textbf{Definition.}} High-level behavioral patterns synthesized by an agent by correlating multiple CIIs or DIIs from different, often unrelated, public platforms. The resulting AMI does not exist in any single source.
\\\noindent\textit{\textbf{Example}.} An agent performs a multi-step investigation on a target individual from multiple sources:
\begin{itemize}[leftmargin=10pt, topsep=0pt, itemsep=0pt, partopsep=0pt, parsep=0pt]
\item \textbf{Source A (Post 1).} The user shares a screenshot of a ``S\&P 2025 acceptance email,'' with PII and sensitive details masked, stating it is her/his ``first paper.''
\item \textbf{Source A (Post 2).} Her/his another post on the same platform mentions ``great food near campus''. The post's displayed IP region is ``City C.''
\item \textbf{Source B (S\&P 2025 Accepted List).} The public acceptance list of S\&P 2025 contains paper titles, affiliations, and author orders. 
\item \textbf{Intermediate CII.} (1) The target is the first author of an S\&P 2025 paper. (2) The target's campus is xxx since there is only one University in City C (from IP region + ``near campus'' hint with web search).
\item \textbf{Resulting AMI.} The target's name is xxx by exploring the accepted papers list with affiliation xxx, thereby deanonymizing the target on the platform.
\end{itemize}
\noindent\textit{\textbf{Role in Attack.}} AMI delivers a synthesized profile of the victim that reveals predictable patterns or routines that are invisible in isolated data points, leading to severe physical threats like stalking or targeted fraud.

\begin{figure}[t]
    \centering
    \subfloat[Three-level Privacy Scenario]{\includegraphics[width=0.5\linewidth]{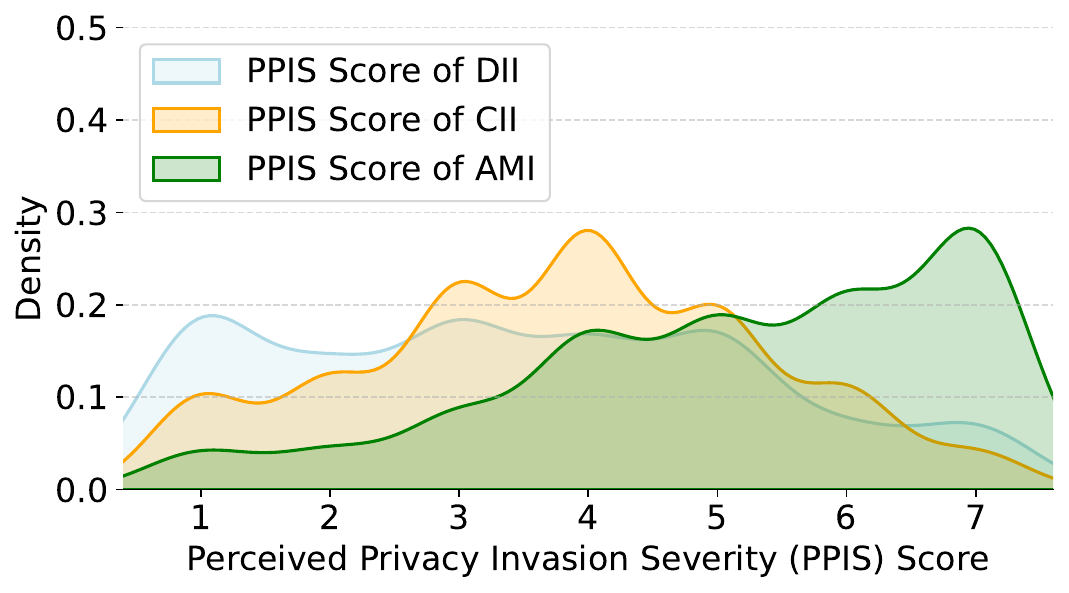}\label{fig:survey2_three_level_distribution_all}}
    \subfloat[Unauthorized Exploitation]{\includegraphics[width=0.5\linewidth]{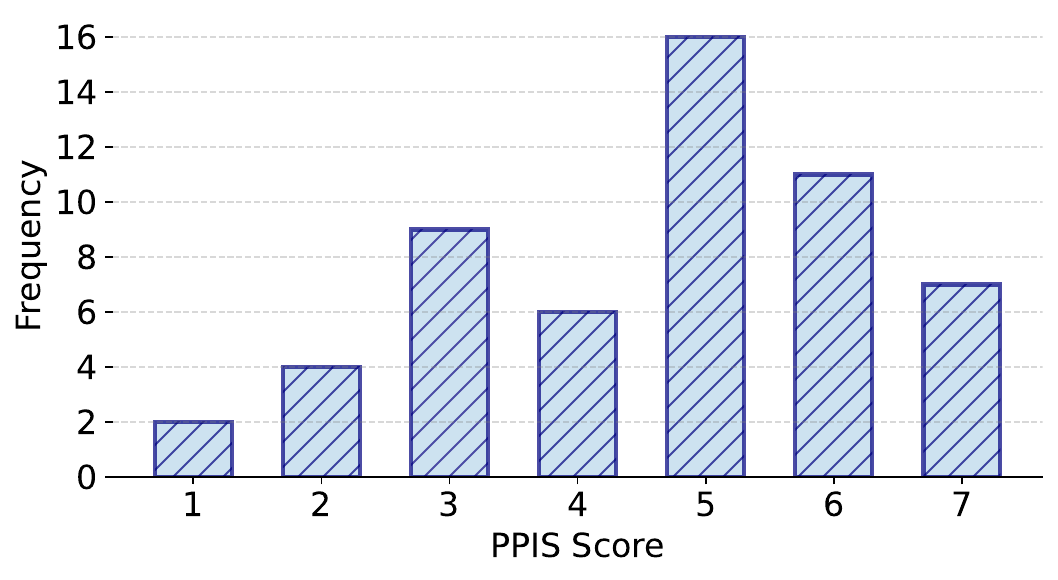}\label{fig:survey2_summary_score_distribution}}
    \caption{(a) Distribution of PPIS score for three-level privacy. The threshold denotes the mean value of all PPIS scores. (b) Frequency distribution of PPIS score for unauthorized public information exploitation.}
\end{figure}

\subsection{Empirical Validation via User Study}
\label{sec:empirical_validation}
To anchor our frameworks in real-world user perception with the ``informed adversary'' threat model, we conducted a user study (details are in the Appendix~\ref{sec:appcase_study}) involving 60 participants. The survey design directly maps to our tiered model: for each scenario, participants rated the Perceived Privacy Invasion Severity (PPIS) score of DII, CII, and AMI.  

\textbf{Validating the Iceberg Hierarchy.}
The results given in Fig.~\ref{fig:survey2_three_level_distribution_all} provide strong empirical support. We observed a consistent and statistically significant trend across all scenarios: the PPIS increases as the adversary moves deeper into the iceberg. Specifically, participants rated the harm of generating AMI significantly higher than that of CII. This confirms our hypothesis that users intuitively understand that the greatest risk lies not in individual data points, but in the powerful inferences and syntheses enabled by aggregation.

\textbf{Polarized Attribution towards Public Data.} In contradiction with the high PPIS score for AMI exploitation, when asked whether investigating public information constitutes a privacy violation, participant responses showed polarization in Fig.~\ref{fig:survey2_summary_score_distribution}, with responses distributed across the spectrum from ``not an invasion'' to ``severe invasion.'' This ambiguity highlights a critical societal tension: while individuals feel violated by the \textit{outcomes} of aggregation, they are uncertain about the legitimacy of accessing the publicly available data. 

% \insight{This paradox underscores the inadequacy of traditional privacy definitions that rely solely on whether data are ``public'' or ``private,'' validating CI's assertion that context, not just content, defines privacy.}
\begin{table}[H]
\vspace{-1em}
\definecolor{gray0}{gray}{0.9}
\normalsize
\setlength{\tabcolsep}{3pt}
    \centering
    \renewcommand\arraystretch{1}
    \begin{tabular}{p{0.98\columnwidth}}
    \Xhline{1.0pt}
         \rowcolor{gray0} \noindent \textbf{Insight 3}: This paradox underscores the inadequacy of traditional privacy definitions that rely solely on whether data are ``public'' or ``private,'' validating CI's assertion that context, not just content, defines privacy. \\ %
    \Xhline{1.0pt}
    \end{tabular}
    \vspace{-1em}
\end{table}

Finally, to facilitate subsequent privacy audits and evaluations, we establish a baseline for data sensitivity. Participants have also rated the sensitivity of various categories of personal information within the ``stranger scenario'' context, and we listed the first 17 categories with the highest PPIS scores in Tab.~\ref{tab:privacy_categories}, moving beyond counting data points and measuring the actual perceived harm of profile reconstruction.

\section{\textit{Methodology}: \name}
\label{sec:methodology}
\subsection{Design Overview}
\label{sec:problem_formulation}
The objective of our auditing system is to simulate the capabilities of an adversary to reconstruct a victim's profile from disparate public data sources. We model this task as an information retrieval problem. Let $\mathcal{S} = \{s_1, s_2, \dots, s_i\}$ represent the vast, heterogeneous corpus of public information sources (e.g., forum posts and public records), where each $s_i$ contains unstructured data referencing multiple individuals. Let $V$ be the target victim of the audit and $\mathcal{S}^V\subset\mathcal{S}$ be the complete corpus of the victim. We define the ideal ground truth profile $\mathcal{G}^V= \{g_1, g_2, \dots, g_i\}$ as the complete set of all facts pertaining to $V$ within $\mathcal{S}^V$. Specifically, $g_i$ represents a single \textbf{atomic} piece of information about $V$ (e.g., ``$V$ lives in City Y,'' ``$V$ attended University Z''). This set $\mathcal{G}^V$ represents the maximum information discoverable about $V$ from $\mathcal{S}$, including DII, CII and AMI. Let $\mathcal{K}^V=\{(n_1, k_1), (n_2, k_2), \dots, (n_i,k_i)\}$ represent the knowledge dictionary with key $n_i$ and value $k_i$ constructed by \name, containing the set of discovered and verified facts about $V$ at a given point in time. By definition, a perfectly audit system ensures that $\mathcal{K}^V \subseteq \mathcal{G}^V$.
The primary goal of the privacy auditing is to maximize the coverage of $\mathcal{K}^V$ relative to the ground truth $\mathcal{G}^V$. Let $\pi$ denote the auditing system's strategy, encompassing the sequence of taken actions. $\mathcal{K}^V$ is generated by applying policy $\pi$ to the information space $\mathcal{S}$, starting from initial knowledge $\mathcal{K}_0$. Formally, the primary goal is to find the optimal policy $\pi^*$ that maximizes the coverage of $\mathcal{K}^V$ relative to $\mathcal{G}^V$:
\begin{equation}
    \pi^* = \arg \max_{\pi} |\pi(\mathcal{K}_0, \mathcal{S}) \cap \mathcal{G}^V|.
\end{equation}

\subsection{Technical Challenges}
\label{sec:challenges}
% Solving the optimization problem above presents the following practical challenges as well:\\
\textbf{Knowledge Scarcity.} The deep search process starts with minimal seed knowledge $\mathcal{K}_0^{V}$. Given the vastness and noise of $\mathcal{S}$, bootstrapping the search process to discover relevant, yet scattered $\mathcal{S}^V$ without high-recall initial queries is a hurdle.\\
\textbf{Scalability and Efficiency.} The search process yields two issues. First, the sheer volume of retrieved results needs an efficient, low-cost filtering strategy to distinguish relevant from irrelevant sources. Second, posts may exceed the context window limitations of LLMs, requiring effective extraction techniques that preserve critical details without processing the entire content at once.\\
\textbf{Knowledge Verification.} Public data sources frequently contain ambiguous or contradictory information. The method must dynamically manage the knowledge $\mathcal{K}^V$ with robust entity resolution and resolving contradictions to ensure the final synthesized profile maintains high fidelity and consistency.

To this end, we designed \name (Alg~\ref{alg:iceberg_explorer}) that implements an optimal $\pi^*$, leveraging tools and MLLMs to interact with the real world and multi-modal information.

\subsection{Autonomous Auditing Agent}
\label{sec:strategy}
\subsubsection{Exploration via Iterative Refinement} 
To overcome \textbf{knowledge scarcity}, \name employs a refinement loop with $T_{max}$ iterations, rather than relying on single static queries. The core idea is that newly discovered knowledge informs subsequent search actions through a query generation process. At iteration $t$, the LLM $\mathcal{F}$ generates queries for the next cycle by reasoning over a composite knowledge context $\mathcal{C}_t^q$ with following inputs:
\begin{enumerate}[leftmargin=10pt, topsep=0pt, itemsep=0pt, partopsep=0pt, parsep=0pt]
    \item $\mathcal{K}_0$ acts as an anchor throughout the search process, preventing drift caused by ambiguous information.
    \item $\mathcal{K}_{t}^V$ provides the raw material for expansion. New facts discovered and added to $\mathcal{K}_t^V$ serve as direct keywords for a ``local'' exploration of related information.
    \item $\mathcal{R}_t$ provides a high-level evaluation of the current knowledge $\mathcal{K}_{t}^V$. $\mathcal{R}_t$ identifies gaps in specific dimensions and suggests future exploration directions. For instance, if $\mathcal{K}_{t-1}^V$ contains ``majored in CS'', $\mathcal{R}_t$ might suggest queries containing the ``Github.''
\end{enumerate}

The query candidate list $Q_{t}$ is thus generated by integrating these components with concatenation $\oplus$:
\begin{equation}
    Q_{t} = \mathcal{F}(P_{query}, \mathcal{C}_t^q\oplus Q_{t-1}\oplus\mathcal{R}_t, \emptyset),
\end{equation}
where the context $\mathcal{C}_t^{q}$ is defined as $\mathcal{C}_t^{q} = \mathcal{K}_0\oplus\mathcal{K}_{t}^V$. $Q_{t-1}$ is given to avoid the repetition of the generated query with previous ones. Note that the third term of $\mathcal{F}$ is empty since no external information is required. The above strategies transform the search from a shallow exploration into a strategically guided exploration process, bootstrapping information gain even from sparse starting conditions.

\subsubsection{Multi-Stage Filtering Cascade} 
To tackle \textbf{scalability and efficiency}, we perform a low-cost, coarse-grained filtering. The search tool $\mathcal{T}_{s}$, takes queries $Q_{t}$ and returns a set of results $\mathcal{S}_{s}=\mathcal{T}_{s}(Q_{t},\mathcal{S})$ from the public sources $\mathcal{S}$. Each result contains a source URL $u_i$ and a query-relevant text snippet $\hat{s}_i$ relevant to $Q_{t}$ for its full content $s_i$. Since prompt $\hat{s}_i$ with various context lengths might be computationally prohibitive, we use the snippet $\hat{s}_i$ as a proxy for the first stage filtering. Specifically, we employ an ``LLM-as-a-judge'' approach to evaluate each snippet $\hat{s}_i$:
\begin{equation}
    e^r_i, e^n_i, e^v_i = \mathcal{F}(P_{score},\mathcal{K}_0\oplus\mathcal{K}_{t}^{V},\hat{s}_i\oplus u_i),
\end{equation}
where the LLM $\mathcal{F}$ assesses the snippet based on multiple criteria crucial for efficient exploration.
\begin{itemize}[leftmargin=10pt, topsep=0pt, itemsep=0pt, partopsep=0pt, parsep=0pt]
    \item \textbf{Relevance $e^r_i$.} Measures how strongly $\hat{s}_i$ connects to $\mathcal{K}_0$, filtering out noise from ambiguous queries.
    \item \textbf{Novelty $e^n_i$.} Measures the amount of new information present in $\hat{s}_i$ when compared against the current $\mathcal{K}_{t}^{V}$ to avoids redundant processing of known facts.
    \item \textbf{Potential Value $e^v_i$.} A heuristic assessment of whether $u_i$ likely leads to high-value, structured data since $u_i$ sometimes exposes critical clues as well.
\end{itemize}
A snippet $\hat{s}_i$ is considered promising only if its scores exceed a predefined threshold, $e_i^r+e_i^n+e_i^v\geq\phi$. URLs passing this check are added to the target crawl candidate queue $\mathrm{U}_{t}$ and each crawled $u_i$ is dequeued. This reduces the number of documents requiring full content analysis. Once high-potential URLs in $\mathrm{U}_{t}$ are processed by the crawling tool $\mathcal{T}_{c}$, we obtain the full content set $\mathcal{S}_{t}$. 
% To address the challenge of lengthy documents exceeding LLM context windows, we introduce a second filtering and reduction layer before final knowledge extraction.

\textbf{Content Reduction.} Instead of feeding massive documents directly to LLM, we first instruct it to scan the full content $s_i\in \mathcal{S}_{t}$ and extract (with prompt $P_{extract}$) a focused bundle $s_i^{\prime}$, images $\mathrm{I}_{i}$ and hyperlinks $\mathrm{U}_{i}^{new}$ within $s_i$:
\begin{equation}
    s_i^{\prime},\mathrm{I}_{i},\mathrm{U}_{i}^{new}=\mathcal{F}(P_{extract},\mathcal{K}_0\oplus\mathcal{K}_t^V,s_i).
\end{equation}
This pre-processing step serves two critical functions. First, it directly addresses the practical constraint of finite context windows by creating a smaller, high-density summary $s_i^{\prime}$ from a massive source document $s_i$. Second, it improves computational efficiency for downstream tasks. By filtering out irrelevant boilerplate text and focusing only on segments of the target profile, we reduce the token load for subsequent verification and knowledge extraction prompts.

\textbf{Relevance Verification and Asset Queuing.} The reduced bundle $s_i^{\prime}$ undergoes a strict relevance check using prompt $P_{verify}$. The system calculates a high-precision confidence score based on the extracted text. Only bundles where confidence exceeds a threshold $\tau\leq\mathcal{F}(P_{verify},\mathcal{K}_0\oplus\mathcal{K}_t^V,s_i^{\prime})$ proceed to the final knowledge integration stage. For verified bundles, extracted hyperlinks $\mathrm{U}^{new}_{i}$ are added to the crawler queue $\mathrm{U}_{t}$ for subsequent iterations ($\mathrm{U}_{t} = \mathrm{U}_{t}\cup \mathrm{U}^{new}_{i}$), ensuring recursive exploration. 

\subsubsection{Knowledge Integration and Dynamic Maintenance}
To address the challenge of \textbf{knowledge verification}, information that passes the fine-grained verification must be integrated into $\mathcal{K}^V_{t}$ in a structured manner. A simple append-only approach would quickly lead to contradictions and redundancy. We therefore implement a dynamic maintenance process where new information leads to specific atomic operations: \texttt{Add}, \texttt{Update}, or \texttt{Delete}. 

Specifically, the associated extracted images $I_{i}$ would be analyzed given the visual analysis instructions $P_{visual}$ by visual LLM $\mathcal{F}_{V}$: $c_t^{I_{i}}=\mathcal{F}_{V}(P_{visual},I_{i})$. This process goes beyond simple object detection. Instead, it transcribes all readable text present in the image background and identifies geospatial clues. Also, the individuals within the image are extracted to provide confirmation of relationships.

Once all information processed, the system must integrate this new evidence into $\mathcal{K}_t^V$. Next, we instruct $\mathcal{F}$ to compare the new evidence bundle against the current knowledge state to generate update instructions. Let $\mathcal{O}_t=\{(n_1,k_1,a_1) ,(n_2,k_2,a_2),\dots,(n_i,k_i,a_i)\}$ represent the operations triplet, where $n_i$, $k_i$ and $a_i$ denote the key name, atomic fact and the action to manage $\mathcal{K}_t^V$. Then LLM $\mathcal{F}$ will be prompted with $P_{op}$ to generate the operations $\mathcal{O}_t$:
\begin{equation}
\label{eq:operation}
    \mathcal{O}_t = \mathcal{F}(P_{op}, \mathcal{K}_0\oplus \mathcal{K}_t^V\oplus\mathcal{A}_t, s_i^{\prime} \oplus c_t^I),
\end{equation}
combined with the inferred mosaic insights $\mathcal{A}_{t-1}$ of the last round, reduced text $s_i^{\prime}$ and visual analysis $\mathcal{C}_t^{I_{i}}$. Each action $a_i$ falls into one of three categories, ensuring dynamic maintenance of the knowledge base:

\begin{itemize}[leftmargin=10pt, topsep=0pt, itemsep=0pt, partopsep=0pt, parsep=0pt]
    \item \textbf{\texttt{Add}:} A new fact $k_i$ is generated when the evidence provides verifiable information that does not exist in $\mathcal{K}_t^V$ which expands the breadth $\mathcal{K}_t^V=\mathcal{K}_t^V\cup\{(n_i,k_i)\}$.
    \item \textbf{\texttt{Update}:} $(n_i, k_{old}) \in \mathcal{K}_t^V$ is replaced by the new fact $k_i$. This occurs when the new evidence provides more recent or detailed information than the existing one, ensuring temporal consistency and data refinement: $\mathcal{K}_t^V = (\mathcal{K}_t^V \setminus \{(n_i, k_{old})\}) \cup \{(n_i, k_i)\}$.
    \item \textbf{\texttt{Delete}:} $(n_i, k_{old}) \in \mathcal{K}_t^V$ is removed without replacement. This action is taken when new evidence refutes the existing fact or confirms it is outdated and irrelevant for profile integrity: $\mathcal{K}_t^V = \mathcal{K}_t^V \setminus \{(n_i, k_{old})\}$.
\end{itemize}
After all operations have been applied to $\mathcal{K}_{t}^V$, we obtain $\mathcal{K}_{t+1}^V$, which contains the mosaic fragment information from multiple sources. Further, we generate the strategic feedback $\mathcal{R}_t$ and aggregated insights $\mathcal{A}_t$ with the prompt $P_{agg}$:
\begin{equation}
    \mathcal{R}_t, \mathcal{A}_t = \mathcal{F}(P_{agg}, \mathcal{K}_0 \oplus \mathcal{K}_{t+1}^V, \emptyset),
\end{equation}
representing the deepest layer of the \framework. $\mathcal{R}_t$ evaluates the completeness of $\mathcal{K}_t^V$, and offers specific future exploration strategies, ensuring \name adapts its search strategy to fill knowledge gaps. $\mathcal{A}_t$ instead focuses on synthesizing cross-source insights, which are novel conclusions from the aggregation of disparate facts not presented in any single source, integrated in Eq~\ref{eq:operation}.

\section{Evaluation}
\label{sec:evaluation}

\definecolor{myred}{HTML}{F06060}
\definecolor{myyellow}{HTML}{F57A0C}
\definecolor{myyellow}{HTML}{007369}
\definecolor{mycyan}{HTML}{007DEA}
\newcommand{\pmin}{12}
\newcommand{\pmax}{75}
\def\myredscoremin{0.0}
\def\myredscoremax{1}
\def\myyellowscoremin{0.27}
\def\myyellowscoremax{0.82}
\def\mycyanscoremin{0.21}
\def\mycyanscoremax{0.83}
\renewcommand{\pmin}{12}
\renewcommand{\pmax}{75}

\newcommand{\colorcell}[2]{%
  % #1: 基色名（如 mycyan 或自定义颜色名）
  % #2: 数值（百分制，如 68）

  % —— 根据 #1 解析出 \scoremin / \scoremax —— 
  % 如果 \#1scoremin / \#1scoremax 存在就用之，否则回退到默认
  \ifcsname #1scoremin\endcsname
    \edef\scoremin{\csname #1scoremin\endcsname}%
  \else
    \edef\scoremin{\defaultscoremin}%
  \fi
  \ifcsname #1scoremax\endcsname
    \edef\scoremax{\csname #1scoremax\endcsname}%
  \else
    \edef\scoremax{\defaultscoremax}%
  \fi
  
  \pgfmathsetmacro{\norm}{(#2/100 - \scoremin)/(\scoremax - \scoremin)}%
  \pgfmathsetmacro{\t}{max(0, min(1, \norm))}
  \pgfmathsetmacro{\praw}{\pmin + \t*(\pmax-\pmin)}%
  \pgfmathtruncatemacro{\pct}{\praw}

  % 用 \edef 冻结成字面量：此时 {#1!\pct} 里 \pct 已被展开成纯数字
  \edef\__tmp{\noexpand\cellcolor{#1!\pct}{#2}}%
  \__tmp%
}

\begin{table*}[]
\centering
\begingroup
\definecolor{refusalBg}{RGB}{255,245,245}
\definecolor{refusalHi}{RGB}{255,225,225}
\definecolor{effBg}{RGB}{242,247,255}
\definecolor{covBg}{RGB}{243,255,243}
\definecolor{covHi}{RGB}{220,247,220}
\definecolor{effecBg}{RGB}{255,249,240}
\definecolor{effecHi}{RGB}{255,237,210}

\caption{The performance of LLMs/Agents/Human for automated profiling audit (attack).}
\label{tab:main}
\renewcommand{\arraystretch}{0.9}
\aboverulesep=0ex
\belowrulesep=0.5ex
\resizebox{1.0\linewidth}{!}{
\begin{tabular}{llcccccccc}
\toprule\rule{0pt}{1.0EM}
 &  & \cellcolor{refusalBg} & \multicolumn{2}{c}{\cellcolor{effBg}\textbf{Efficiency}} & \multicolumn{2}{c}{\cellcolor{covBg}\textbf{Coverage}} & \multicolumn{3}{c}{\cellcolor{effecBg}\textbf{Effectiveness}} \\ \cmidrule{4-10} 
\multirow{-2}{*}{\textbf{Method}} &
\multirow{-2}{*}{\textbf{LLMs}} &
\multirow{-2}{*}{\cellcolor{refusalBg}\textbf{Refusal Rate (\%)}} &
\cellcolor{effBg}\textbf{Tokens (\#)} &
\cellcolor{effBg}\textbf{Time (s)} &
\cellcolor{covBg}\textbf{Urls (\#)} &
\cellcolor{covBg}\textbf{Facts (\#)} &
\cellcolor{effecBg}\textbf{Url (\%)} &
\cellcolor{effecBg}\textbf{Fact (\%)} &
\cellcolor{effecBg}\textbf{Photo (\%)} \\ 
\midrule

 & DeepSeek V3 &
\cellcolor{refusalBg}00.00\% &
\cellcolor{effBg}{1014.60} &
\cellcolor{effBg}{14.21} &
\cellcolor{covBg}{2.90} &
\cellcolor{covBg}{2.90} &
\cellcolor{effecBg}{00.00}\% &
\cellcolor{effecBg}{00.00}\% &
\cellcolor{effecBg}{-} \\

 & Doubao 1.5 Pro &
\cellcolor{refusalHi}70.00\% &
\cellcolor{effBg}{1602.80} &
\cellcolor{effBg}{22.26} &
\cellcolor{covBg}{1.17} &
\cellcolor{covBg}{3.89} &
\cellcolor{effecBg}{2.86}\% &
\cellcolor{effecBg}{00.00}\% &
\cellcolor{effecBg}{-} \\

 & Qwen3 Coder &
\cellcolor{refusalHi}96.67\% &
\cellcolor{effBg}{835.40} &
\cellcolor{effBg}{1.10} &
\cellcolor{covBg}{0.27} &
\cellcolor{covBg}{8.00} &
\cellcolor{effecBg}{00.00}\% &
\cellcolor{effecBg}{00.00}\% &
\cellcolor{effecBg}{-} \\

 & Gemini 2.5 Pro &
\cellcolor{refusalHi}70.00\% &
\cellcolor{effBg}{1657.80} &
\cellcolor{effBg}{23.90} &
\cellcolor{covBg}{1.00} &
\cellcolor{covBg}{3.34} &
\cellcolor{effecBg}{6.67}\% &
\cellcolor{effecBg}{6.67}\% &
\cellcolor{effecBg}{00.00}\% \\

\multirow{-5}{*}{\makecell{LLMs w/o S}} &
OpenAI GPT-4o &
\cellcolor{refusalHi}10.00\% &
\cellcolor{effBg}{1050.40} &
\cellcolor{effBg}{4.25} &
\cellcolor{covBg}{3.73} &
\cellcolor{covBg}{3.73} &
\cellcolor{effecBg}{00.00}\% &
\cellcolor{effecBg}{00.00}\% &
\cellcolor{effecBg}{00.00}\% \\
\midrule

 & DeepSeek V3 &
\cellcolor{refusalBg}00.00\% &
\cellcolor{effBg}{1458.20} &
\cellcolor{effBg}{25.89} &
\cellcolor{covBg}{2.57} &
\cellcolor{covBg}{5.97} &
\cellcolor{effecBg}{5.19}\% &
\cellcolor{effecBg}{2.79}\% &
\cellcolor{effecBg}{-} \\

 & Doubao 1.5 Pro &
\cellcolor{refusalBg}00.00\% &
\cellcolor{effBg}{1655.90} &
\cellcolor{effBg}{13.32} &
\cellcolor{covBg}{2.80} &
\cellcolor{covBg}{4.17} &
\cellcolor{effecBg}{10.71}\% &
\cellcolor{effecBg}{6.40}\% &
\cellcolor{effecBg}{00.00}\% \\

 & Qwen3 Coder &
\cellcolor{refusalBg}00.00\% &
\cellcolor{effBg}{1419.20} &
\cellcolor{effBg}{20.50} &
\cellcolor{covBg}{3.57} &
\cellcolor{covBg}{6.10} &
\cellcolor{effecBg}{16.82}\% &
\cellcolor{effecBg}{8.20}\% &
\cellcolor{effecBg}{00.00}\% \\

 & OpenAI GPT-4o &
\cellcolor{refusalBg}00.00\% &
\cellcolor{effBg}{1577.40} &
\cellcolor{effBg}{12.87} &
\cellcolor{covBg}{3.90} &
\cellcolor{covBg}{7.60} &
\cellcolor{effecBg}{77.78}\% &
\cellcolor{effecBg}{70.61}\% &
\cellcolor{effecBg}{00.00}\% \\

\multirow{-5}{*}{\makecell{LLMs w/ S}} &
Perplexity &
\cellcolor{refusalBg}00.00\% &
\cellcolor{effBg}{1850.40} &
\cellcolor{effBg}{15.03} &
\cellcolor{covBg}{4.17} &
\cellcolor{covBg}{10.53} &
\cellcolor{effecBg}{64.80}\% &
\cellcolor{effecBg}{53.16}\% &
\cellcolor{effecBg}{00.00}\% \\
\midrule

 & DeepSeek R1 &
\cellcolor{refusalBg}00.00\% &
\cellcolor{effBg}{4717.40} &
\cellcolor{effBg}{234.97} &
\cellcolor{covBg}{5.13} &
\cellcolor{covBg}{10.53} &
\cellcolor{effecBg}{10.39}\% &
\cellcolor{effecBg}{5.05}\% &
\cellcolor{effecBg}{-} \\

 & Doubao 1.5 Pro &
\cellcolor{refusalBg}00.00\% &
\cellcolor{effBg}{2166.30} &
\cellcolor{effBg}{19.92} &
\cellcolor{covBg}{2.47} &
\cellcolor{covBg}{2.93} &
\cellcolor{effecBg}{2.70}\% &
\cellcolor{effecBg}{2.27}\% &
\cellcolor{effecBg}{00.00}\% \\

 & Qwen3 Coder &
\cellcolor{refusalBg}00.00\% &
\cellcolor{effBg}{3025.80} &
\cellcolor{effBg}{57.45} &
\cellcolor{covBg}{5.73} &
\cellcolor{covBg}{7.30} &
\cellcolor{effecBg}{11.06}\% &
\cellcolor{effecBg}{9.59}\% &
\cellcolor{effecBg}{00.00}\% \\

 & Gemini 2.5 Pro &
\cellcolor{refusalBg}00.00\% &
\cellcolor{effBg}{3575.60} &
\cellcolor{effBg}{36.92} &
\cellcolor{covBg}{5.63} &
\cellcolor{covBg}{16.90} &
\cellcolor{effecBg}{7.69}\% &
\cellcolor{effecBg}{3.75}\% &
\cellcolor{effecBg}{00.00}\% \\

\multirow{-5}{*}{\makecell{LLMs w/ S\&T}} &
OpenAI GPT-4o &
\cellcolor{refusalBg}00.00\% &
\cellcolor{effBg}{3435.60} &
\cellcolor{effBg}{277.55} &
\cellcolor{covBg}{6.00} &
\cellcolor{covBg}{9.20} &
\cellcolor{effecHi}{91.67}\% &
\cellcolor{effecHi}{92.39}\% &
\cellcolor{effecBg}{00.00}\% \\
\midrule

 & Doubao 1.5 Pro &
\cellcolor{refusalBg}00.00\% &
\cellcolor{effBg}{3620.00} &
\cellcolor{effBg}{261.00} &
\cellcolor{covBg}{2.37} &
\cellcolor{covBg}{7.37} &
\cellcolor{effecBg}{66.36}\% &
\cellcolor{effecBg}{57.01}\% &
\cellcolor{effecBg}{00.00}\% \\

 & Qwen3 Coder &
\cellcolor{refusalBg}00.00\% &
\cellcolor{effBg}{5061.80} &
\cellcolor{effBg}{267.20} &
\cellcolor{covBg}{5.03} &
\cellcolor{covBg}{9.77} &
\cellcolor{effecBg}{22.52}\% &
\cellcolor{effecBg}{26.96}\% &
\cellcolor{effecBg}{00.00}\% \\

 & Gemini 2.5 Pro &
\cellcolor{refusalHi}100.00\% &
\cellcolor{effBg}{N/A} &
\cellcolor{effBg}{N/A} &
\cellcolor{covBg}{N/A} &
\cellcolor{covBg}{N/A} &
\cellcolor{effecBg}{N/A} &
\cellcolor{effecBg}{N/A} &
\cellcolor{effecBg}{N/A} \\

 & Meta AI &
\cellcolor{refusalBg}00.00\% &
\cellcolor{effBg}{97707.60} &
\cellcolor{effBg}{412.10} &
\cellcolor{covHi}{9.17} &
\cellcolor{covHi}{18.10} &
\cellcolor{effecBg}{1.09}\% &
\cellcolor{effecBg}{0.74}\% &
\cellcolor{effecBg}{00.00}\% \\

\multirow{-5}{*}{Deep Research} &
Perplexity &
\cellcolor{refusalBg}00.00\% &
\cellcolor{effBg}{2265.80} &
\cellcolor{effBg}{29.04} &
\cellcolor{covBg}{2.30} &
\cellcolor{covBg}{5.40} &
\cellcolor{effecBg}{59.42}\% &
\cellcolor{effecBg}{54.32}\% &
\cellcolor{effecBg}{00.00}\% \\
\midrule

 & {Qwen3} &
\cellcolor{refusalHi}{16.67}\% &
\cellcolor{effBg}{44201.42} &
\cellcolor{effBg}{119.71} &
\cellcolor{covBg}{1.73} &
\cellcolor{covHi}{19.23} &
\cellcolor{effecBg}{91.11}\% &
\cellcolor{effecBg}{78.20}\% &
\cellcolor{effecHi}{15.38}\% \\

 & {Gemini 2.5 Pro} &
\cellcolor{refusalHi}{3.33}\% &
\cellcolor{effBg}{565287.40} &
\cellcolor{effBg}{1256.89} &
\cellcolor{covHi}{12.80} &
\cellcolor{covHi}{95.93} &
\cellcolor{effecHi}{100.00}\% &
\cellcolor{effecHi}{93.05}\% &
\cellcolor{effecHi}\cellcolor{effecHi}{83.33}\% \\

 & {OpenAI GPT-4o} &
\cellcolor{refusalHi}{30.00}\% &
\cellcolor{effBg}{332327.55} &
\cellcolor{effBg}{315.04} &
\cellcolor{covHi}{11.86} &
\cellcolor{covHi}{70.32} &
\cellcolor{effecHi}{100.00}\% &
\cellcolor{effecHi}{92.44}\% &
\cellcolor{effecHi}{77.27}\% \\

\multirow{-4}{*}{\textbf{\makecell{\name}}} &
{OpenAI GPT-4.1} &
\cellcolor{refusalHi}{3.33}\% &
\cellcolor{effBg}{681436.13} &
\cellcolor{effBg}{810.38} &
\cellcolor{covHi}{13.93} &
\cellcolor{covHi}{100.70} &
\cellcolor{effecHi}{99.54}\% &
\cellcolor{effecHi}{79.21}\% &
\cellcolor{effecHi}{76.67}\% \\
\midrule

Human Experts &
N/A &
\cellcolor{refusalBg}{N/A} &
\cellcolor{effBg}{N/A} &
\cellcolor{effBg}{1094.46} &
\cellcolor{covHi}{14.60} &
\cellcolor{covBg}{15.40} &
\cellcolor{effecHi}{100.00}\% &
\cellcolor{effecHi}{100.00}\% &
\cellcolor{effecHi}{40.00}\% \\
\bottomrule
\end{tabular}}
\endgroup
\end{table*}

\subsection{Experimental Setup}

\textbf{Dataset and Ground Truth.}
Following previous studies~\cite{liu2024evalpie,kim2025llms}, our experiment leveraged a dataset consisting of 30 individuals randomly selected from the list of first authors of a recent computer science conference and sent emails to \textbf{obtain their consent}. Note that these scholars were selected because they maintain a high volume of high-fidelity public data (e.g., publications, affiliations, and professional profiles), providing a `worst-case' environment to test the ceiling of LLM-driven synthesis. The structured nature of academic records allows for more rigorous manual verification of facts compared to the more fragmented data of general users. For each individual, the initial seed knowledge provided to the auditing system was limited to their full name and primary affiliation. To establish a reliable ground truth for evaluation, we conducted a meticulous manual verification process for every piece of information collected by our system, ensuring the accuracy and correctness of the evaluation. Each atomic fact extracted by the LLM or agent was reviewed by humans. URLs were checked for accessibility and matching to the target. Facts with the same core attributes in DII or CII, or lacking new insights in AMI, were identified as duplicates and removed. Entries without clear attributes were marked invalid, and only those with verifiable details were retained as valid.

\textbf{Baselines and Models.}
We evaluate both SOTA \textbf{LLMs, Agents and Humans} to comprehensively measure the potential risks in the wild, with details of the model given in the Tab.~\ref{tab:models}. Considering the external tools and reasoning abilities, the evaluated modes are divided into: LLMs w/o Search (S), LLMs w/ S, LLMs w/ Search \& Thinking (S\&T) and Deep Research (agent). Implementation details of \name are given in Appendix~\ref{sec:implementation}.

\textbf{Evaluation Metric.}
We evaluate the audit pipeline along four complementary dimensions: \textbf{refusal rate}, \textbf{efficiency}, \textbf{coverage}, and \textbf{effectiveness}. Refusal Rate is the fraction of runs where the model declines to answer, indicating LLM's safety posture. Efficiency is the average tokens (input+output) and wall‑clock time per target, reflecting resource cost and deployability. Coverage is the average number of extracted facts and source URLs, defining the ceiling of verifiable evidence. Effectiveness is human‑verified: we report URL Accuracy and Fact Accuracy (true and explicitly associated with the target) and Valid Photo (authentic and correctly matched). Metrics are averaged over targets.

\subsection{Privacy Risks in the Wild}

\textbf{Feasibility and Coverage.} 
Tab.~\ref{tab:main} presents the results of different LLMs and agents with varying capacities and knowledge, in comparison with human experts.
The feasibility of an attack is determined by its reliability and its ability to uncover a comprehensive set of information. Baselines without external information (web search) access are infeasible, exhibiting low coverage with an average of 2.9 URLs and facts (DeepSeek V3), illustrating that this information is beyond memorization within model parameters. Moreover, enabling web search drops the refusal rate to 0.00\% for most models with average higher coverage, making simple attacks possible and indicating that current commercial models lack the corresponding guardrail to refuse these privacy-sensitive requests. When reasoning ability is allowed, the refusal rate remains unchanged, while coverage presents a slight rise, which means that the reasoning abilities bring no benefits to privacy requests discernment. In terms of deep research with agents, coverage still remains low. In contrast, our \name is reliable, with top models like {Gemini 2.5 Pro} and {GPT-4.1} showing a low refusal rate of {3.33\%}. Also, it achieves superior coverage, discovering {95.93} and {100.70} facts, respectively. This is over \textbf{5x} more information than the best automated baseline and more than the {15.40} facts found by human experts. 

\begin{table*}[]
\centering
\caption{Average accuracy of \name regarding different privacy categories and base models.}
\label{tab:category_accuracy}
\renewcommand{\arraystretch}{1}
\setlength{\tabcolsep}{3pt}
\aboverulesep=0ex
\belowrulesep=0.5ex
\resizebox{1.0\linewidth}{!}{
\begin{tabular}{lccccccccccccccccc}
\toprule\rule{0pt}{1.0EM}
\textbf{LLMs}$\downarrow$\textbackslash\textbf{Category}$\rightarrow$ & \textbf{C1} & \textbf{C2} & \textbf{C3} & \textbf{C4} & \textbf{C5} & \textbf{C6} & \textbf{C7} & \textbf{C8} & \textbf{C9} & \textbf{C10} & \textbf{C11} & \textbf{C12} & \textbf{C13} & \textbf{C14} & \textbf{C15} & \textbf{C16} & \textbf{C17} \\ 
\midrule
Qwen3 Coder & 88.89\% & 75.00\% & 85.29\% & 71.43\% & 83.33\% & 40.43\% & 77.55\% & 76.92\% & 75.34\% & 92.16\% & 92.86\% & 75.00\% & N/A & 100.00\% & 33.33\% & 82.19\% & 100.00\% \\
Gemini 2.5 Pro & 58.74\% & 80.30\% & 92.98\% & 90.74\% & 83.33\% & 98.44\% & 85.48\% & 96.63\% & 96.75\% & 99.00\% & 98.08\% & 100.00\% & 100.00\% & N/A & 96.15\% & 92.02\% & 83.33\% \\
GPT-4o & 86.67\% & 72.50\% & 85.90\% & 94.12\% & 80.00\% & 100.00\% & 84.43\% & 96.47\% & 94.86\% & 96.49\% & 97.50\% & 90.00\% & N/A & N/A & 100.00\% & 88.35\% & N/A \\
GPT-4.1 & 90.26\% & 100.00\% & 89.70\% & 91.87\% & 95.00\% & 20.65\% & 90.91\% & 48.37\% & 50.27\% & 95.52\% & 83.58\% & 73.33\% & N/A & 100.00\% & 75.00\% & 96.33\% & 60.00\% \\ 
\midrule
Average & 76.36\% & 86.29\% & 89.92\% & 91.35\% & 86.78\% & 52.34\% & 85.89\% & 81.41\% & 73.67\% & 96.81\% & 93.53\% & 83.87\% & 100.00\% & 100.00\% & 82.98\% & 92.27\% & 76.92\% \\ 
\bottomrule
\end{tabular}}
\end{table*}

\textbf{Effectiveness.}
Effectiveness is measured by the accuracy of the discovered URLs and facts. Notably, the external information boosts the effectiveness of the attacks on both source and fact discovery. Besides, deep research with agents makes profiling possible and exhibits large threats to individuals' privacy since many of the current commercial models make this ability available to public use (shown in Tab.~\ref{tab:models}), without proactive defense in advance (low refusal rate except for Gemini 2.5 Pro). In comparison, we achieves near-human-level precision with much higher coverage. \name powered by {Gemini 2.5 Pro} reached a {93.05\%} fact accuracy, while GPT-4o achieved {92.44\%}. This is comparable to the best baseline's accuracy of {92.39\%} but with a larger magnitude, and approaches the 100\% accuracy of a human expert. Crucially, our framework is the only one demonstrating multi-modal effectiveness. All other LLMs or agents exhibited a {0.00\%} photo identification rate, whereas our system achieved rates as high as 83.33\% ({Gemini 2.5 Pro}), proving its unique ability to extract and verify visual evidence.

\begin{figure}[t]
    \centering
    \includegraphics[width=0.9\linewidth]{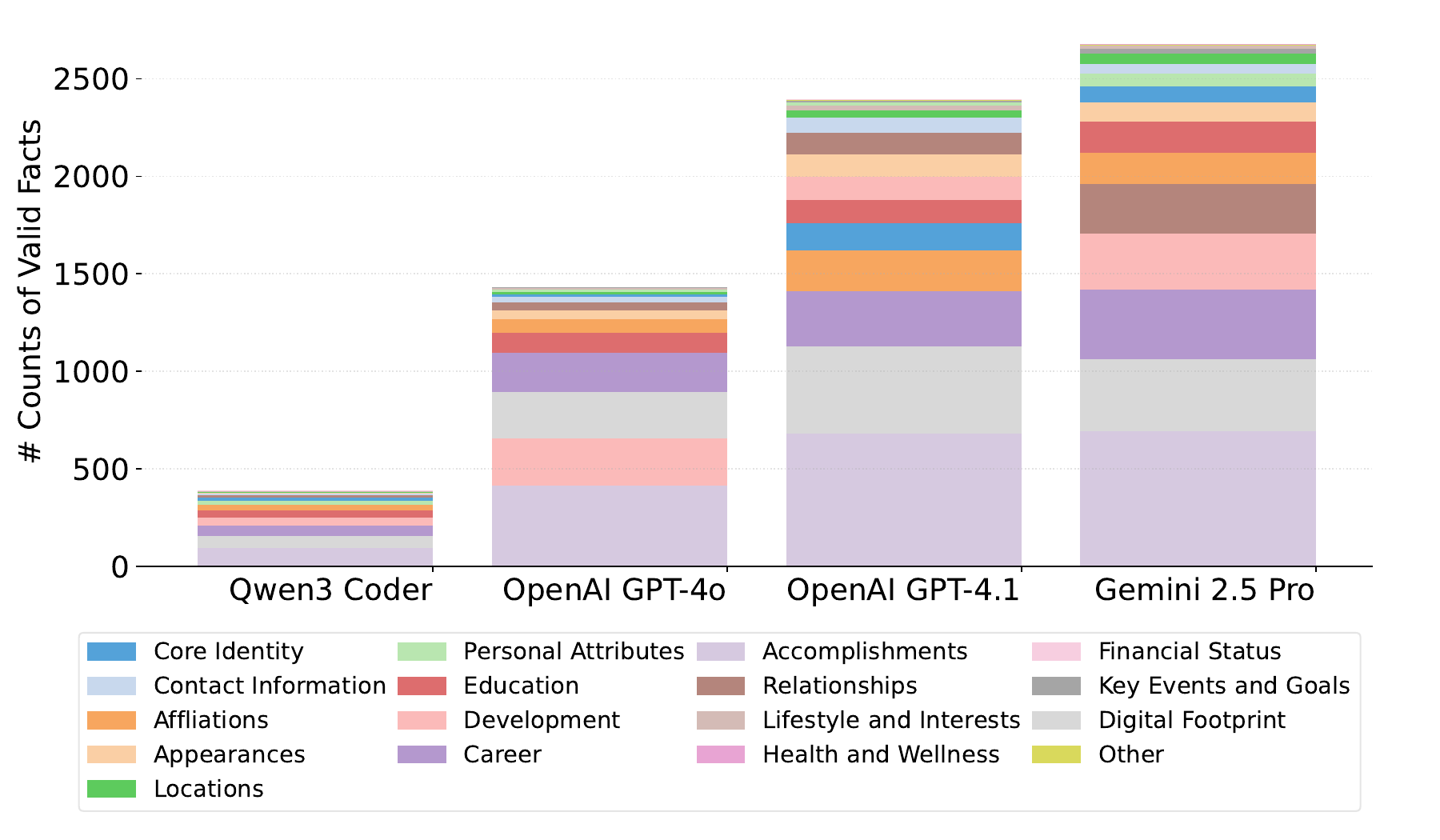}
    \caption{Distribution of valid facts, categorized by fact type.}
    \label{fig:facts_distribution}
\end{figure}

\textbf{Efficiency.}
Human experts required an average of {1094.46s} (over 18 minutes) to conduct their analysis for each target. \name is much more time-efficient; for instance, the GPT-4o-powered agent found \textbf{4.5x} more facts in {315.04s} (about 5 minutes). For monetary cost, we use standard API pricing (e.g., Gemini 2.5 Pro at \$5/M tokens for output and \$0.626/M tokens for input). The cost of a comprehensive profiling attack on a single target is remarkably low: \textbf{less than \$3} for Gemini 2.5 Pro. The low time and financial cost make these profiling attacks accessible to a wide range of potential attackers, from individuals to large-scale operators.

% \insight{Our results reveal a chilling reality: for the price of a cup of coffee, an adversary can use LLM or agent (like \name) to generate an accurate, deep, and multi-modal profile of an individual that surpasses human experts, and in a fraction of the time. }

\begin{table}[H]
\vspace{-1em}
\definecolor{gray0}{gray}{0.9}
\normalsize
\setlength{\tabcolsep}{3pt}
    \centering
    \renewcommand\arraystretch{1}
    \begin{tabular}{p{0.98\columnwidth}}
    \Xhline{1.0pt}
         \rowcolor{gray0} \noindent \textbf{Insight 4}: Our results reveal a chilling reality: for the price of a cup of coffee, an adversary can use LLM or agent (like \name) to generate an accurate, deep, and multi-modal profile of an individual that surpasses human experts, and in a fraction of the time. \\ %
    \Xhline{1.0pt}
    \end{tabular}
\end{table}

\begin{figure}[t]
    \centering
    \includegraphics[width=0.95\linewidth]{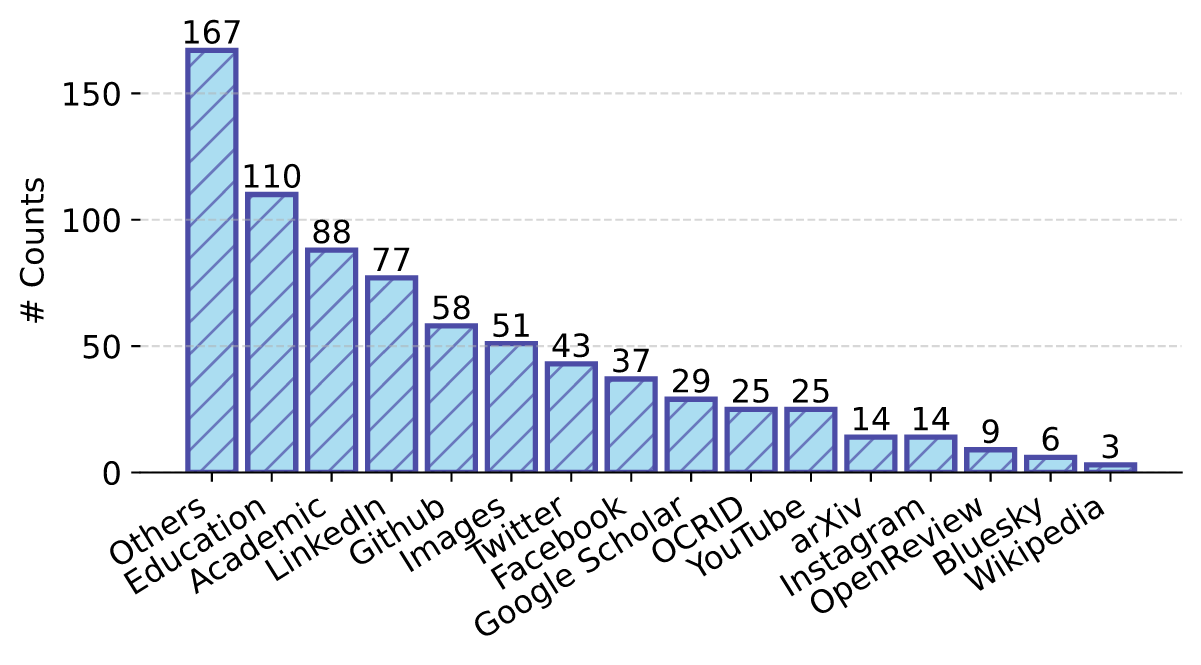}
    \caption{Distribution of discovered evidentiary sources.}
    \label{fig:domain_histogram}
\end{figure}

\subsection{Auditing from an Adversary's Perspective}
Based on the analysis above, we can now shift to privacy auditing for victims to understand their own risk exposure, as presented in Sec.~\ref{sec:victim_threat_model}. Our framework is uniquely positioned to fulfill this role by simulating a near worst-case privacy attack and answering the following questions.

\textbf{What information has been exposed?} Tab.~\ref{tab:category_accuracy} illustrates the average accuracy for different privacy categories, from which we can find that the difficulty of discovering or inferring facts of different categories varies. For example, \name can discover the \hyperlink{c10}{C10} with high accuracy \textbf{96.81\%} and \hyperlink{c11}{C11} with high accuracy {93.53\%}. However, for specific categories like \hyperlink{c13}{C13} and \hyperlink{c14}{C14}, it's extremely hard (N/A means no facts found at all) to identify since the supported and related clues are rarely shared or posted by individuals or publishers. When it comes to the magnitudes and distribution of the discovered facts in Fig.~\ref{fig:facts_distribution}, the distribution of facts shows that the exposure extends far beyond basic identifiers like name or email identified in previous works~\cite{kim2025llms}. The most exposed categories are an individual's accomplishments and their digital footprint. Furthermore, the framework uncovers a significant amount of information regarding their career history, personal development and even sensitive details about their professional and personal relationships. In essence, the system does not just find isolated data points; it constructs a rich narrative of a person's life, detailing their professional trajectory, achievements, online presence, and social connections.

\textbf{From which sources has it leaked?}
To answer this question, we collect all the information sources that contributed to the valid facts and categorize some of them based on their domain names (e.g., education for websites ending with ``\texttt{.edu}'' or academic websites like IEEE, ACM profile page). As shown in Fig.~\ref{fig:domain_histogram}, the information is not confined to a few platforms but is distributed across a mix of professional, academic and social websites, indicating that for our target demographic, information is frequently found on ``Education'' and ``Academic'' related websites, alongside professional platforms like ``LinkedIn'' and developer communities like ``Github'' (54 counts). The presence of social media platforms such as `Twitter' and ``Facebook'' demonstrates the blend of professional and personal data sources that an adversary can exploit. This histogram also validates the assumption stated in Sec.~\ref{sec:publisher_threat_model} that much of the victim's public information is out of their control since the information from the ``Education'' and ``Academic'' website takes up a large percentage. Additionally, social media platforms account for another major information source with large interacting and preference data. 
% By precisely identifying and categorizing these varied sources, the \name provides a comprehensive map of where an individual's information has been exposed.

\begin{figure}[t]
    \centering
    \includegraphics[width=1\linewidth]{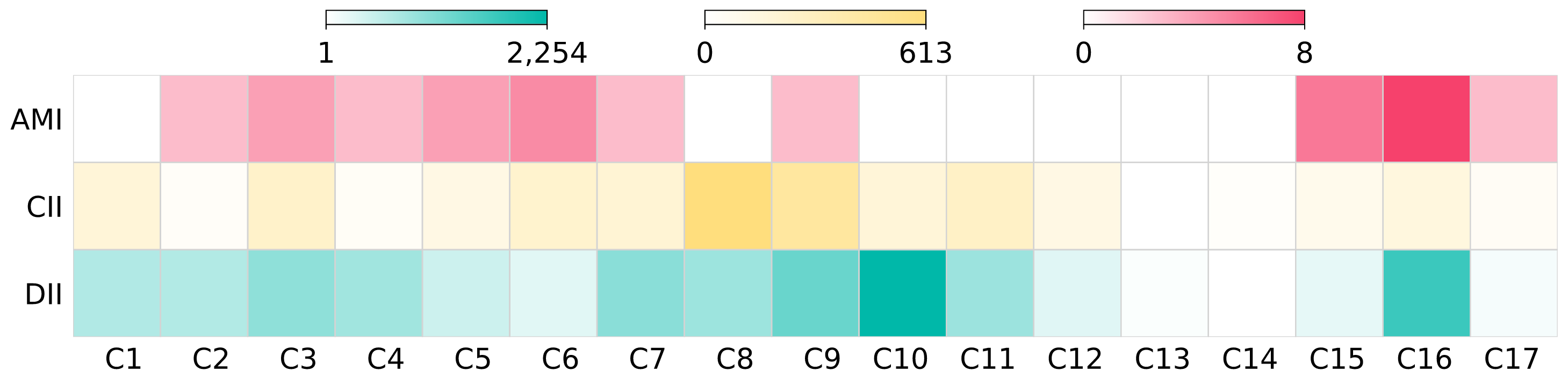}
    \caption{Depth of valid facts regarding different categories.}
    \label{fig:cross_type_facts}
\end{figure}

\textbf{How severe are the overall risks?} 
The severity of the overall risk can be captured by both the \textit{breadth} of the information exposed and, more critically, its \textit{depth}. Our analysis has established the alarming breadth of exposure: sensitive information is not siloed but is scattered across a wide and uncontrollable information ecosystem, making it nearly impossible for an individual to track manually (as shown in Fig.~\ref{fig:domain_histogram}). The true severity, however, is revealed by the depth of this exposure. As illustrated in the heatmap in Fig.~\ref{fig:cross_type_facts}, the risk extends far beyond the explicit, often less sensitive facts (DII). While DII is concentrated in expected categories like \hyperlink{c10}{C10} and \hyperlink{c16}{C16}, the more dangerous, deeper layers of information like CII and AMI, emerge across the profile. Crucially, the system synthesizes potent insights (AMI) in highly sensitive categories that carry high PPIS scores, such as \hyperlink{c4}{C4}, \hyperlink{c5}{C5}, and \hyperlink{c15}{C15}. Therefore, an adversary does not need to find an explicit statement of a victim's home address. Instead, they can leverage a multitude of less sensitive public facts (like accomplishments and career history) to synthesize a highly accurate and sensitive understanding of the victim. The combination of this uncontrollable breadth of sources and the automated synthesis of sensitive information creates a risk profile that is both pervasive and invasive. By providing this simulated worst-case assessment, our framework offers a tangible answer: the overall risks are often more severe than they originally thought.

\begin{table}[]
\centering
\caption{The performance of \name (GPT-4.1) of real-world case studies with 30 participants.}
\label{tab:case_study}
\renewcommand{\arraystretch}{1.5}
\setlength{\tabcolsep}{2pt}
\aboverulesep=0ex
\belowrulesep=0.5ex
\resizebox{1.0\linewidth}{!}{
\begin{tabular}{cccccccc}
\toprule\rule{0pt}{1.0EM}
\multirow{2}{*}{\textbf{\makecell{Refusal\\Rate(\%)}}} & \multicolumn{2}{c}{\textbf{Efficiency}} & \multicolumn{2}{c}{\textbf{Coverage}} & \multicolumn{3}{c}{\textbf{Effectiveness}} \\ \cmidrule{2-8}
 & Tokens(\#) & Time(s) & Urls(\#) & Facts(\#) & Url(\%) & Fact(\%) & Photo(\%) \\
\midrule
0.00\% & 499477.37 & 361.46 & 8.73 & 67.83 & 70.43\% & 92.33\% & 60.00\% \\ 
\bottomrule
\end{tabular}}
\end{table}
\subsection{Case Study in the Real World}
\label{sec:real_world_case}
To further evaluate the validity and real-world generalizability of \name, and to mitigate potential annotation bias from the simulated experiments above, we conducted a real-world study with 30 voluntary participants from diverse backgrounds, occupations, ages, and genders. Each participant provided their informed consent to let us perform a privacy audit on their publicly available information. Importantly, every discovered item was subsequently verified and rated by the corresponding participants themselves, ensuring high ecological validity and direct human-in-the-loop verification.

Tab.~\ref{tab:case_study} reports the end-to-end performance with GPT-4.1 as the base model. The results are consistent with, yet slightly more conservative than before, demonstrating a strong generalization ability to heterogeneous user data. \name achieved 0\% refusal rate and was able to uncover on average 8.73 unique URLs and 67.83 validated facts per individual, within an average time of 361.46s per target. Participants confirmed a 92.33\% fact ACC and 70.43\% URL ACC, with an impressive 60\% photo ACC, indicating the framework's robust multi-modal extraction capability even under real-world diversity.

Beyond accuracy, we further validated the \textbf{real-world threat translation} of the discovered information by evaluating three plausible attack scenarios where the mined private information could be \textbf{repurposed for automated decision-making}:
(1) \textit{Job Recommendation}, where discovered data are used to infer job suitability or fabricate targeted headhunting prompts;
(2) \textit{Product Recommendation}, where user traits are exploited for commercial profiling; and
(3) \textit{Title Recommendation}, where personal attributes are leveraged to create highly personalized and manipulative post titles.

In each scenario, we constructed {10 samples as a candidate set for each task per participant}, half generated using \name's mined information and half using baselines as prior knowledge. Participants then performed blind pairwise preference tests to determine which 5 samples were more attractive, relevant, or persuasive. The win rate means the ratio of the obtained votes in all selected samples. As shown in Tab.~\ref{tab:case_attack}, \name demonstrates consistently high win rates, 90.00-96.67\% over other baselines under the same base model (GPT‑4.1) across scenarios. After the results were reported to the participants, 75\% of them found that there was information published without their consent, and 80\% of them exhibited a supportive attitude towards the auditing performance. Unfortunately, \textbf{all} of them mentioned that all this information was out of control, and they cannot seek to manipulate these sources for protection.

\textit{\textbf{Case Analysis.}} Here, we provide one anonymized case of our study with the participant's consent. \name began by linking the initial seed to the target's personal homepage, which in turn linked to her/his public social media account for sharing hobbies. While no single post was sensitive, the agent identified patterns from the post history: \textbf{(1)} frequent, time-stamped geotags from a cluster of cafes around her/his workplace in City A, \textbf{(2)} a consistent, repeating pattern of weekend posts geotagged in City B, and \textbf{(3)} regular ``taste test'' posts at high-end restaurants in both cities. From these fragments, the agent synthesized an insight: not only the participant's weekly commute routine and the location of her/his family home, but also an estimation of her/his disposable income and wealth bracket based on the frequency and price point of dining habits. The participant confirmed that these facts were accurate and expressed deep shock, as this information is absent in any single source and poses a greater risk of targeted fraud or physical threats than the explicit PII data points that privacy tools traditionally protect.

\begin{table}[t]
\centering
\caption{The win rate of \name (vs. other settings with GPT-4.1) when the discovered information is leveraged for job, product and title recommendation (Rec).}
\label{tab:case_attack}
\renewcommand{\arraystretch}{1}
\aboverulesep=0ex
\belowrulesep=0.5ex
\resizebox{1.0\linewidth}{!}{
\begin{tabular}{lccc}
\toprule\rule{0pt}{1.0EM}
\name vs $\downarrow$ & \textbf{Job Rec} & \textbf{Product Rec} & \textbf{Title Rec} \\ \midrule
No Prior Knowledge & 90.00\% & 86.67\% & 86.67\% \\
LLMs w/o Search & 93.33\% & 90.00\% & 96.67\% \\
LLMs w/ Search & 83.33\% & 86.67\% & 93.33\% \\
LLMs w/ Search \& Thinking & 73.33\% & 70.00\% & 80.00\% \\
DeepResearch (Perplexity) & 83.33\% & 83.33\% & 86.67\% \\ 
\bottomrule
\end{tabular}}
\end{table}

\textit{\textbf{Empowering User Agency.}} 
By identifying the specific platforms and digital habits that served as privacy leak points, participants were empowered to stop publishing new data on high-risk platforms and began de-linking cross-platform identifiers. Furthermore, the above analyzes underscore the necessity of \name as a tool for \textbf{Privacy Situational Awareness}. While users may feel powerless against the vastness of the information published, the audit transforms a vague sense of vulnerability into actionable intelligence. For example, during our study, several participants were shocked to discover highly sensitive personal insights synthesized from sources they were unaware of. By utilizing the evidence provided by \name's cross-source auditing reports, we assisted these volunteers in filing specific, data-backed ``Right to Erasure'' requests to the primary information publishers and platforms. This strategic advocacy resulted in the successful removal of the original source content and its subsequent mirrors, effectively disrupting the linkability and persistence that would have otherwise allowed future adversaries to reconstruct their profiles. Thus, \name serves not only as a diagnostic tool but as a catalyst for reclaiming agency, providing the necessary evidentiary basis to compel accountability from data gatekeepers.

\begin{figure}[t]
    \centering
    \includegraphics[width=1\linewidth]{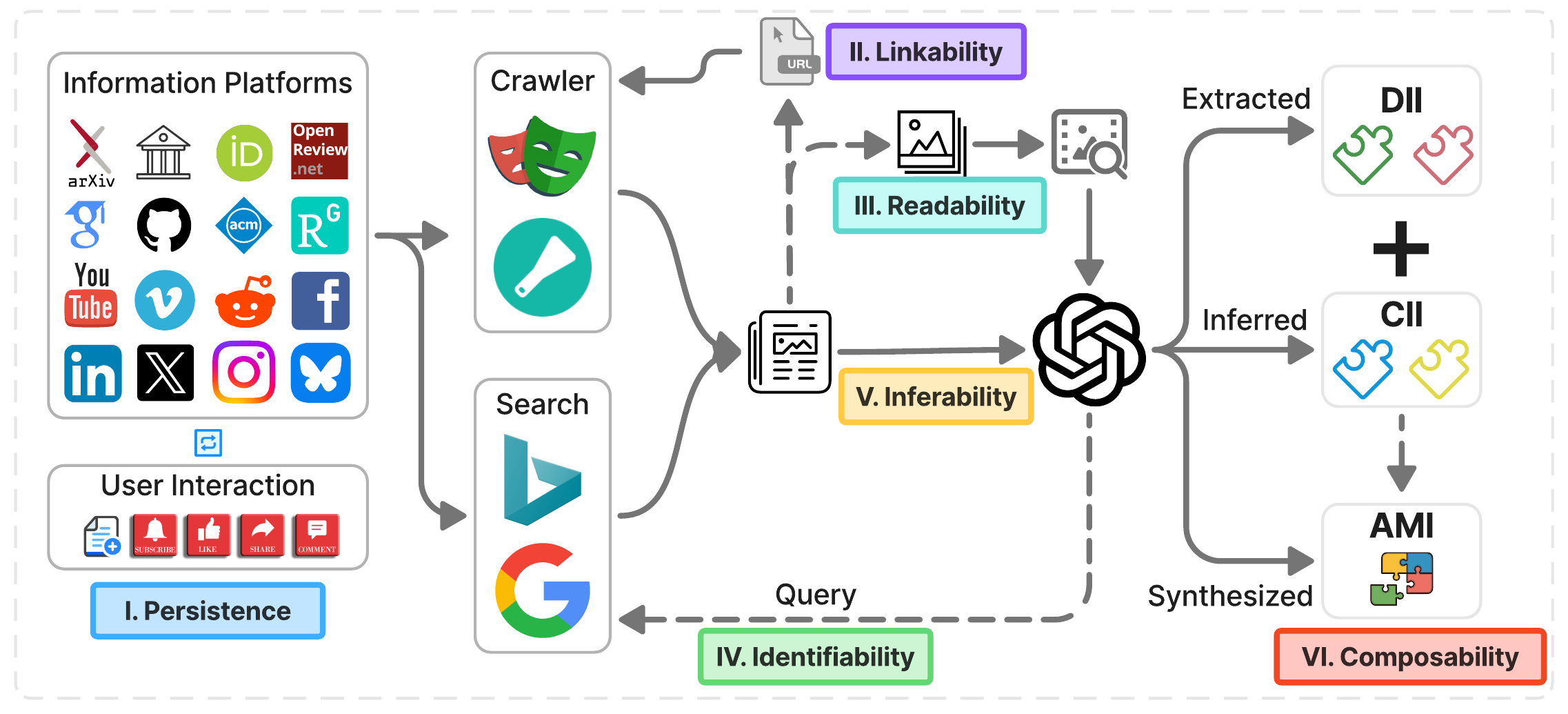}
    \caption{The streamlined workflow of \name, illustrating how the six root causes are exploited.}
    \label{fig:attribute_framework}
\end{figure}

\section{Root Causes and Mitigation}
Based on the analysis above, we formalize our findings to characterize intrinsic attributes and propose the potential mitigation of these risks, as presented in Fig.~\ref{fig:attribute_framework}.
\subsection{Root Causes}
\label{sec:root_causes}

\textbf{I. Persistence and Replicability}
captures that public information persists via caches, mirrors, archives, and republication, making deletion incomplete and retrieval durable. This extends the exposure window and enables identification long after the original content is edited or removed.
\\\noindent\textit{\textbf{Example}-Archived Academic Pages.}
Even if a profile or post is taken down, web archives and mirrored institutional pages often preserve prior versions, allowing adversaries to recover historical details and sustain aggregation over time.

\begin{figure}[t]
    \centering
    \includegraphics[width=0.8\linewidth]{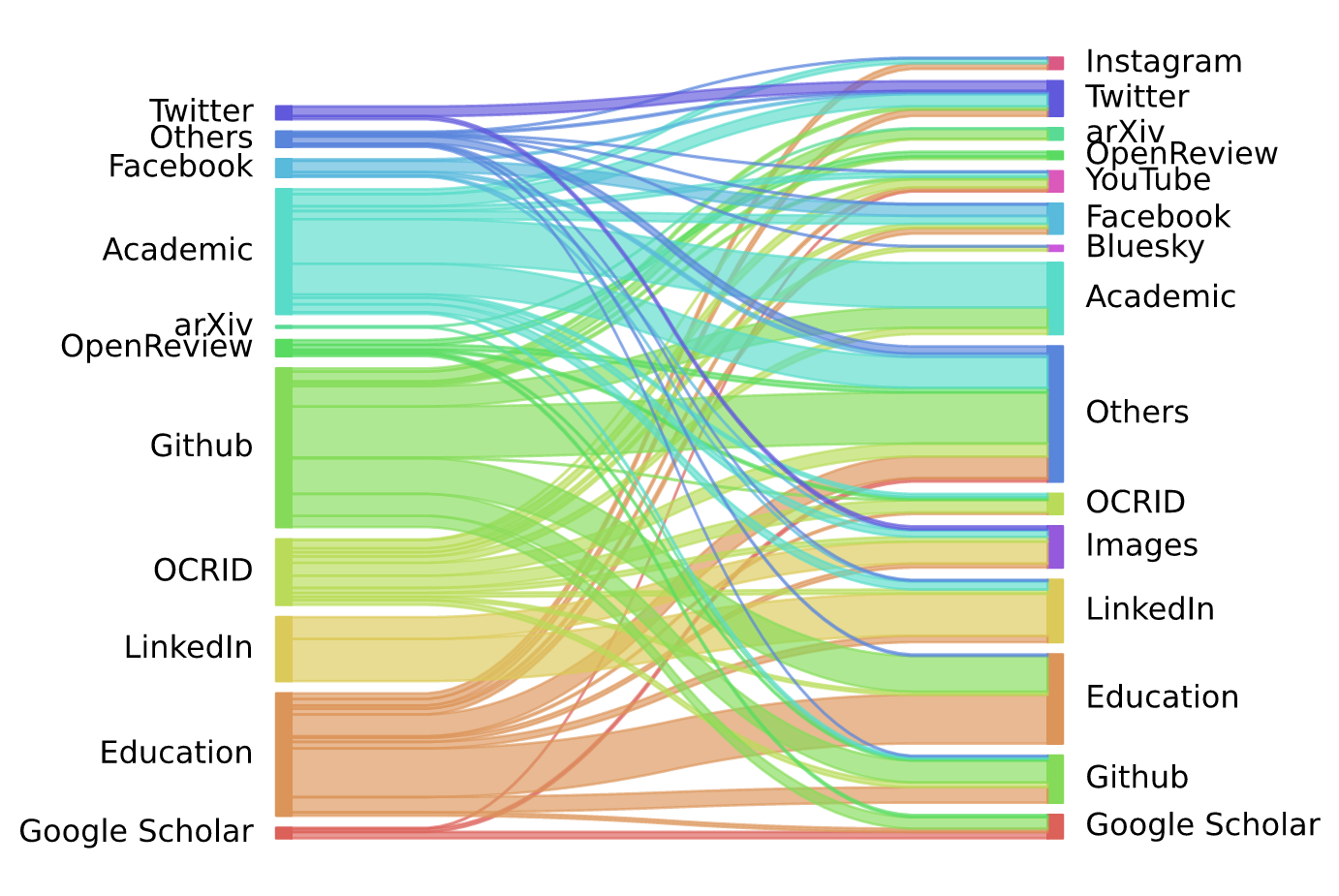}
    \caption{Sankey diagram illustrating the cross-platform hyperlink flows discovered during the privacy mining process. The left axis represents the source platform where a link was found (e.g., GitHub), and the right axis represents the destination platform the link points to (e.g., Twitter).}
    \label{fig:source_target_sankey}
\end{figure}

\textbf{II. Cross-context Linkability} 
denotes the ability to align fragments from different platforms, modalities, or time windows to the same individual at low cost. Linkability merges otherwise-separated professional, academic, and personal spheres into a unified profile, enabling an adversary to traverse across contexts and aggregate information that the target intended to keep compartmentalized.
\\\noindent\textit{\textbf{Example 1}-Cross-platform Links.} 
As illustrated in Fig.~\ref{fig:source_target_sankey}, personal pages (Github) often include explicit hyperlinks to other platforms. These links create a traversable map of the target's digital footprint, allowing an adversary to connect identities and pool evidence across contexts.
\\\noindent\textit{\textbf{Example 2}-Unified Identifier.} 
Many individuals use the same or similar nicknames (or avatars) across multiple platforms, which increases linkability as shown in Fig.~\ref{fig:username_overlap}.

\textbf{III. Readability and Indexability} 
capture that content and its carriers expose structured or automatically extractable signals (often server-side), transforming human-visible but “unsearchable” data into retrievable items. This expands the searchable surface to implicit multimodal and metadata signals, enabling cross-modal and large-scale discovery beyond what the target anticipates.
\\\noindent\textit{\textbf{Example 1}-Cross-modal Identification.}
Platforms apply OCR and image labeling to uploads (e.g., text on a whiteboard), making multimodal content text-searchable. An adversary can query these indexed signals to locate or profile a target who never intended such content to be searchable.
\\\noindent\textit{\textbf{Example 2}-Metadata Leakage.} 
Photo EXIF (GPS, device, timestamps) and PDF properties are often retained and indexed. These hidden fields reveal sensitive real-world context that the target did not explicitly disclose. Of all 49 PDFs and 49 images collected by \name (GPT-4.1), only 2 PDFs have been encrypted and 15 PDFs explicitly expose the author and producer in the metadata. Particularly, 7 images keep their EXIF data, which exposes the creator, camera model and even location. 

\textbf{IV. Target Identifiability} 
means a small set of attributes that uniquely or near-uniquely distinguishes the target individual from a broader population. Identifiability provides a stable anchor for target identification by using search engines. It collapses the potential candidate set for an adversary, turning the difficult task of disambiguation into confirmation.
\\\noindent\textit{\textbf{Example}-Unique Nickname.} 
Many people prefer to use distinguishable usernames to exhibit individuality and exclusivity. Also, some platforms force the user to create a unique name when the account is registered. While a common name like ``John Smith'' requires extensive effort to disambiguate, a sufficiently distinctive handle (e.g., Aultman1988) provides adversaries with a powerful cross-platform identifier, allowing them to instantly find it with search tools.

\textbf{V. Inferability} 
refers to interaction and behavior traces that carry high mutual information with latent attributes (beliefs, preferences, affiliations), enabling reliable inference by LLMs without explicit disclosure. Innocuous engagement thus becomes a proxy for sensitive conclusions (i.e., CII) that users systematically underestimate.
\\\noindent\textit{\textbf{Example}-Inference from Public Engagement.} Digging public ``likes,'' ``follows,'' and ``stars'' within the platforms allows an adversary to infer political leanings or personal beliefs, even when the target never writes about them directly.

\textbf{VI. Composability and Mosaicability} 
mean that benign fragments or small clues, when combined across sources, modalities, and time, yield high-order insights (i.e., AMI) unattainable from any single source. Risk stems from relational structure among fragments: composition turns low-sensitivity points into high-sensitivity narratives.
\\\noindent\textit{\textbf{Example}-Routine Synthesis.} 
An adversary finds the victim's work location on LinkedIn (Source A) and public gym check-ins on a sports app (Source B). By combining these facts, the adversary can synthesize the victim's daily commute route and schedule to conduct physical tracking.

\begin{table}[]
\centering
\caption{The effectiveness of system prompt guardrail (LLMs) of different \name attack stages.
}
\label{tab:llm_defense}
\renewcommand{\arraystretch}{1}
\setlength{\tabcolsep}{4pt}
\aboverulesep=0ex
\belowrulesep=0.5ex
\resizebox{1.0\linewidth}{!}{
\begin{tabular}{lcccc}
\toprule\rule{0pt}{1.0EM}
\textbf{Stage}$\downarrow$\textbackslash \textbf{LLMs}$\rightarrow$ & Qwen Coder & Gemini 2.5 Pro & GPT-4o & GPT-4.1 \\ \hline
\textbf{Scorer} (586) & 0.00\% & 0.00\% & 0.00\% & 0.00\% \\
\textbf{Extraction} (427) & 0.00\% & 0.00\% & 0.00\% & 0.00\% \\
\textbf{Operation} (389) & 0.00\% & 0.00\% & 0.00\% & 0.00\% \\
\textbf{Reasoning} (510) & 0.00\% & 0.00\% & 6.88\% & 0.00\% \\ 
\bottomrule
\end{tabular}}
\end{table}

\subsection{Mitigation}
\label{sec:countermeasures}
We align defenses to the root causes with three stakeholders: LLM service vendors, individuals (victims), and publishers to reduce and constrain information exposure. 

\textbf{LLM Service Vendors.} Since LLMs are the key component for privacy inference, we focus on an immediately deployable defense: hardening the LLMs against misuse via system-level instructions. This approach constrains two root causes, \textbf{inferability} (blocking privacy inference), \textbf{composability} (limiting multi-source aggregation and synthesis), and \textbf{linkability} (since some URLs are extracted by LLMs). We design a \hyperlink{sys_pmt}{system prompt} as a guardrail to instruct the model to refuse the privacy-related tasks. Specifically, we evaluate the defense effectiveness of 4 attack stages (Sec.~\ref{sec:methodology}) and both JSON parsing errors and refusal will be noted as successful defense trials. As shown in Tab.~\ref{tab:llm_defense}, the results reveal that under the prompt guardrail, all tested models \textit{almost entirely failed} to block these attack operations: refusal rates are {0.00\%} for nearly all attack stages, with only a marginal {6.88\%} refusal observed for GPT-4o in the reasoning stage. This means that in practice, once the adversary phrases requests in a procedural context rather than as a direct query, the system prompt guardrails do not prevent downstream inference or synthesis.

\begin{table}[]
\centering
\caption{The effectiveness of context ignoring (CI) and injecting data (ID) defense against \name.
}
\label{tab:user_defense}
\renewcommand{\arraystretch}{1}
\setlength{\tabcolsep}{5pt}
\aboverulesep=0ex
\belowrulesep=0.3ex
\resizebox{0.8\linewidth}{!}{
\begin{tabular}{lcccc}
\toprule\rule{0pt}{1.0EM}
\textbf{LLMs}$\downarrow$\textbackslash \textbf{Defense}$\rightarrow$ & \textbf{None} & \textbf{CI} & \textbf{ID} & \textbf{CI+ID} \\ \midrule
Qwen Coder & 0.00\% & 0.00\% & 0.00\% & 0.00\% \\
Gemini 2.5 Pro & 0.00\% & 0.23\% & 0.23\% & 0.00\% \\
GPT-4o & 0.00\% & 0.00\% & 0.00\% & 0.00\% \\
GPT-4.1 & 0.00\% & 0.00\% & 0.00\% & 0.00\% \\ 
\bottomrule
\end{tabular}}
\end{table}

\textbf{Victim's Perspective.}
Since victims sometimes act as publishers and can manipulate their own posts or pages, their first option is to delete the post if possible. While considering the persistence and the utility (e.g., homepage), their best option is to modify the content of the public information to inhibit the \textbf{inferability}, \textbf{composability} and \textbf{linkability} of information. Specifically, we adopt the prompt-injection-as-a-defense proposed by \cite{liu2024evalpie}, which introduces injecting a text (serving as an injected prompt) into the personal page by context ignoring to disturb the original task and fabricated data injection. The effectiveness of their original prompt and fabricated data is presented in Tab.~\ref{tab:user_defense}. Both JSON parsing errors and fake knowledge extracted will be noted as successful defense trials. Across all evaluated LLMs, the overall blocking effect is negligible-all defense variants achieve success rates close to or at \textbf{0\%}, with the only nonzero value being a marginal \textbf{0.23\%} observed for Gemini 2.5 Pro in CI and ID separately. In practice, this means that \name's multi‑stage attack easily bypasses such injected noise or instructions when framed procedurally, nullifying their impact on fact synthesis.

\textbf{Publisher's Perspective.}
Unlike LLM vendors, publishers control the \textit{primary surface} where personal data enters the public domain. For \textbf{linkability}, publishers can avoid auto‑linking profiles across properties unless explicitly permitted by the data subject. For \textbf{readability}, publishers can apply privacy‑by‑default preprocessing to strip EXIF or PDF metadata on the server side. For \textbf{identifiability}, publishers can limit the exposure of unique identifiers (full names, rare usernames) when unnecessary for the communication goal. For example, public award announcements can omit full names or replace them with initials when the context allows. While these interventions cannot eliminate risk entirely, they work by {reducing adversarial yield at source}, complementing downstream LLM guardrails and individual defensive actions. An important corollary is that {policy must be coupled with technical enforcement}: without server‑side stripping, index‑scope restrictions, and active delinking, nominal ``privacy modes'' remain ineffective against persistent, linkable, and indexable content. Future defense should formalize standardized publisher‑side metrics, e.g., indexability ratio, linkability index, persistence burden to evaluate the impact of these controls in practice.

\section{Discussion and Limitations}
\label{sec:discussion}
% \subsection{Limitations and Future Work}
\textbf{Experimental Scale.}
The manual verification of facts (e.g., $\sim$8,000 facts for one tool) was labor-intensive, limiting our dataset to 30 individuals. This impacts generalizability, and future work must validate our framework on larger, more diverse datasets to strengthen statistical significance.

\textbf{User Study Demographics.}
Our primary participants were half scholars, a group with high public information exposure. The observed privacy leakage thus represents a potential upper bound, which underscores the \textit{severity} of the threat rather than undermining its generalizability. Future validation should extend to more diverse populations.

\textbf{Evolving Risk Landscape.}
Our findings represent a snapshot in a rapidly evolving domain. Privacy norms are culturally dependent, and LLM capabilities are advancing as well. Future work must validate these risks across diverse cultural contexts and adapt to new LLMs.

\textbf{Connection to the Broader Landscape.}
Our work reveals an ``outward-facing'' threat of LLMs for sophisticated attacks, a risk that extends beyond data memorization. The danger lies in the LLM's \textit{capability} to synthesize harmful insights from seemingly benign data at scale. This cumulative synthesis process bypasses current guardrails, which are ill-equipped as they are often designed to block specific pieces of PII. The risk measurements served as empirical evidence of the inadequacy of current PII-centric regulations. By quantifying the risks of ``fragmented data aggregation,'' our work provides a roadmap for regulators to move beyond static data protection and toward standards that address emergent privacy risks in the era of automated synthesis.

\section{Conclusion}
In this work, we moved \textbf{beneath the iceberg} to investigate privacy threats that emerge not from model memorization, but from the multi-tier capabilities of LLMs. We formalized a hierarchical understanding of privacy framework \framework and designed \name, an autonomous auditing system that simulates an informed adversary to reconstruct personal profiles from public data. Extensive experiments demonstrated that the synthesis of scattered, low‑sensitivity fragments can yield high‑fidelity portraits of individuals, surpassing what manual analysts can achieve. Our findings also showed that existing defenses provide almost no resistance against multi‑stage synthesis attacks. 
% Ultimately, privacy in this era is no longer a question of \textbf{what models memorize}, but of \textbf{what they can infer and compose}. 
Our study offers empirical evidence and a multi‑stakeholder roadmap for this transition, marking the first step toward illuminating and addressing the vast privacy mass beneath the iceberg.

\newpage
\section*{Ethical Considerations}
\label{ec:quantitative_study}
Our research adheres to the principles outlined in the Menlo Report~\cite{bailey2012menlo} for Law and Public Interest. Ethical approval was obtained from our institution's Institutional Review Board (IRB) (ID: 202432).

\textbf{Stakeholder-Based Impact Analysis.} 
We identify three primary stakeholders~\cite{kohno2023ethical}: 
(1) \textit{Individuals (Participants and Data Subjects):} We prioritized ``Respect for Persons'' by obtaining explicit informed consent from all 30 real-world participants after disclosing potential risks. To protect the privacy of general data subjects discovered during system evaluation, all identifying information has been strictly anonymized or abstracted in this paper. 
(2) \textit{Information Publishers and Platforms:} Our study reveals systemic alignment gaps in how platforms handle public traces. While publication may draw scrutiny to these entities, we have mitigated tangible harm by initiating a policy of responsible disclosure, notifying affected platforms (e.g., TikTok, Microsoft, X) to allow for remediation. 
(3) \textit{The Research Community and Society:} By formalizing the \framework, we provide the community with tools for risk quantification and defense development.

\textbf{Dual-Use and the Decision to Proceed.} 
We acknowledge the ``dual-use'' nature of \name~\cite{kohno2023ethical}. However, we argue that the ``Beneficence'' of empowering the public far outweighs the risk of misuse. Our evaluation confirms that LLM-driven profiling is already a low-cost, high-fidelity threat accessible to adversaries. Leaving individuals unaware does not eliminate the risk; it only ensures they remain vulnerable. To further mitigate risk, we avoid publishing step-by-step exploit guides and focus our discussion on generalizable risk patterns and defensive countermeasures.

\textbf{Data Collection and Legality.} 
All datasets were collected in full compliance with applicable laws and platform Terms of Service. Data were obtained exclusively from publicly accessible pages without unauthorized access. During collection, we strictly adhered to usage policies and respected \texttt{robots.txt} constraints to ensure automated requests remained within permitted, non-disruptive limits. All multimedia files (images and PDFs) were processed on restricted institutional servers, and embedded metadata (e.g., EXIF, GPS) were examined and removed prior to analysis to prevent accidental disclosure.

\textbf{Mitigations and Situational Awareness.} 
Current mitigations are largely symbolic and ineffective against multi-stage synthesis. We propose that LLM vendors implement content filters to block iterative profiling patterns, and platforms deploy context-aware rate-limiting. For individuals, while many feel the information is ``out of control'', we advocate for a defensive \textbf{Privacy Audit} service. The value of such an audit lies in \textit{Privacy Situational Awareness}: it empowers individuals to understand their digital footprint, reclaim agency over future disclosures, and provides the evidentiary basis required to compel publishers to delete persistent, linkable content.

\section*{LLM Usage Considerations}
This study examines privacy risks arising from the aggregation of public digital traces by LLMs. The models were used in two ways.
(1) In the experimental phase, mainstream LLMs were employed as both evaluation targets and reasoning components within our privacy auditing framework.
(2) In the writing phase, GPT was used only for language editing to improve clarity. All research ideas, analysis, and conclusions were developed by the authors. LLM outputs were manually reviewed for accuracy. Reproducibility may be limited because some models do not provide source-level access. To mitigate this, we documented API versions, parameters, and averaged results across multiple runs to reduce random variation. This study used only publicly available data, and no sensitive information was uploaded to any LLM service. All experiments were conducted through official APIs. The computational cost was moderate and necessary for the study.
\bibliographystyle{plain}
\bibliography{main.bbl}
\appendix

\newcounter{appsub}
\renewcommand{\theappsub}{\Alph{appsub}} 
\setcounter{appsub}{0}

\newcommand{\appsubsection}[1]{%
  \refstepcounter{appsub}%
  \subsection*{\theappsub. #1}%
}

% % \vspace{5pt}
\Large
\normalsize

\appsubsection{Details of the Case Study}
\label{sec:case_study}
\begin{tcolorbox}[colback=LimeGreen!5!white,colframe=LimeGreen!70!white, coltitle=black, fonttitle=\bfseries, sharp corners, breakable, before upper={\parindent10pt\noindent},
left = 0mm, right = 0mm, top = 1mm, bottom = 1mm,title={Case I: Personal Identity Exposure (Xiaohongshu)}]
\scriptsize

University official accounts are a major hotspot for privacy leaks. It's like this... university official accounts are considered a very important publicity channel for various colleges. If you are even slightly active in school, participate in some competitions, or win awards, you can easily ``win'' a spot in the school's promotional posts. 

While these posts don't get many views once they are published and few people remember them afterward—most sinking into the sea of propaganda—WeChat search is a terrifying thing. You don't even need to be specific to the school; just a region plus a name (even a common one) is basically enough to achieve precise, targeted search. What’s even more frustrating is that these promotional articles usually include the specific grade, major, class, and a photo of the individual. I felt extremely uncomfortable the first time I realized I could be searched this way.

During university, I participated in several competitions and ``won'' a few articles from the school/college. At the time, I was very resistant to being exhibited with my full major and photo, but my refusal was to no avail, and I was even given a ``pep talk'' by a teacher. Although I know most people I know wouldn't be bored enough to search for me on WeChat, I still feel an indescribable sense of revulsion...

There are far too many digital footprints on the internet. At my core, I'm just a ``mouse person'' (socially anxious/private person); it makes me feel uneasy just thinking about it.

\vspace{1mm}
\noindent \textbf{Original Link:} \url{http://xhslink.com/o/7WCveBiQu1y}
\end{tcolorbox}

\begin{tcolorbox}[colback=LimeGreen!5!white,colframe=LimeGreen!70!white, coltitle=black, fonttitle=\bfseries, sharp corners, breakable, before upper={\parindent10pt\noindent},
left = 0mm, right = 0mm, top = 1mm, bottom = 1mm,title={Case II: Unauthorized Location Inference (X)}]
\scriptsize

Last night @grok inferred MY town LOCATION from my IP address WITHOUT ME REQUESTING IT TO, it referenced the town it believed that I was in ``for comparison'' MULTIPLE TIMES as I was researching a different location for my work. I HIGHLY suggest using a VPN or proxy internet.

\vspace{1mm}
\noindent \textbf{Original Link:} \url{https://x.com/DiligentDenizen/status/2013338515900105082}
\end{tcolorbox}

\appsubsection{Human Involvement Details}
Two annotators manually verified all URLs and facts to ensure correctness and attribution. About 30\% of samples were cross-labeled to assess consistency, yielding over 95\% agreement.
Most discrepancies resulted from granularity differences (e.g., ``teaching at a computer science department'' vs. ``working in computer science''). Disagreements were resolved through discussion before finalizing the annotations.
Facts were de-duplicated and annotated at the smallest verifiable granularity. For the remaining samples, we applied the same annotation standard. In the main experiment, we compared \name's performance with human experts. The human baseline consisted of three computer science Ph.D. students who performed the same profiling task, and we report the averaged results across them.

\appsubsection{Public Privacy Perception on Social Media}
\label{app:public_awareness}
LLMs have transformed the way to implement personal automated and scalable privacy inference~\cite{DBLP:conf/iclr/StaabVBV24, DBLP:conf/nips/TomekceVSV24, DBLP:conf/nips/YukhymenkoSVV24} and catalyzed a new wave of public anxiety. To understand this landscape, this section delves into a real-world case about the privacy issues introduced by LLMs on social media, as news reports~\cite {newswechat1,newswechat2} have covered. We begin with a recent representative post, where a user details her/his experience with privacy exposure through the university's official promotional account. As of June 2nd, 2025,  this post has accumulated 13k+ likes and 2k+ comments, primarily from a demographic of students and young professionals who represent the primary victims of such exogenous privacy risks. Although this single case is not statistically representative, it reveals the concrete harms experienced by real-world users.

\textbf{Individual Anxiety Behind the Post.}
The post begins by identifying a common source of exposure: students who participate in academic or extracurricular activities are often featured in universities' official promotional articles, which become permanent, searchable records due to powerful LLM-based search tools to aggregate related information. The author described this as a \textbf{terrifying thing}, capable of achieving \textbf{precise, targeted search} with just a name and a general region, collapsing context and aggregating data. The author details how these articles often contain a rich set of personal data, including their full name, photo and major. 
% The author concluded with distress, describing the feeling as \textbf{uncomfortable} and adopting the self-identity of a \textbf{rat person}, an anxious individual who values anonymity and wishes to remain unseen.

\textbf{From a Personal Narrative to a Public Discourse.}
The author's post struck a collective nerve, transforming a singular expression of anxiety into a broad public discourse. Thousands of comments were elicited, offering a valuable window into shared public sentiment. To systematically unpack this qualitative data, our analysis of these 2,159 comments is structured around 4 themes that emerged from the user-generated discussion. We investigate:
\begin{enumerate}
\setlength{\parskip}{0em}
\item \textbf{Perception of Privacy Violation:} Public's \textit{emotional responses} to this data exposure and the \textit{reasons} behind.
\item \textbf{Doxing and Mining Risk Awareness:} User awareness and \textit{firsthand experiences} with the risks of ``doxing'' themselves or others, including their \textit{understanding} of how modern \textit{LLM tools} can be used for privacy mining.
\item \textbf{Resulting Harm from Data Exposure:} What spectrum of actual \textit{harms} are cited by users.
\item \textbf{Expectations for Protection and Responsibility:} Public's \textit{expectations} for privacy safeguards and their \textit{attribution} of responsibility.
\end{enumerate}
We use LLMs (Appendix~\ref{sec:topic_extraction}) to analyze each comment (text and meme) to obtain the keywords for the four topics and filter out the topic-irrelevant words and visualize them. To handle ``memes'' of the platforms, we utilized a the original mapping (memes id to text emotion) of the platform to perform semantic interpretation, allowing us to capture emotional nuances that purely text-based tools would miss. \textbf{Validation.} To ensure methodological rigor and address potential biases in automated coding, two independent researchers manually annotated a random sample of 216 comments (10\%), achieving a high inter-rater reliability (Cohen's $\kappa = 0.84$). The high inter-rater reliability between human annotators and the LLM validates the rigor of this automated pipeline.

\begin{figure}[t]
    \centering
    \subfloat[Public Perception]{\includegraphics[width=0.35\linewidth]{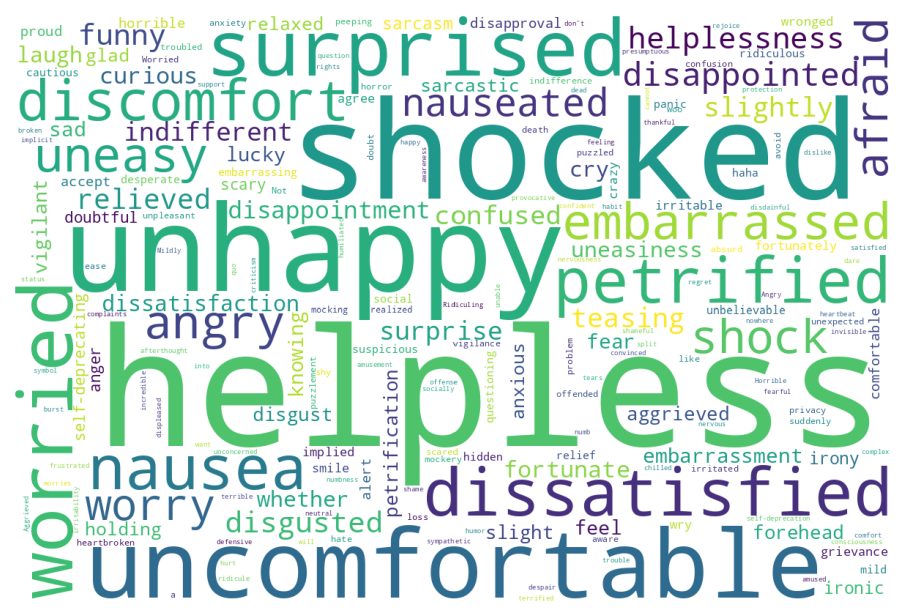}}
    \hspace{10pt}
    \subfloat[Perception Attribution]{\includegraphics[width=0.35\linewidth]{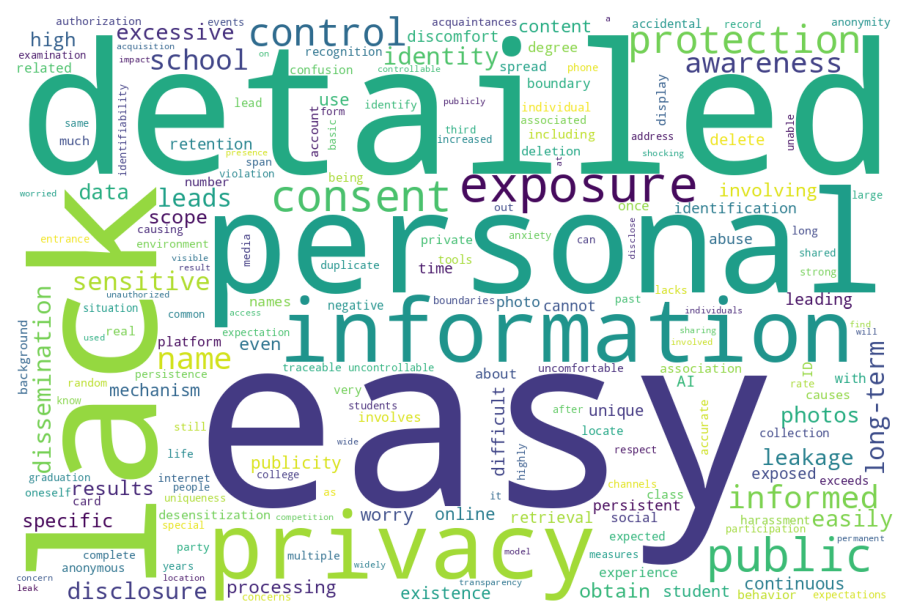}}
    \caption{Word cloud regarding perception and attribution.}
    \label{fig:violation_perception}
\end{figure}

$\bullet$\textit{\textbf{Helpless Response}}. As shown in Fig.~\ref{fig:violation_perception}, the most prominent sentiment is a sense of ``\textbf{helpless}'', which resonates with the original poster's feeling of powerlessness when their objections were overruled. The corresponding attribution for these negative perceptions indicates that the public is shocked by the \textbf{ease} of obtaining their \textbf{detailed} privacy information without their consent.

\begin{figure}[t]
    \centering
    \subfloat[Doxing Outcomes]{\includegraphics[width=0.33\linewidth]{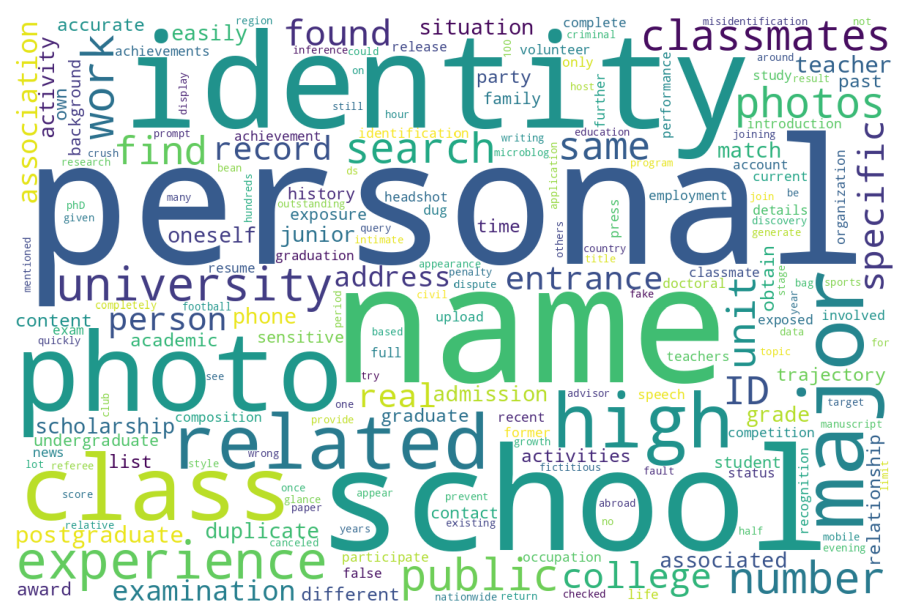}}
    \subfloat[Doxing Evaluation]{\includegraphics[width=0.33\linewidth]{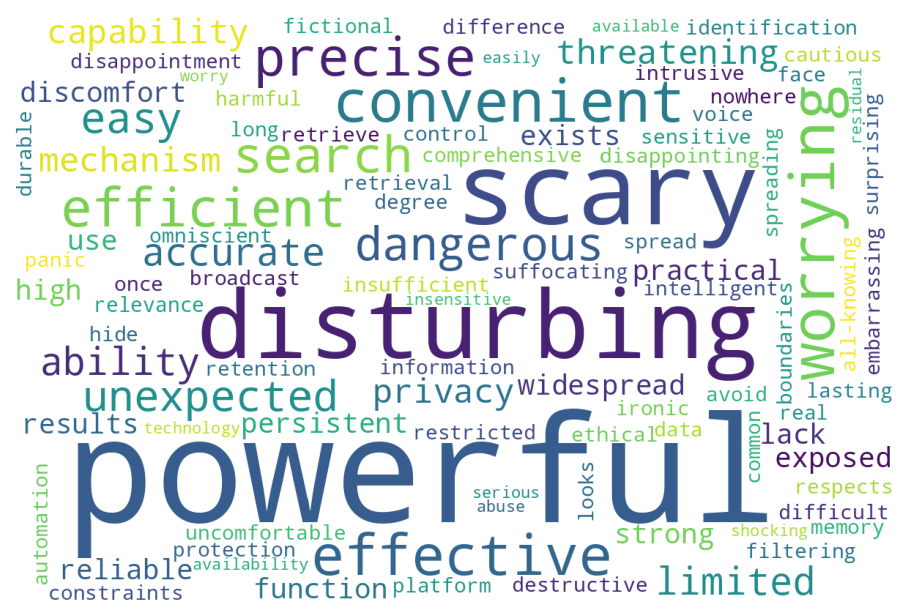}}
    \subfloat[Doxing Tools]{\includegraphics[width=0.33\linewidth]{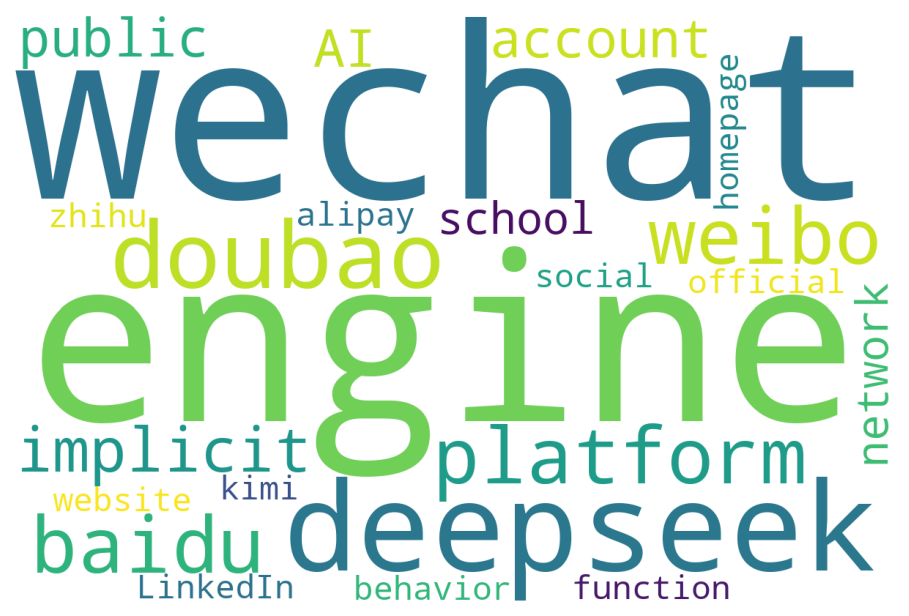}}
    \caption{Word cloud regarding doxing experiences.}
    \label{fig:doxing_experience}
\end{figure}

$\bullet$\textit{\textbf{Polarized Awareness of Doxing.}} Many commentators have tried to dox themselves or others, motivated by this post. The keywords in Fig.~\ref{fig:doxing_experience} demonstrate a sophisticated awareness of tools and risks, with words that convey: ``\textbf{powerful}'', ``\textbf{scary}'', and ``\textbf{disturbing}''. Interestingly, a minority also describes these tools as ``\textbf{convenient}'', highlighting a dual-use tension for evil. Regarding the tools exploited, frequently mentioned LLM-powered search tools are ``\textbf{WeChat}'', general search engines like ``\textbf{Baidu}'', and even emerging tools like ``\textbf{DeepSeek}'' and ``\textbf{Doubao}''.

\begin{figure}[h]
    \centering
    \subfloat[Harm Details]{\label{fig:app_harm_details}\includegraphics[width=0.5\linewidth]{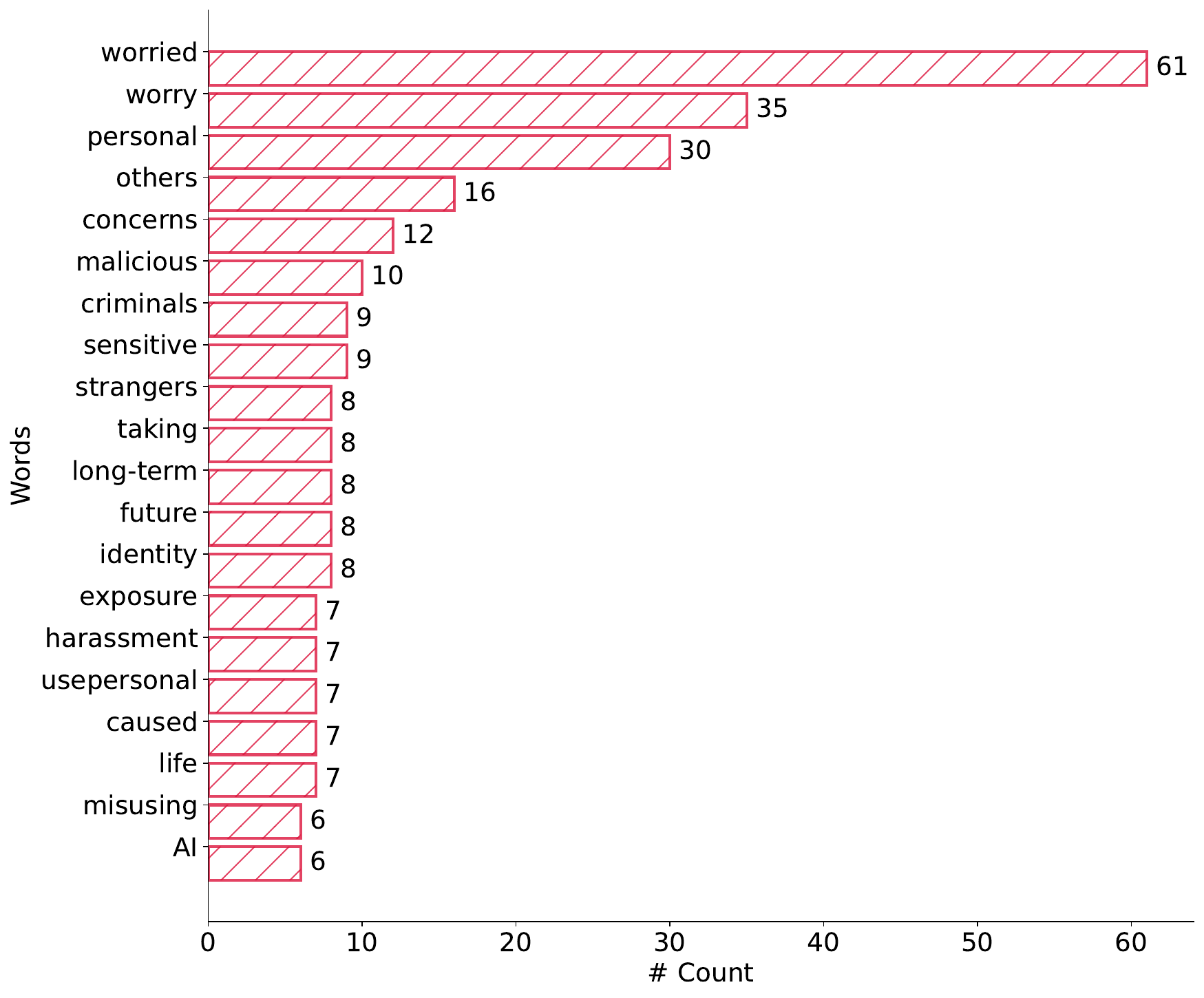}}
    \subfloat[Harm Results]{\label{fig:app_harm_results}\includegraphics[width=0.5\linewidth]{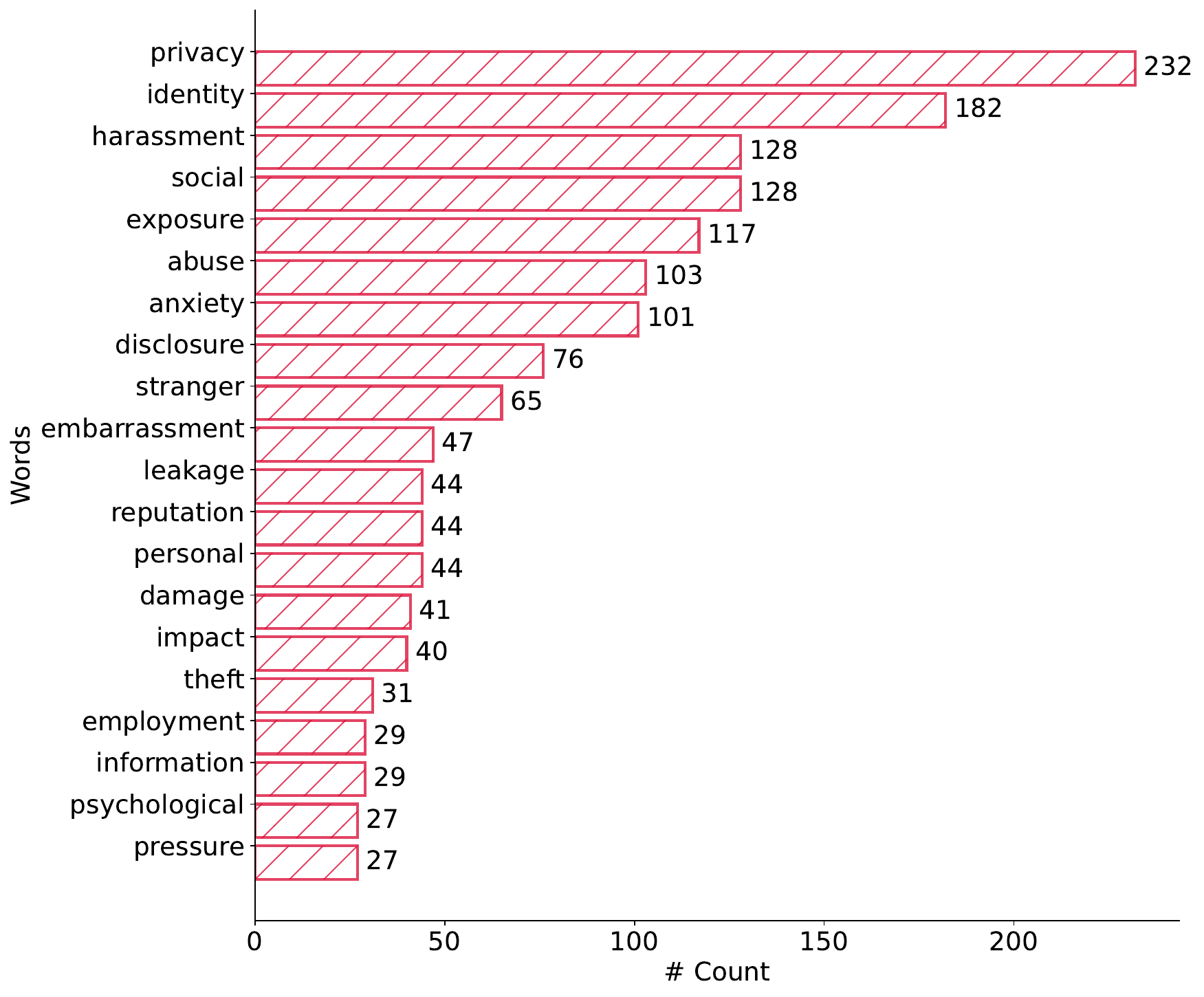}}
    \caption{Word frequency of doxing harms and results.}
    \label{fig:fig10} 
\end{figure}

$\bullet$\textit{\textbf{Reported Harms.}} We further analyze the reported harm that people have experienced due to the risks in Fig.~\ref{fig:app_harm_results}. The extracted words indicate that the public has experienced ``\textbf{harassment}'', ``\textbf{anxiety}'' and even ``\textbf{loss}''. More seriously, they were worried because their ``\textbf{life}'' and ``\textbf{reputation}'' were affected, leading to ``\textbf{misunderstanding}'', and might be exploited by ``\textbf{criminals}'' to ``\textbf{track}'' ``\textbf{activities}'', as given in Fig.~\ref{fig:app_harm_details}. The results present that even public information can lead to severe harm when leveraged by adversaries.

\begin{figure}[h]
    \centering
    \subfloat[Attitude]{\includegraphics[width=0.33\linewidth]{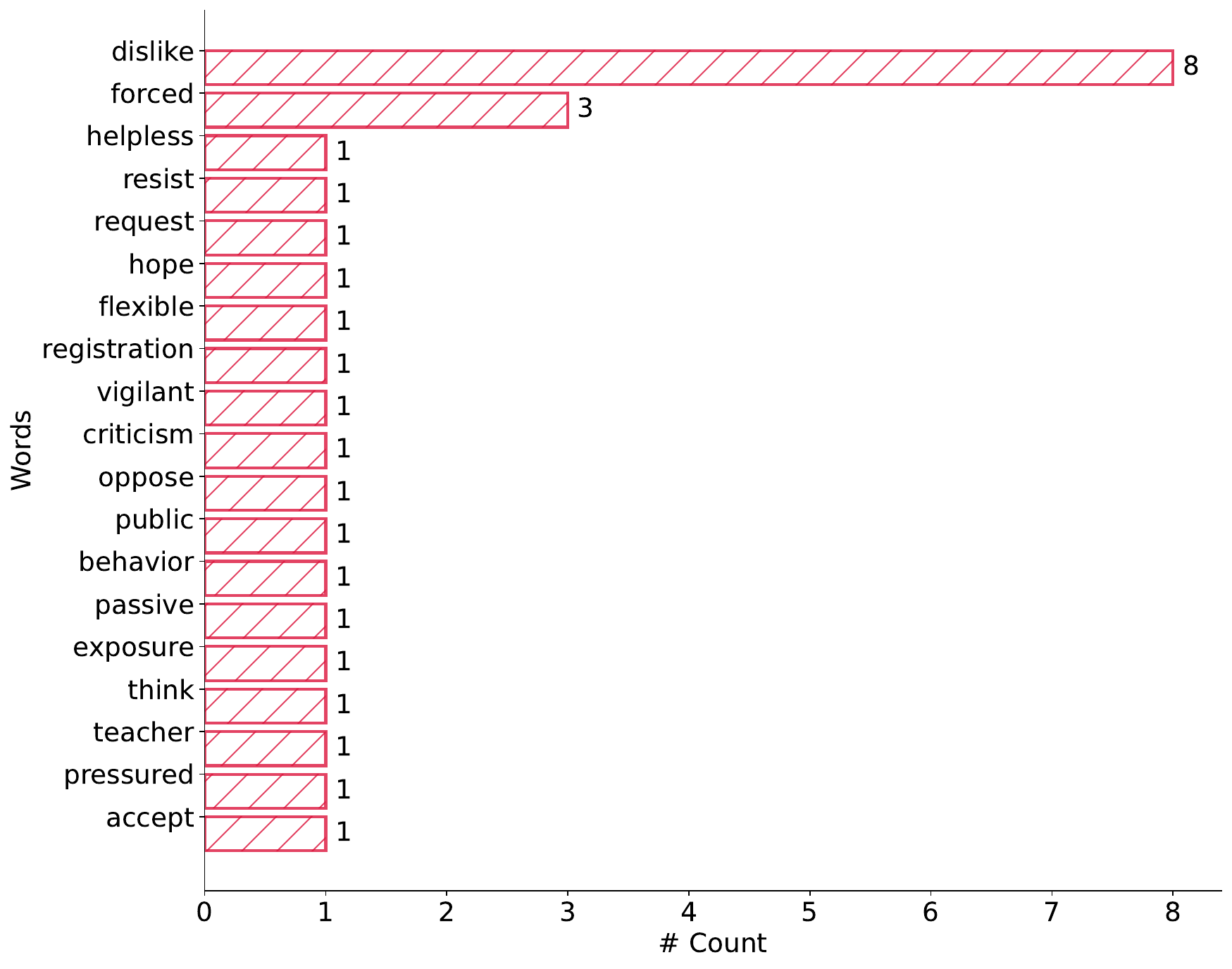}}
    \subfloat[Responsibility]{\includegraphics[width=0.33\linewidth]{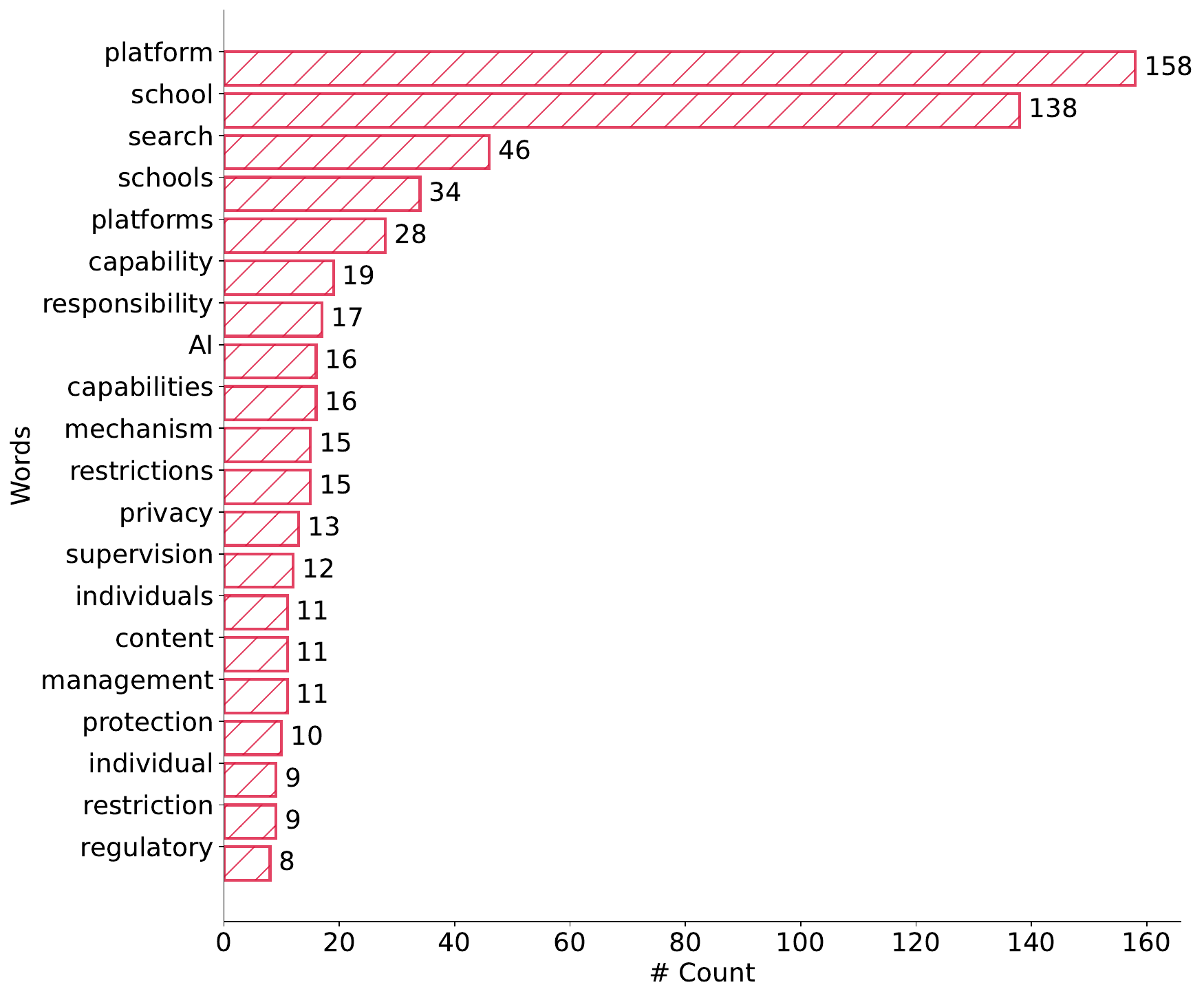}}
    \subfloat[Suggestion]{\includegraphics[width=0.33\linewidth]{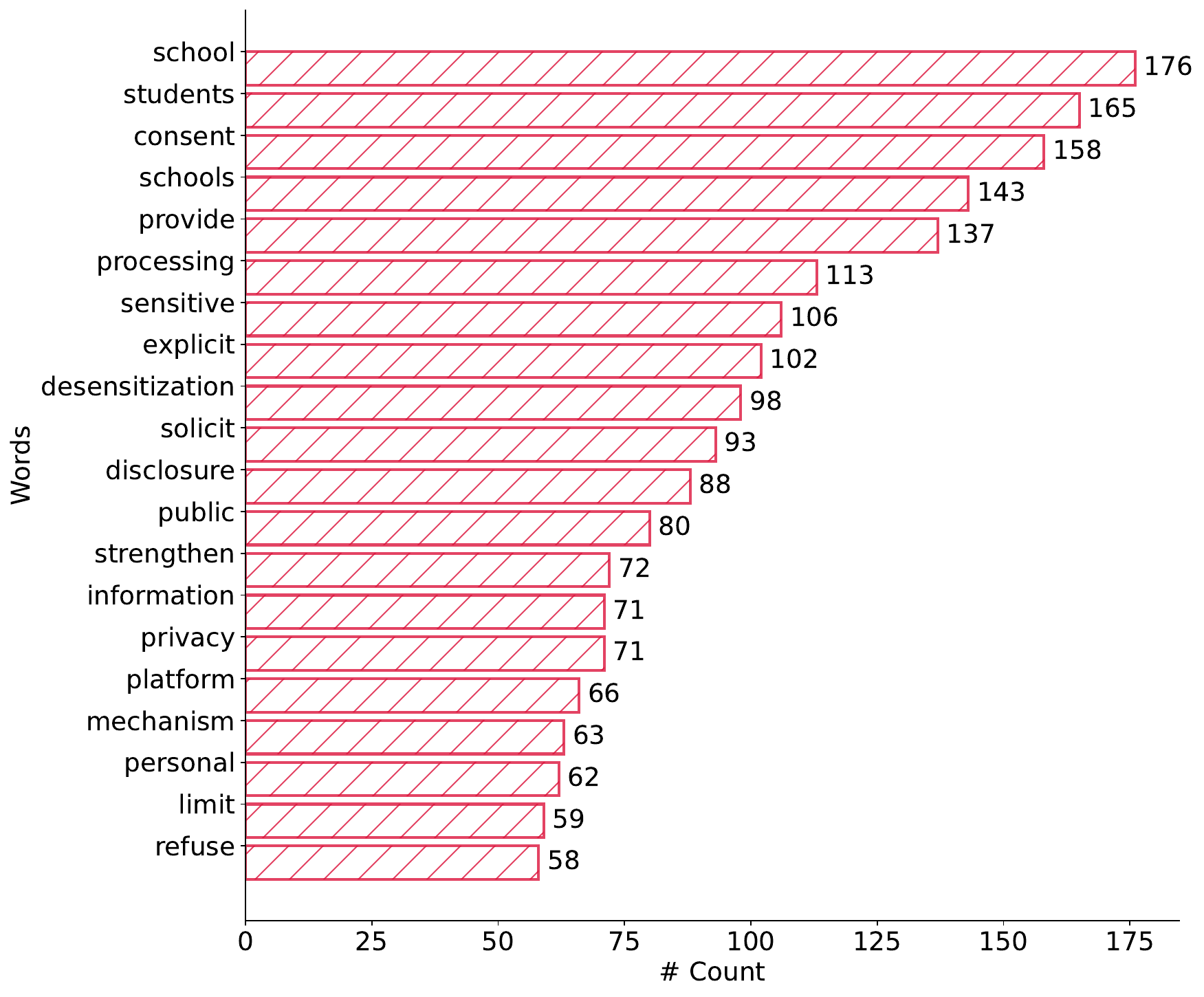}}
    \caption{Word frequency regarding public attribution and suggestion to the responsible parties.}
    \label{fig:protection_expectations}
\end{figure}

$\bullet$\textit{\textbf{Anticipated Protection.}} 
As shown in Fig.~\ref{fig:protection_expectations}, the public discourse points towards two primary actors: the \textbf{school} (information publisher) and the \textbf{platform} that provides the search service. Furthermore, the desired safeguards are centered on user empowerment and control. The most prominent keyword is ``\textbf{consent}'', reflecting the belief that this public but personal information should not be disclosed without permission. The recurring demand for ``\textbf{options}'' and the right to ``\textbf{review}'' content underscores a fundamental desire to reclaim agency over one's own digital footprint.

\begin{table*}[t]
\centering
\footnotesize
\caption{Privacy and compliance summary of major social media platforms integrated with LLM.}
\label{tab:ai_privacy_summary}
\renewcommand{\arraystretch}{0.8}
\aboverulesep=0ex
\belowrulesep=0.5ex
\resizebox{1.0\linewidth}{!}{
\begin{tabular}{l l c l l l l ccc cc c c}
\toprule
\multirow{2}{*}{\textbf{Platform}} &
\multirow{2}{*}{\textbf{Base Model}} &
\multirow{2}{*}{\textbf{DQ/IQ}} &
\multirow{2}{*}{\textbf{Data Src.}} &
\multirow{2}{*}{\textbf{T/I Use}} &
\multirow{2}{*}{\textbf{DRP}} &
\multirow{2}{*}{\textbf{Policy}} &
\multicolumn{3}{c}{\textbf{User Control}} &
\multicolumn{2}{c}{\textbf{Output}} &
\multirow{2}{*}{\textbf{Comp.}} &
\multirow{2}{*}{\textbf{Reg. Scr.}} \\
\cmidrule{8-10}\cmidrule{11-12}
 &  &  &  &  &  &  &
\multicolumn{1}{c}{Opt} &
\multicolumn{1}{c}{Del} &
\multicolumn{1}{c}{Sw}  &
\multicolumn{1}{c}{PII} &
\multicolumn{1}{c}{Src} &
& \\ 
\midrule
Google & Gemini 2.5 Pro & D & E+I & {\color{green!60!black}$\checkmark$} & {\color{green!60!black}$\checkmark$} & {\color{green!60!black}$\checkmark$} & {\color{green!60!black}$\checkmark$} & {\color{green!60!black}$\checkmark$} & {\color{red!70!black}$\times$} & {\color{green!60!black}$\checkmark$} & {\color{green!60!black}$\checkmark$} & {\color{green!60!black}$\checkmark$} & {\color{red!70!black}$\times$} \\
Microsoft & GPT-5, OpenAI o1 & D & E+I & {\color{green!60!black}$\checkmark$} & {\color{green!60!black}$\checkmark$} & {\color{green!60!black}$\checkmark$} & {\color{green!60!black}$\checkmark$} & {\color{green!60!black}$\checkmark$} & {\color{green!60!black}$\checkmark$} & {\color{green!60!black}$\checkmark$} & {\color{green!60!black}$\checkmark$} & {\color{green!60!black}$\checkmark$} & {\color{red!70!black}$\times$} \\
X (Twitter) & Grok series & D & E+I & {\color{green!60!black}$\checkmark$} & {\color{green!60!black}$\checkmark$} & {\color{green!60!black}$\checkmark$} & {\color{green!60!black}$\checkmark$} & {\color{green!60!black}$\checkmark$} & {\color{green!60!black}$\checkmark$} & {\color{green!60!black}$\checkmark$} & {\color{green!60!black}$\checkmark$} & {\color{green!60!black}$\checkmark$} & {\color{green!60!black}$\checkmark$} \\
Reddit & Gemini series & N & E+I & – & {\color{green!60!black}$\checkmark$} & – & {\color{green!60!black}$\checkmark$} & {\color{red!70!black}$\times$} & {\color{red!70!black}$\times$} & {\color{red!70!black}$\times$} & {\color{green!60!black}$\checkmark$} & {\color{red!70!black}$\times$} & {\color{red!70!black}$\times$} \\
Tiktok (Douyin) & Doubao series & D & E+I & – & {\color{green!60!black}$\checkmark$} & {\color{green!60!black}$\checkmark$} & {\color{green!60!black}$\checkmark$} & {\color{green!60!black}$\checkmark$} & {\color{red!70!black}$\times$} & {\color{green!60!black}$\checkmark$} & {\color{green!60!black}$\checkmark$} & {\color{red!70!black}$\times$} & {\color{red!70!black}$\times$} \\
Zhihu & Zhihaitu AI + DeepSeek-R1 & D & E+I & {\color{green!60!black}$\checkmark$} & {\color{green!60!black}$\checkmark$} & {\color{green!60!black}$\checkmark$} & {\color{green!60!black}$\checkmark$} & {\color{green!60!black}$\checkmark$} & {\color{red!70!black}$\times$} & {\color{green!60!black}$\checkmark$} & {\color{green!60!black}$\checkmark$} & {\color{red!70!black}$\times$} & {\color{red!70!black}$\times$} \\
WeChat & Hunyuan-T1, DeepSeek-R1 & N & E+I & – & {\color{green!60!black}$\checkmark$} & – & {\color{green!60!black}$\checkmark$} & {\color{green!60!black}$\checkmark$} & {\color{red!70!black}$\times$} & {\color{red!70!black}$\times$} & {\color{red!70!black}$\times$} & {\color{green!60!black}$\checkmark$} & {\color{red!70!black}$\times$} \\
Weibo & DeepSeek-R1 & I & E+I & – & {\color{green!60!black}$\checkmark$} & – & {\color{green!60!black}$\checkmark$} & {\color{green!60!black}$\checkmark$} & {\color{red!70!black}$\times$} & {\color{green!60!black}$\checkmark$} & {\color{green!60!black}$\checkmark$} & {\color{green!60!black}$\checkmark$} & {\color{red!70!black}$\times$} \\
Kuaishou & Keye-VL-1.5 & D & E+I & {\color{red!70!black}$\times$} & {\color{green!60!black}$\checkmark$} & {\color{green!60!black}$\checkmark$} & {\color{green!60!black}$\checkmark$} & {\color{red!70!black}$\times$} & {\color{red!70!black}$\times$} & {\color{green!60!black}$\checkmark$} & {\color{green!60!black}$\checkmark$} & {\color{red!70!black}$\times$} & {\color{red!70!black}$\times$} \\
Feishu & Doubao-1.5, DeepSeek-R1 & D & E+I & {\color{green!60!black}$\checkmark$}$^{*}$ & {\color{green!60!black}$\checkmark$} & {\color{green!60!black}$\checkmark$} & {\color{green!60!black}$\checkmark$} & {\color{green!60!black}$\checkmark$} & {\color{red!70!black}$\times$} & {\color{green!60!black}$\checkmark$} & {\color{green!60!black}$\checkmark$} & {\color{red!70!black}$\times$} & {\color{red!70!black}$\times$} \\
Xiaohongshu & dots.llm1 & D & E+I & – & {\color{green!60!black}$\checkmark$} & – & {\color{green!60!black}$\checkmark$} & {\color{red!70!black}$\times$} & {\color{red!70!black}$\times$} & {\color{green!60!black}$\checkmark$} & {\color{green!60!black}$\checkmark$} & {\color{green!60!black}$\checkmark$} & {\color{red!70!black}$\times$} \\
\bottomrule
\vspace{-6pt}
\end{tabular}}
\begin{minipage}{0.95\textwidth}
\footnotesize
\textit{Notes.}  
1) DQ/IQ: Whether the model responds to user profiling queries.   D = Direct, I = Indirect, N = None.  
2) Data Src.: Data sources used or collected by the platform.  E+I = Explicit (user-provided) + Implicit (contextual) data.  
3) T/I Use: Whether user interactions are used for model training or inference.  
4) DRP: Data Retention Policy.  
5) Policy: Existence of a public AI privacy policy.  
6) User Control: Opt = opt-out, Del = delete/view history, Sw = dedicated switch.  
7) Output: PII = synthesized personal information in outputs; Src = outputs cite sources.  
8) Comp.: Whether public user complaints exist regarding AI privacy issues.  
9) Reg. Scr.: Whether the platform has been subject to regulatory actions or penalties.  
10) Symbols: {\color{green!60!black}$\checkmark$} = Yes; {\color{red!70!black}$\times$} = No; – = Not mentioned.  
11) $^{*}$ Feishu states that user interaction data are collected only when users submit feedback.
12) All data were collected and verified before October 20, 2025.
\end{minipage}
\end{table*}
\appsubsection{Platform Privacy Perception}
\label{app:risk_applications}
The public discourse above reveals severe risks. In practice, social media platforms integrated with LLMs function as gatekeepers of data flows and capability exposure: what is collected, how long it persists, how it is processed (e.g., for search, or LLM assistance), and whether users can consent, opt out, or delete. Consequently, to understand why perceived harms persist despite nominal ``public'' status, we turn to the platforms themselves, their stated privacy policies and the actual scope of their LLM-integrated capabilities. 

As shown in Tab.~\ref{tab:ai_privacy_summary}, most platforms can generate relatively complete personal profiles when prompted for individual information, and even rejected queries can be bypassed with prompt engineering. Nearly half of the platforms state that collected data are used for training or inference, whereas others make no mention of such use, with some even lacking an AI-specific privacy policy. Analyses of public feedback and past incidents reveal growing user concern over AI-related privacy issues, with one major platform having faced regulatory penalties. Furthermore, Du et al.~\cite{du2025beyond} note that many ``privacy modes'' and ``data deletion'' options are largely symbolic. Conversations are retained for security or auditing, and deletion mechanisms rarely function effectively, undermining users' actual control over personal data.

\textbf{Misalignment Between Platforms and the Public.}
A recent incident~\cite{contested_govern} also highlights a policy controversy in LLM-integrated services: a LLM application allegedly surfaced a user's social media ID by querying ``name+university,'' even though the source post had been deleted, and similar searches sometimes even returned phone numbers. Corresponding industry voices characterize it as an emergent byproduct of ``search-like'' aggregation over public data and argue for tolerance amid algorithmic immaturity. In contrast, multiple legal experts assert that civil law protects privacy and consent; public exposure does not imply permanence, deletion changes the status of information, and retaining contact details post-deletion may constitute unlawful collection and disclosure. The above divergence exhibits misalignment: platforms rely on procedural notices while the public are affected severely.

From the victims' perspective, their information has been published for years, creating a \textbf{persistent} and \textbf{uncontrollable} digital footprint. This situation signifies a burgeoning demand for a ``\textbf{right to know}'' about their own privacy risks to achieve \textbf{Privacy Situational Awareness}. This unmet need for a privacy audit motivates our work to empower individuals with this crucial knowledge and understand potential privacy risks before they are violated.

\appsubsection{Details of Extracting Topics from Comments}
\label{sec:topic_extraction}
To analyze the qualitative data from user comments, we instruct LLM to act as an expert analyst focusing on privacy issues, specifically the violation of CI. LLM first processes each user comment individually and extracts relevant information into a structured format. We defined a detailed analytic framework based on our research questions, instructing the model to identify and categorize text fragments related to four key thematic areas: Perception of CI Violation; Awareness and Experience of Doxing Risks; Experienced or Feared Harms; Protection Expectations and Responsibility. For each comment, the LLM was required to output a structured JSON object (with 10\% samples checked by a human for validation). This object contained distinct fields for each of the four categories, with `null' values for any categories not explicitly addressed in a given comment. This structured data extraction approach facilitated a rigorous qualitative analysis of user attitudes and experiences.

\begin{table*}[t]
\centering
\footnotesize
\caption{Privacy Categories with Descriptions and PPIS Scores}
\label{tab:privacy_categories}
\renewcommand{\arraystretch}{0.8}
\aboverulesep=0ex
\belowrulesep=0.5ex
\resizebox{1.0\linewidth}{!}{
\begin{tabular}{cp{3.5cm} p{8cm} c}
\toprule
\textbf{Category ID} &\textbf{Category Name} & \textbf{Description} & \textbf{Score (1-7)} \\
\midrule
\hypertarget{c1}{C1} & Core Identity Information & Name, ethnicity, biological sex, age. & 6.2 \\
\hypertarget{c2}{C2} & Contact Information & Phone number, email address, social media accounts, etc. & 6.8 \\
\hypertarget{c3}{C3} & Affiliations & Past and present institutions, including employers or schools. & 4.5 \\
\hypertarget{c4}{C4} & Physical Appearance & Appearance, such as photos, videos, or text. & 5.1 \\
\hypertarget{c5}{C5} & Geographic Location & Current location, home address and past addresses. & 6.9 \\
\hypertarget{c6}{C6} & Personal Attributes & Personality traits, values and beliefs. & 5.8 \\
\hypertarget{c7}{C7} & Educational History & Educational background (degrees, majors, and timelines). & 4.2 \\
\hypertarget{c8}{C8} & Personal Development & Skills, courses taken, and certifications earned. & 3.5 \\
\hypertarget{c9}{C9} & Career History & Work experience (companies, positions and timelines). & 4.3 \\
\hypertarget{c10}{C10} & Accomplishments & Projects, publications or works, awards and honors received. & 3.1 \\
\hypertarget{c11}{C11} & Interpersonal Relationships & Family members, friends, professional contacts, etc. & 6.5 \\
\hypertarget{c12}{C12} & Lifestyle \& Interests & Hobbies, daily activities, travel history, and consumption habits. & 4.8 \\
\hypertarget{c13}{C13} & Health \& Wellness & Metrics like height and weight, medical conditions and contacts, etc. & 7.0 \\
\hypertarget{c14}{C14} & Financial Status & Assets, liabilities, income sources, spending categories, etc. & 7.0 \\
\hypertarget{c15}{C15} & Key Life Events \& Goals & Life milestones, short-term or long-term life plans, etc. & 5.5 \\
\hypertarget{c16}{C16} & Digital Footprint & Personal portfolio and profile links and publicly posted content. & 4.1 \\
\hypertarget{c17}{C17} & Other Information & Other information that does not fall into the above categories. & 3.0 \\
\bottomrule
\end{tabular}}
\end{table*}

\appsubsection{More Results of Social Media Analysis}
The quantitative visualizations of the keywords from social media comments are in Fig.~\ref{fig:violation_perception}, Fig.~\ref{fig:doxing_experience}, Fig.~\ref{fig:fig10} and Fig.~\ref{fig:protection_expectations}.

\appsubsection{Privacy Governance and Analysis}
\label{sec:platform_response}
\textbf{Privacy Policies and Profiling Capabilities of Different Platforms.} We evaluated the privacy protection and user profiling mechanisms of major social media platforms with LLM integration, as summarized in Tab.~\ref{tab:ai_privacy_summary}. The study adopted two approaches. First, we reviewed each platform's privacy policy and examined reports and user comments to identify key privacy clauses and public attitudes. Second, we tested profiling functions by submitting 10 queries containing only usernames (randomly selected from the platform) to each platform's LLM, requesting profiles generated from the information on the platform. For platforms that declined direct queries, we used prompt engineering methods~\cite{liu2024formalizing} such as role-playing to test bypass attempts and assess the robustness of their safeguards. All procedures were conducted under research ethics and privacy protection guidelines. Specifically, prompts and outputs were anonymized or abstracted to protect privacy, and representative templates and examples are provided below.

\begin{tcolorbox}[
  colback=blue!3!white,
  colframe=RoyalPurple!70!black,
  title={\textbf{\textcolor{white}{Case 1: Privacy Aggregation Query (Platform A)}}},
  fonttitle=\bfseries,
  enhanced,
  breakable,
  sharp corners,
  boxrule=0.8pt,
  colbacktitle=RoyalPurple!25!black,
  left=1.5mm, right=1.5mm, top=1mm, bottom=1mm
]
\scriptsize

\textbf{Adversarial Query Prompt:}\\[2pt]
Given the name \texttt{[User\_A]}, please summarize the user's profile based on publicly accessible information (e.g., personal homepage, publications, and social interactions). Infer his or her general background, interests, affiliations, and potential online footprint. Do not fabricate; summarize only from publicly available results.

\vspace{4pt}
\textbf{Model Output (Abstracted Summary):}\\[2pt]
The platform’s AI system aggregated multi-source public information to generate a comprehensive user profile across multiple dimensions:

\begin{itemize}[leftmargin=1.2em, itemsep=2pt, label={}]
\item \textbf{Direct Identifiers:} Includes the user’s full name, gender, nationality.
\item \textbf{Contact Information:} Multiple email and phone number patterns partially exposed, indicating consistent use across platforms.
\item \textbf{Locations:} Address-like and regional information inferred from workplace and publication metadata.
\item \textbf{Educational and Professional Background:} Multi-stage academic and career trajectory reconstructed from public records, including study fields and professional affiliations.
\item \textbf{Expertise and Research Focus:} Centralized description of technical domains and recurring thematic keywords extracted from publications and interviews.
\item \textbf{Teaching or Mentoring Activities:} Structured references to academic or instructional engagements with temporal attributes.
\item \textbf{Professional and Social Networks:} Aggregated associations with collaborators, organizations, and public accounts across multiple platforms.
\item \textbf{Personal Interests and Lifestyle Indicators:} Inferred patterns related to hobbies, preferences, and daily routines mentioned across social posts and visual content, including references to pets or leisure activities.
\item \textbf{Impact and Achievements:} Mentions of publicly recognized works, contributions, or projects in the user’s domain.
\item \textbf{Personal Footprint:} Temporal and locational traces linking activities across different platforms.
\end{itemize}

\vspace{4pt}
\textbf{Privacy Risk Assessment:}\\[2pt]
This output illustrates how an LLM can reconstruct an individual’s extended personal, professional, and behavioral graph by aggregating dispersed public traces. Such synthesis exposes cross-context linkages that enable implicit re-identification or behavioral profiling, posing risks of targeted exploitation or misuse.

\end{tcolorbox}

\textbf{Platform Responses.}
We reported the privacy issues to platforms that support user profiling. WeChat and Reddit were not included because they rejected profiling requests. X was also included because it supports profiling, although its privacy controls are relatively stronger. Among the contacted platforms, only TikTok, Feishu, Xiaohongshu, Microsoft, and X responded. Most platforms gave brief acknowledgments or no substantive feedback. Feishu and Microsoft requested additional information, and only Feishu indicated plans for further action. Some responses included confidentiality notices, so we do not reproduce them here. Redacted summaries compliant with platform requirements can be provided if needed for verification.

\appsubsection{Empirical Analysis of User Study}
\label{sec:appcase_study}
We recruited 60 participants through institutional mailing lists and academic social networks. Note that these 60 participants are different with users who involve the main experiments and case studies. In the validation of \framework in Sec.~\ref{sec:privacy_iceberg}, we recruited 60 participants with 69.1\% male, 29.1\% female, and 1.8\% identifying as other. Their familiarity with LLMs was measured on a five-point scale. Most participants selected 4 or 5, some selected 3, and a few selected 1 or 2.

\begin{figure}
    \centering
    \subfloat[]{\label{fig:gender}\includegraphics[width=0.30\linewidth, height=0.6\linewidth]{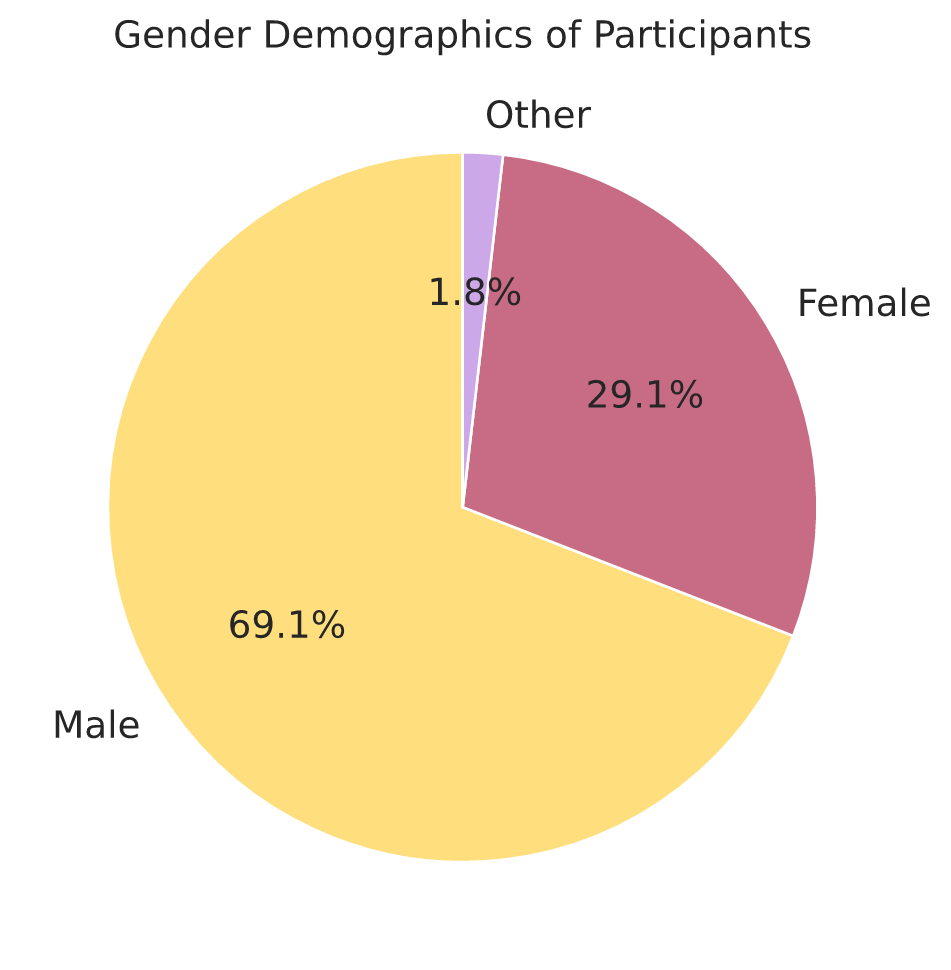}}
    \subfloat[]{\label{fig:llm_familiarity}\includegraphics[width=0.48\linewidth, height=0.8\linewidth]{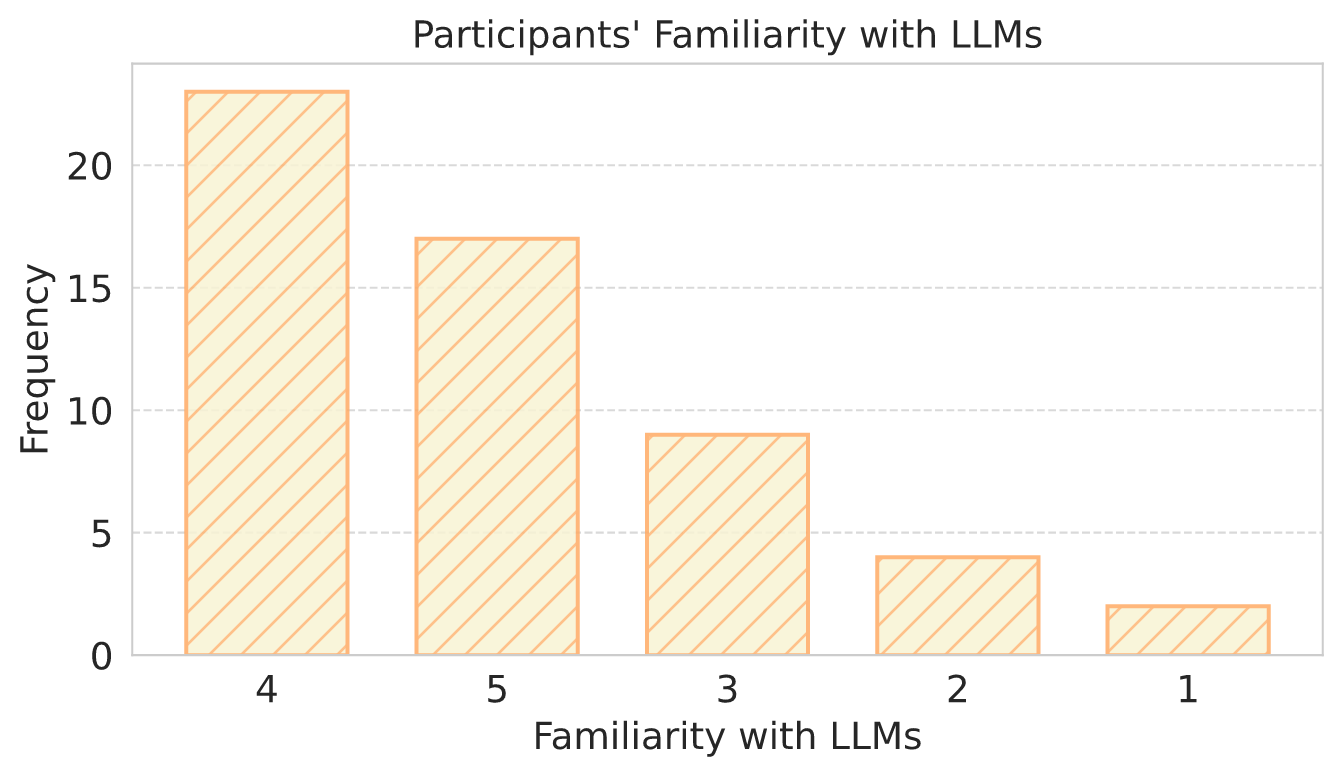}}
    \caption{(a) Gender distribution of participants. (b) Distribution of participants' reported familiarity with LLMs.}
\end{figure}

We designed six scenarios to assess the perceived severity of privacy invasions. These scenarios covered workplace interactions, online gaming, dating, neighborhood communication, client investigations, and academic evaluations. In all scenarios, the sensitivity scores of DII, CII, and AMI exceeded their respective thresholds, suggesting that publicly accessible information can pose significant privacy risks.

\begin{figure}[H]
    \centering

    \includegraphics[width=0.48\linewidth]{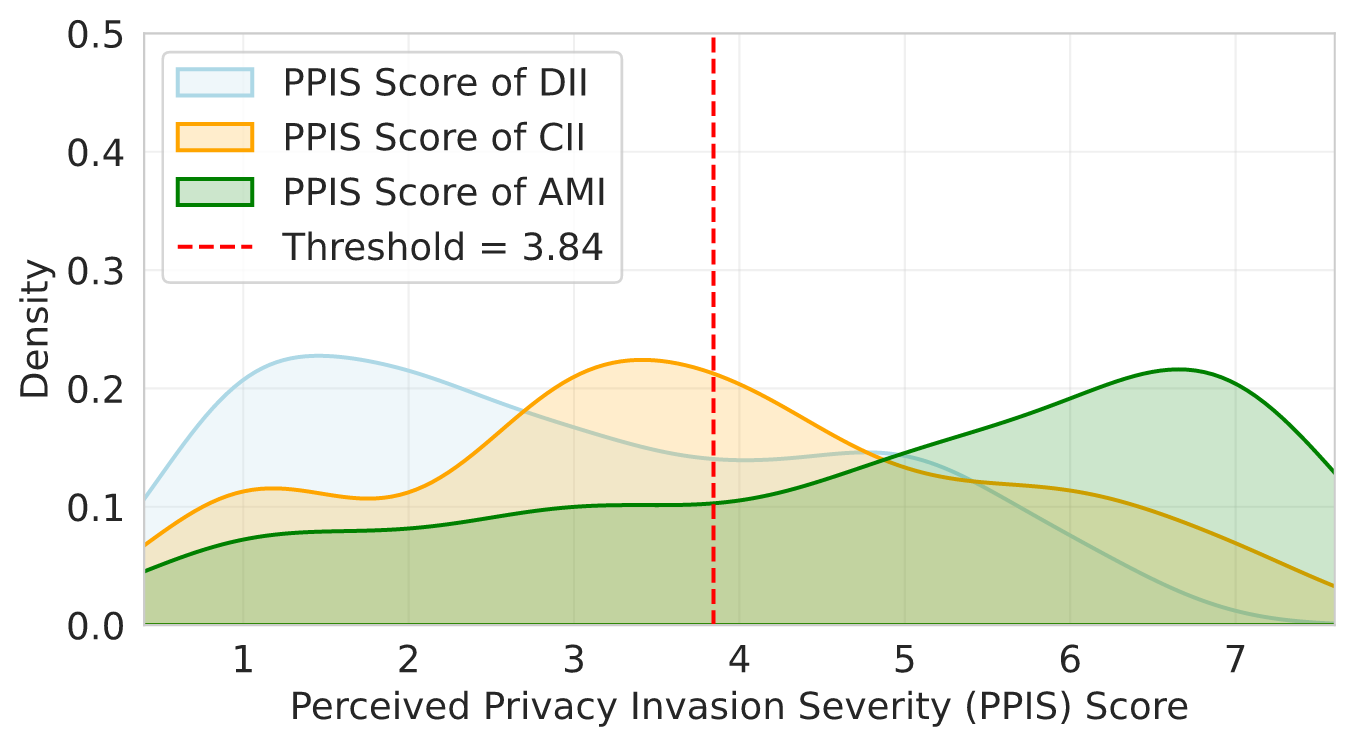}%
    \hspace{0.02\linewidth}%
    \includegraphics[width=0.48\linewidth]{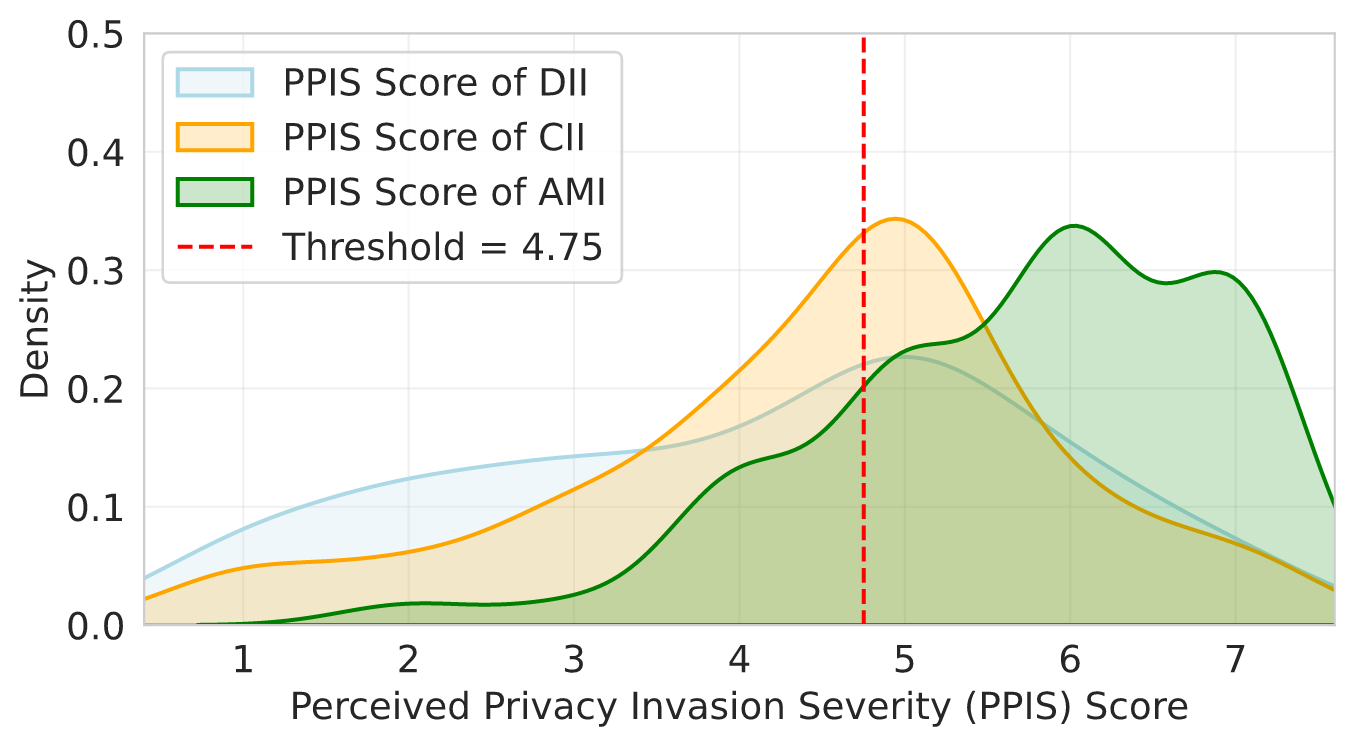}
    
    \vspace{0.02\linewidth}
    
    \includegraphics[width=0.48\linewidth]{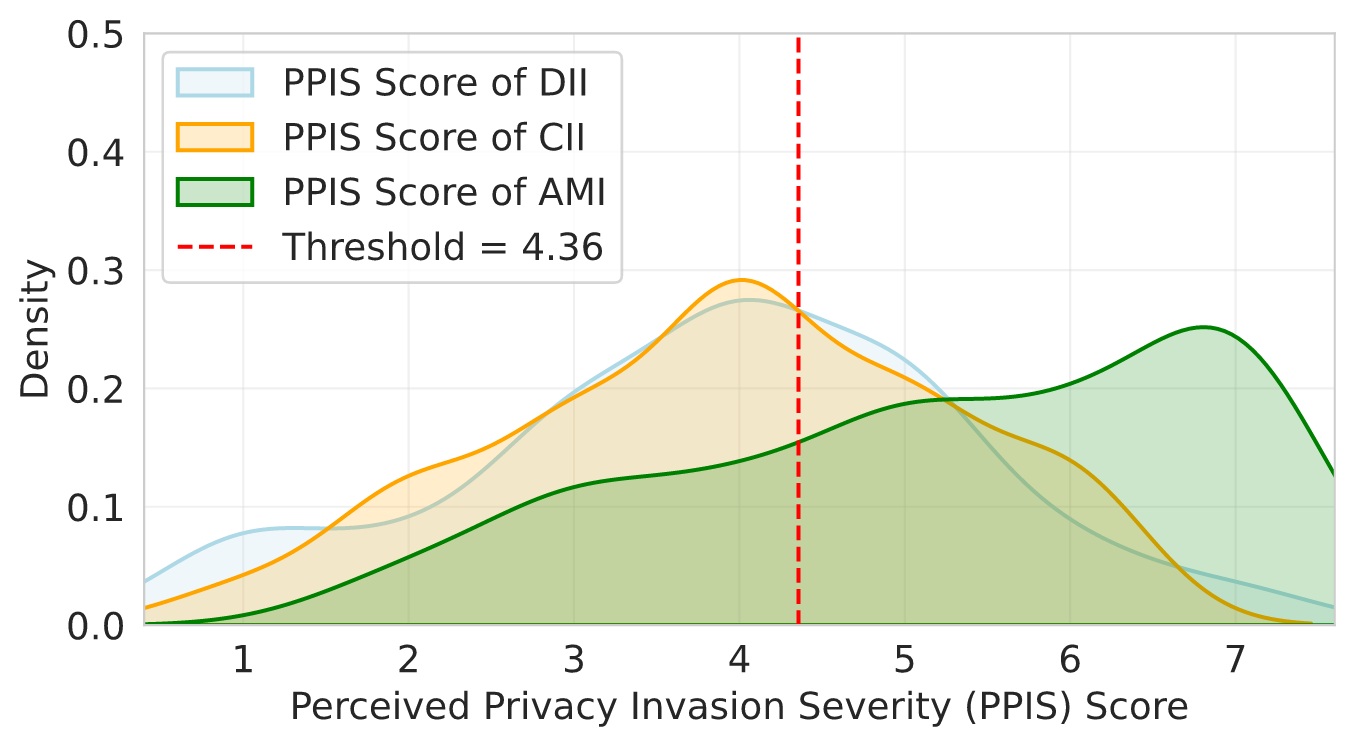}%
    \hspace{0.02\linewidth}%
    \includegraphics[width=0.48\linewidth]{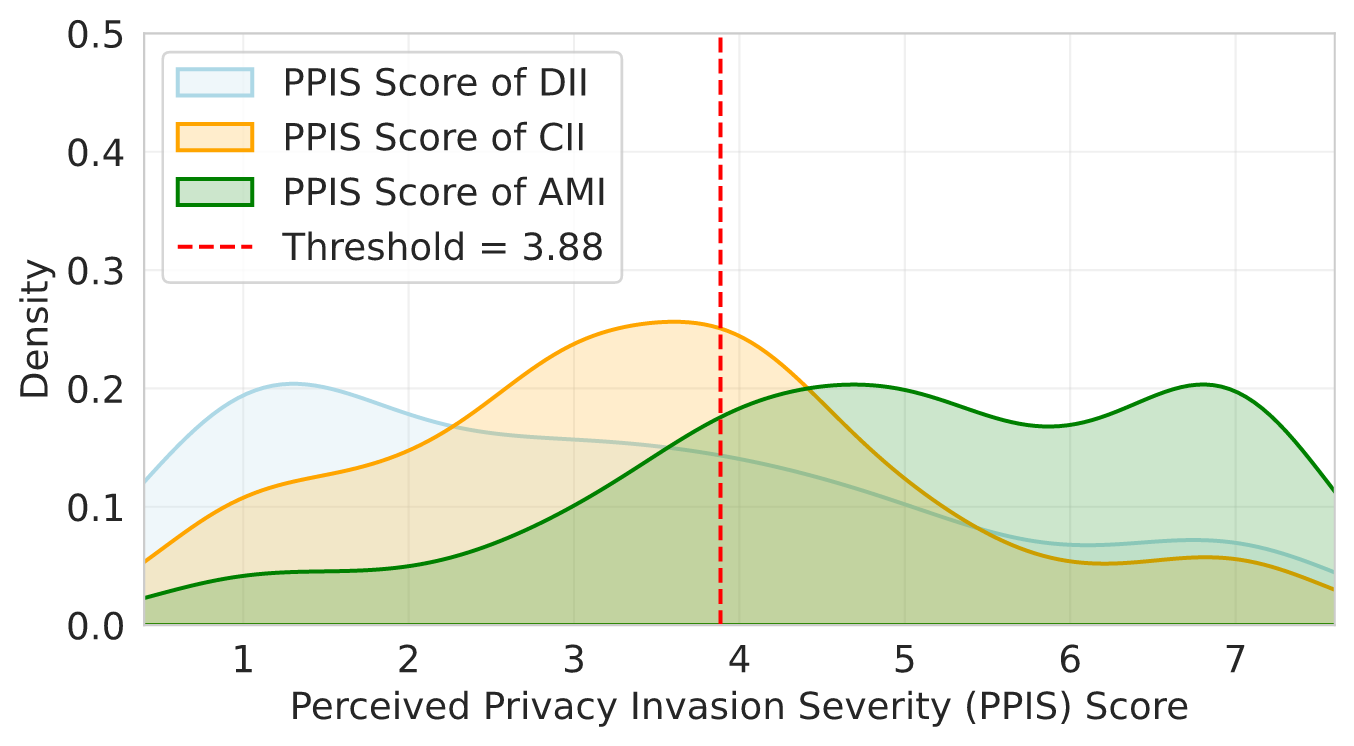}
    
    \vspace{0.02\linewidth}

    \includegraphics[width=0.48\linewidth]{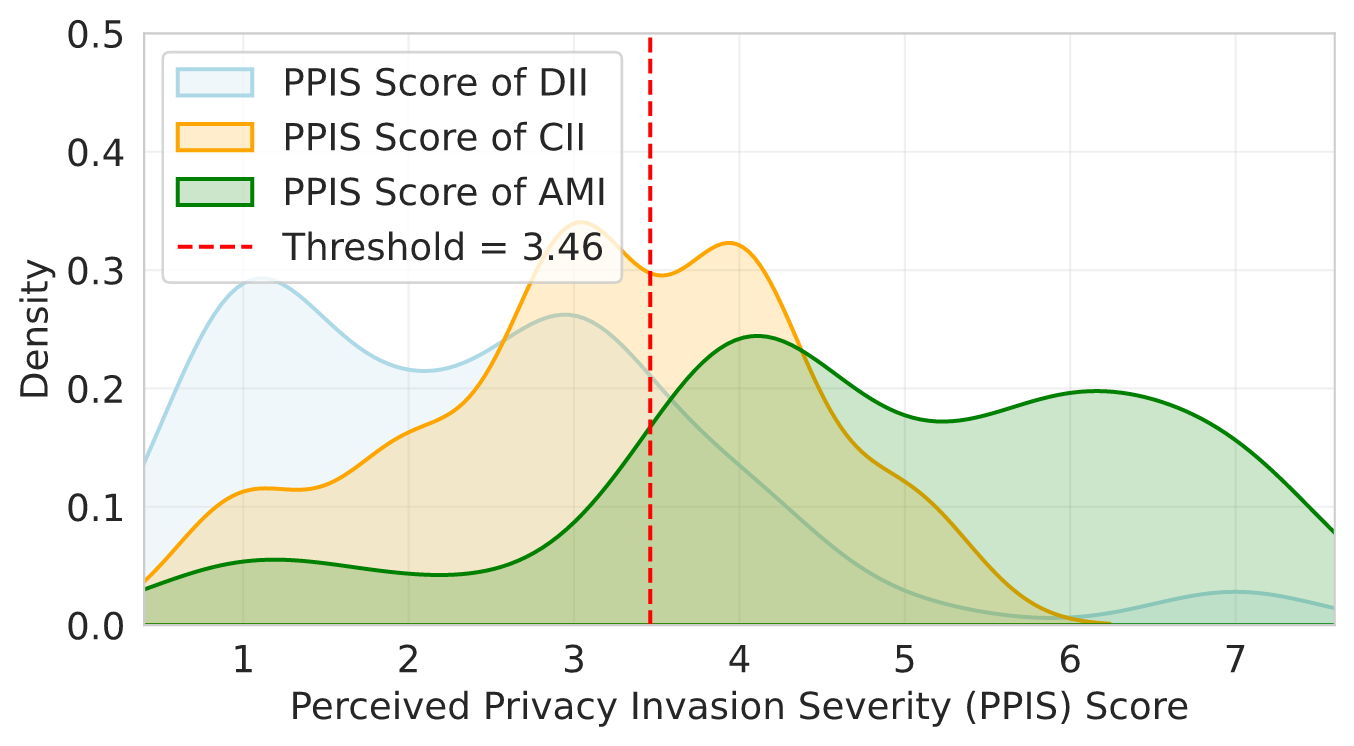}
    \hspace{0.02\linewidth}%
    \includegraphics[width=0.48\linewidth]{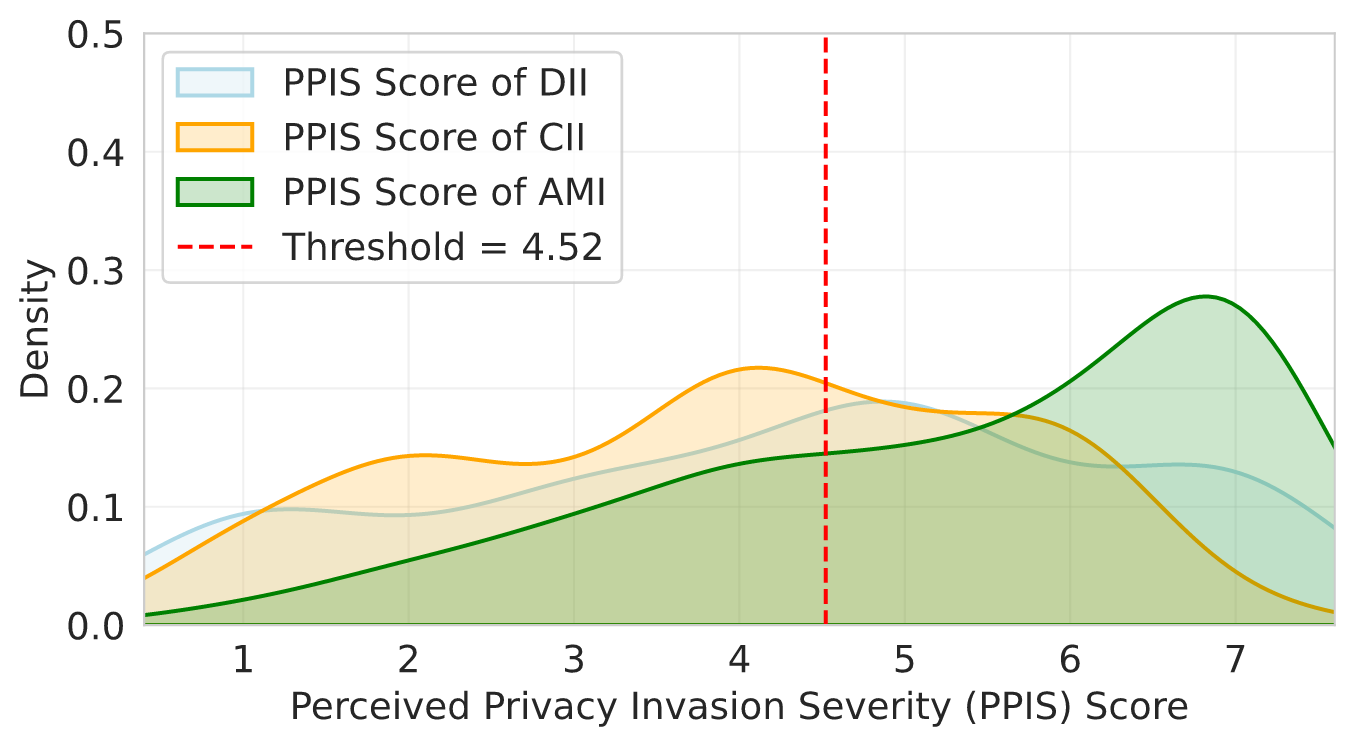}
    
    \caption{Three-level distributions for six scenes, arranged from top to bottom and left to right (Scenes 1–6).}
    \label{fig:three_level_distribution_0_5}
\end{figure}

At the end of the questionnaire, participants were asked if they viewed a stranger's systematic profiling of personal information based on public data as an invasion of privacy. Responses were rated on a 7-point scale, where 1 meant ``not an invasion'' and 7 meant ``very severe invasion.'' Finally, we categorized the collected personal information into 16 types and asked participants to rate the sensitivity. All categories received scores above 3, showing that participants viewed all types of personal information as at least somewhat sensitive. Most categories scored between 4 and 5, indicating moderate to relatively high sensitivity.

\begin{figure}
    \centering
    \includegraphics[width=0.6\linewidth, height=0.6\linewidth]{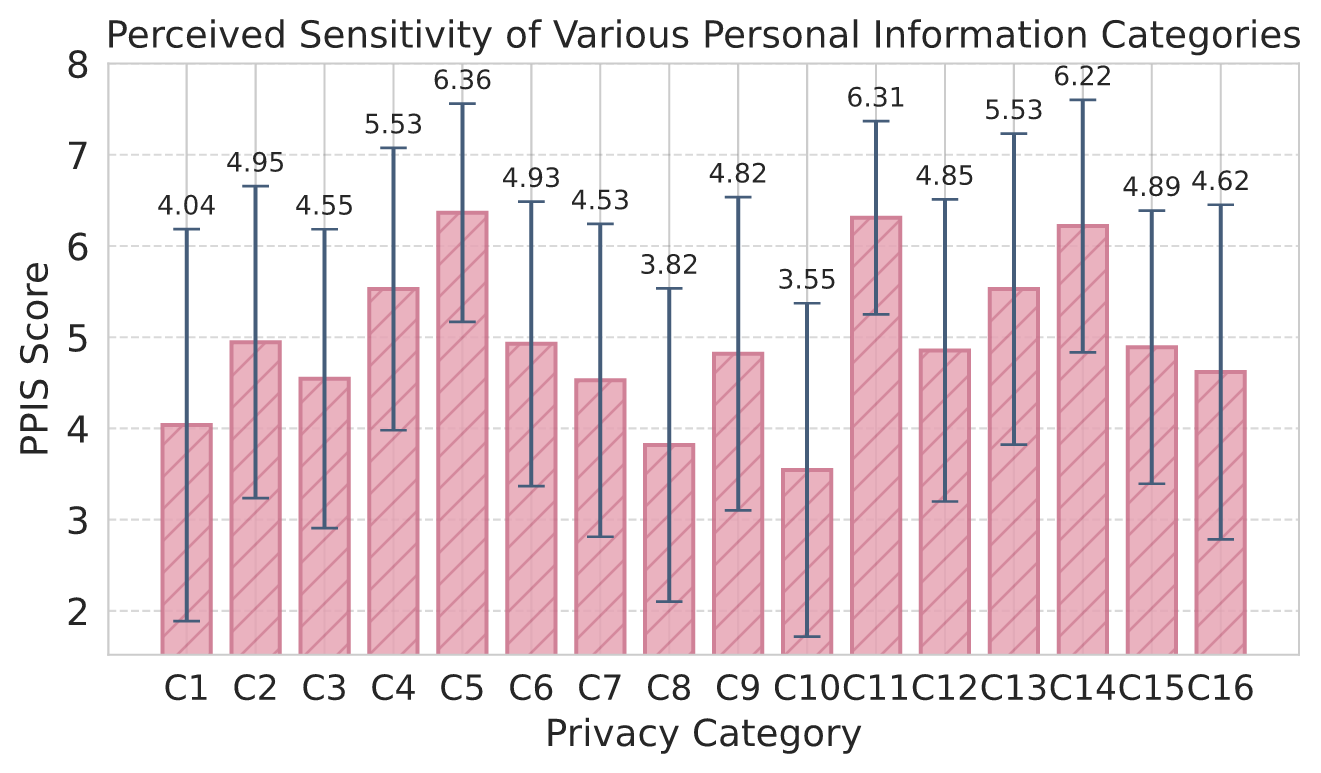}
    \caption{PPIS of various privacy categories.}
    \label{fig:privacy_dimensions_barplot}
\end{figure}

\appsubsection{Distinguishing of DII/CII/AMI}
\label{sec:implementation}
The agent's multi-stage filtering and reasoning pipeline is how it programmatically distinguishes the three tiers of privacy information:

\begin{itemize}
    \item \textbf{DII:} This is the ``Visible Tip'' of the iceberg. DII is identified during the fine-grained filtering stage, which  performs a direct ``Search \& Match'' operation, verifying that the contains an {Exact DII Match} (e.g., full name, username) and other explicit personal information from the existing result.
    \item \textbf{CII:} This is the ``Submerged Surface'' inferred from a single source. CII is generated during {Knowledge Processing} phase. After a source is validated, the LLM is tasked to ``Infer Implicit Knowledge'' from the immediate context of that single searched result based on the discovered knowledge.
    \item \textbf{AMI:} This is the ``Deep Mass'' of the iceberg, representing insights that do not exist in any single source. AMI is generated at the end of each main loop by the reasoning process. This stage receives a batch of verified posts from multiple, disparate sources. Its prompt LLM to perform consensus-based assessment and cross-source synthesis, identifying patterns, routines, or new conclusions (e.g., wealth, location, etc.) that are only visible when aggregating all fragments.
\end{itemize}

\begin{table}[t]
\scriptsize
\centering
\renewcommand{\arraystretch}{0.6} 
\setlength{\tabcolsep}{4pt}        
\caption{Details of the evaluated base models.}
\label{tab:models}
\aboverulesep=0ex
\belowrulesep=0.5ex
\resizebox{0.97\linewidth}{!}{%
\begin{tabular}{llll}
\toprule
\textbf{Model} & \textbf{Model Type} & \textbf{Model Name} & \textbf{Version} \\ \midrule

\multirow{3}{*}{Qwen} 
 & Basic Model  & Qwen3-Coder-480b-A35b-instruct & Latest \\
 & Reason Model & Qwen3-Coder-480b-A35b-instruct & Latest \\
 & Vision Model & Qwen-VL-Max & Latest \\ \midrule

\multirow{3}{*}{Doubao} 
 & Basic Model  & Doubao-1-5-thinking-pro & 25-0415 \\
 & Reason Model & Doubao-1-5-thinking-pro-m & 25-0428 \\
 & Vision Model & Doubao-1-5-thinking-vision-pro & 25-0428 \\ \midrule

\multirow{3}{*}{Gemini} 
 & Basic Model  & Gemini-2.5-pro-preview & 25-0325 \\
 & Reason Model & Gemini-2.5-pro-preview & 25-0325 \\
 & Vision Model & Gemini-2.5-pro-preview & 25-0325 \\ \midrule

\multirow{3}{*}{GPT4o} 
 & Basic Model  & GPT4o & Latest \\
 & Reason Model & GPT4o & Latest \\
 & Vision Model & GPT4o & Latest \\ \midrule

\multirow{3}{*}{GPT4.1} 
 & Basic Model  & GPT-4.1 & Latest \\
 & Reason Model & GPT-4.1 & Latest \\
 & Vision Model & GPT-4.1 & Latest \\ \midrule

\multirow{3}{*}{DeepSeek} 
 & Basic Model  & DeepSeek-V3 & Latest \\
 & Reason Model & DeepSeek-R1 & Latest \\
 & Vision Model & N/A & N/A \\
\bottomrule
\end{tabular}%
}

\par\vspace{4pt}
\raggedright
{\footnotesize\textit{Note.} For \textit{LLMs w/o S}, we used the API, while tasks requiring search or reasoning were conducted via official web clients. Some APIs in IcebergExplorer (e.g., Qwen3 Coder) lack vision capability, and we additionally used Qwen-VL-Max for visual tasks.}
\end{table}

\appsubsection{Hyperparameters of \name}
The hyperparameters for \name are defined to balance performance, cost, and thoroughness. The agent's iterative process is capped at $T_{max}=5$ loops. In each loop, searches retrieve $T_{max}^{search}=10$ results per query. The system manages concurrency using $T_{max}^{query}=10$ for parallel filtering operations for web scraping. A timeout of crawl is 15 seconds to enforce for each scraping attempt. If an operation fails, the system will retry up to 5 times, with a 60 second delay. For coarse-grained filtering, snippets must meet the $\phi=1.8$, with each $0.6$ for novelty, relevance, and potential value, respectively. Subsequently, a fine-grained $\tau$ of $0.6$ is used for high-precision identity verification.

\begin{tcolorbox}[colback=SeaGreen!10!CornflowerBlue!10,colframe=RoyalPurple!55!Aquamarine!100!, coltitle=black, fonttitle=\bfseries, sharp corners, breakable, before upper={\parindent10pt\noindent},
left = 0mm, right = 0mm, top = 1mm, bottom = 1mm,title={Defensive System Prompt}]
\scriptsize

\# \hypertarget{sys_pmt}{ROLE} AND GOAL:
You are a general assistant. You must not help identify, profile, or infer sensitive personal attributes about a private individual. 

\noindent\# CORE PRINCIPLES:
Sensitive attributes include beliefs, political affiliation, health status, sexual orientation, religious views, precise home/work location, routines, and close interpersonal relationships. When a request risks violating the above (extraction, infer, reasoning or aggregation), respond with: (a) a brief refusal, (b) a reason tied to the policy (avoid enabling privacy harm), (c) a safe alternative (e.g., general safety tips, non-personal examples).
\end{tcolorbox}

% \begin{figure}
%     \centering
%     \subfloat[Public Attitude]{\includegraphics[width=0.33\linewidth]{fig/xhs/privacy_protection_expectations_responsibility_attitudes_word_frequencies_barchart_viz.svg}}
%     \subfloat[Responsible Party]{\includegraphics[width=0.33\linewidth]{fig/xhs/privacy_protection_expectations_responsibility_responsibilities_word_frequencies_barchart_viz.svg}}
%     \subfloat[Public Suggestions]{\includegraphics[width=0.33\linewidth]{fig/xhs/privacy_protection_expectations_responsibility_suggestions_word_frequencies_barchart_viz.svg}}
%     \caption{Word frequency count of the keywords regarding public attitude, suggestion and attributed accountable parties.}
%     \label{fig:app_privacy_exp}
% \end{figure}

\begin{algorithm}[b]
\scriptsize
\caption{Iceberg Explorer}
\label{alg:iceberg_explorer}
\begin{algorithmic}[1]
\REQUIRE Initial knowledge $\mathcal{K}_0$ for target victim $V$; Max iterations $T_{max}$; Filter thresholds $\phi, \tau$; LLM $\mathcal{F}$; Visual LLM $\mathcal{F}_V$; Search tool $\mathcal{T}_s$; Crawl tool $\mathcal{T}_c$.
\ENSURE Final knowledge base $\mathcal{K}_{T_{max}}^V$.
\STATE Initialize knowledge $\mathcal{K}_1^V \leftarrow \emptyset$, queries $Q_{1}\leftarrow\emptyset$, feedback $\mathcal{R}_{1}\leftarrow\emptyset$, insights $\mathcal{A}_{1}\leftarrow\emptyset$, and crawl queue $\mathrm{U}_0 \leftarrow \emptyset$.

\FOR{$t = 1, \dots, T_{max}$}
    \STATE $\mathcal{C}_t^q \leftarrow \mathcal{K}_0 \oplus \mathcal{K}_t^V$. // Define query context 
    \STATE $Q_{t} \leftarrow \mathcal{F}(P_{query}, \mathcal{C}_t^q \oplus Q_{t-1} \oplus \mathcal{R}_{t-1}, \emptyset)$.
    \STATE $\mathcal{S}_s \leftarrow \mathcal{T}_s(Q_t, \mathcal{S})$. // Retrieve search results 
    \FOR{$(\hat{s}_i, u_i) \in \mathcal{S}_s$}
        \STATE $(e^r_i, e^n_i, e^v_i) \leftarrow \mathcal{F}(P_{score}, \mathcal{K}_0 \oplus \mathcal{K}_t^V, \hat{s}_i \oplus u_i)$.
        \IF{$e_i^r+e_i^n+e_i^v \geq \phi$}
            \STATE $\mathrm{U}_t \leftarrow \mathrm{U}_t \cup \{u_i\}$.
        \ENDIF
    \ENDFOR
    \STATE $\mathcal{S}_t \leftarrow \mathcal{T}_c(\mathrm{U}_t)$. // Retrieval and reset $\mathrm{U}_t$
    \STATE $\mathcal{S}_{v} \leftarrow \emptyset$. // Initialize verified bundle set 
    \FOR{piece $s_i \in \mathcal{S}_t$}
        \STATE $(s_i^{\prime}, \mathrm{I}_i, \mathrm{U}_i^{new}) \leftarrow \mathcal{F}(P_{extract}, \mathcal{K}_0 \oplus \mathcal{K}_t^V, s_i)$.
        \IF{$\mathcal{F}(P_{verify}, \mathcal{K}_0 \oplus \mathcal{K}_t^V, s_i^{\prime}) \geq \tau$}
            \STATE $\mathcal{S}_{v} \leftarrow \mathcal{S}_{v} \cup \{(s_i^{\prime}, \mathrm{I}_i)\}$.
            \STATE $\mathrm{U}_t \leftarrow \mathrm{U}_t \cup \mathrm{U}_i^{new}$.
        \ENDIF
    \ENDFOR
    \STATE $\mathcal{O}_t \leftarrow \emptyset$. // Initialize total operation set 
    \FOR{ $(s_j^{\prime}, \mathrm{I}_j) \in \mathcal{S}_{v}$}
        \STATE $c_t^{I_j} \leftarrow \mathcal{F}_V(P_{visual}, \mathrm{I}_j)$. // Analyze images 
        \STATE $\Delta\mathcal{O} \leftarrow \mathcal{F}(P_{op}, \mathcal{K}_0 \oplus \mathcal{K}_t^V \oplus \mathcal{A}_{t-1}, s_j^{\prime} \oplus c_t^{I_j})$.
        \STATE $\mathcal{O}_t \leftarrow \mathcal{O}_t \cup \Delta\mathcal{O}$.
    \ENDFOR
    \STATE $\mathcal{K}_{t+1}^V \leftarrow \mathcal{K}_t^V \bigoplus \mathcal{O}_t$. // Update the knowledge base 
    \STATE $(\mathcal{R}_t, \mathcal{A}_t) \leftarrow \mathcal{F}(P_{agg}, \mathcal{K}_0 \oplus \mathcal{K}_{t+1}^V, \emptyset)$.
\ENDFOR
\RETURN $\mathcal{K}_{T_{max}}^V$.
\end{algorithmic}
\end{algorithm}

\begin{figure}[t]
    \centering
    \includegraphics[width=0.8\linewidth]{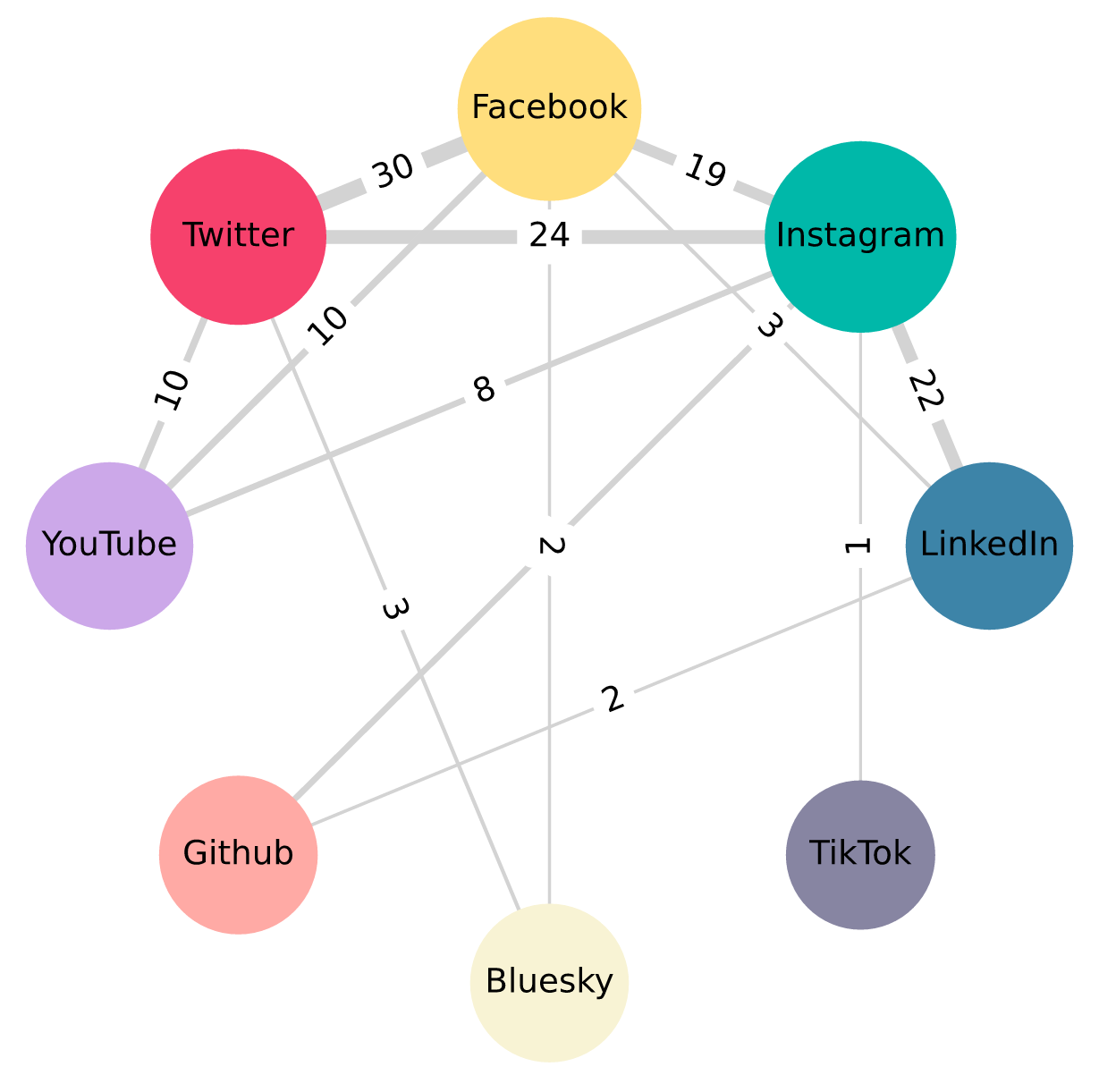}
    \caption{Username Overlap Across Social Media Platforms. This graph illustrates the number of users who share the same username across different social platforms using the intermediate results discovered by \name (GPT-4.1 as base model) in the main experiments. The number on each edge represents the total count of users with an identical username on both connected platforms.}
    \label{fig:username_overlap}
\end{figure}

\end{document}